\documentclass[showpacs,amssymb,preprint,preprintnumbers 
,nofootinbib,superscriptaddress]{revtex4-2}

\usepackage{amsmath}
\usepackage{graphicx}
\graphicspath{{Figures/}}
\usepackage{latexsym}
\usepackage{amsfonts}
\usepackage{url,hyperref}
\usepackage{bm}
\usepackage{textcomp}
\usepackage[dvipsnames]{xcolor} 
\usepackage{bbm}
\usepackage{slashed}
\usepackage{caption}
\usepackage{epstopdf}
\usepackage{subcaption}
\usepackage{placeins}
\usepackage{array}
\usepackage{cancel}
\usepackage{soul}       
\usepackage{ulem}       
\setstcolor{blue}    	
\usepackage{verbatim}

\newcommand{\beq}{\begin{equation}}
\newcommand{\eeq}{\end{equation}}
\def\bea{\begin{eqnarray}}
\def\eea{\end{eqnarray}}

\hypersetup{colorlinks=true,linkcolor=Mulberry,citecolor=OrangeRed,filecolor=OrangeRed,
  urlcolor=Mulberry}
\renewcommand{\theequation}{\arabic{section}.\arabic{equation}}
\counterwithin*{equation}{section}

\usepackage{etoolbox}

\makeatletter
\patchcmd{\frontmatter@abstract@produce}
  {\vskip200\p@\@plus1fil
   \penalty-200\relax
   \vskip-200\p@\@plus-1fil}
  {}
  {}
  {}
\makeatother

\usepackage{amstext} 
\usepackage{array}   
\newcolumntype{C}{>{$}c<{$}} 
\newcolumntype{L}{>{$}l<{$}} 
\usepackage{multirow}

\begin{document}

\title{Asymptotic quasinormal frequencies of different spin fields\\ in $d$-dimensional spherically-symmetric black holes}


\author{Chun-Hung Chen}
\email{chun-hungc@nu.ac.th}
\affiliation{The Institute for Fundamental Study, \\
Naresuan University, Phitsanulok 65000, Thailand.}

\author{Hing-Tong Cho}
\email{htcho@mail.tku.edu.tw}
\affiliation{Department of Physics, Tamkang University,\\
Tamsui District, New Taipei City, Taiwan 25137.}

\author{Anna Chrysostomou}
\email{annachrys97@gmail.com}
\affiliation{Department of Physics, University of Johannesburg, \\
PO Box 524, Auckland Park 2006, South Africa.}
\affiliation{Universit{\'e} de Lyon, F-69622 Lyon, France: Universit{\'e} Lyon 1,\\
Villeurbanne CNRS/IN2P3, UMR5822, Institut de Physique des 2 Infinis de Lyon.}

\author{Alan S. Cornell}
\email{acornell@uj.ac.za}
\affiliation{Department of Physics, University of Johannesburg, \\
PO Box 524, Auckland Park 2006, South Africa.}


\begin{abstract}
\par While Hod's conjecture is demonstrably restrictive, the link he observed between black hole (BH) area quantisation and the large overtone ($n$) limit of quasinormal frequencies (QNFs) motivated intense scrutiny of the regime, from which an improved understanding of asymptotic quasinormal frequencies (aQNFs) emerged. A further outcome was the development of the ``monodromy technique", which exploits an anti-Stokes line analysis to extract physical solutions from the complex plane. Here, we use the monodromy technique to validate extant aQNF expressions for perturbations of integer spin, and provide new results for the aQNFs of half-integer spins within higher-dimensional Schwarzschild, Reissner-Nordstr{\"o}m, and Schwarzschild (anti-)de Sitter BH spacetimes. Bar the Schwarzschild anti-de Sitter case, the spin-1/2 aQNFs are purely imaginary; the spin-3/2 aQNFs resemble spin-1/2 aQNFs in Schwarzschild and Schwarzschild de Sitter BHs, but match the gravitational perturbations for most others. Particularly for Schwarzschild, extremal Reissner-Nordstr{\"o}m, and several Schwarzschild de Sitter cases, the application of $n \rightarrow \infty$ generally fixes $\mathbb{R}e \{ \omega \}$ and allows for the unbounded growth of $\mathbb{I}m \{ \omega \}$ in fixed quantities.      
\end{abstract}


\date{\today}

\maketitle

%
%
\section{\label{sec:intro}Introduction}

\par Within the highly-damped regime, the classical oscillations of a perturbed black hole (BH) known as quasinormal frequencies (QNFs) exhibit a linear, unbounded growth in the imaginary component $\left( \mathbb{I}m \{\omega \} \right)$ accompanied by a finite value for the real component $\left( \mathbb{R} e \{ \omega \} \right)$ \cite{refNollert1999,refKokkotasRev,refBertiCardoso,refKonoplyaZhidenkoReview,refLeaver1985,refLeaver1986a,refLeaver1986b,refCardosoLemos2001,CardosoLemosYoshida,Berti2003,refAnderssonLinnaeus,refAndersson1993,refAnderssonHowls,refMotl,refMotlNeitzke,refNatarioSchiappa,Das2005,Ghosh2006,Birmingham2003,Daghigh2005generic,Daghigh2006universal,Daghigh2006RNsmallQ,Daghigh2007eRN,Daghigh2011QMBH,Cho2006,Lopez-Ortega2006,Casals2018,Dreyer2003,refMaggiore2008,Skakala2010}. To first order, we can express this as 
\begin{equation} \label{eq:aQNFoffset}
\lim_{n \rightarrow +\infty} \omega \thicksim \left[{\text{offset}} \right] + in \left[ {\text{gap}} \right] \;,
\end{equation}
\noindent where the ``offset" refers to the frequency of the emitted radiation and the ``gap" represents a quantised increment in the inverse relaxation time corresponding to the surface gravity. 

\par This expression for the asymptotic quasinormal frequency (aQNF) garnered specific interest for its speculated link to a quantum theory of gravity, initiated by Hod in Ref. \cite{Hod1998}. Motivated by Bohr's correspondence principle and the quantised BH area spectrum proposed by Bekenstein and Mukhanov \cite{refBekenstein1972,refMukhanov1986,refBekenstein1995}, Hod interpreted Nollert's numerical result of
\begin{equation}
\mathbb{R} e \bigg \{ \lim_{n \rightarrow +\infty} \omega \bigg \} \; \approx \; \frac{0.0437123}{M} \;\; \; \; \xrightarrow{\text{Hod}}\; \frac{\ln 3}{8 \pi M} 
\end{equation}
\noindent for the Schwarzschild BH under Planck units ($c=\hbar=G=1$) \cite{refNollert1993} as fundamental to the scaling of a ``quantum Schwarzschild BH" area. Refs. \cite{refMotl,refAnderssonHowls,refMotlNeitzke,Birmingham2003,Daghigh2005generic,Daghigh2006universal}, with the application of different analytical methods, confirmed this $\ln 3/8\pi M$ result for 4D and higher-dimensional Schwarzschild BHs. On the basis of statistical arguments and the established relationships between BH entropy and surface area, Hod also derived a minimum equidistant spacing of 
\begin{equation}
\Delta S = \ln 3
\end{equation}
\noindent for the Bekenstein-Hawking entropy spectrum. Hod considered his analysis applicable to all BHs of the Kerr-Newman ``family" \cite{Hod1998}.

\par That classical oscillations could provide insight into quantum behaviour appeared to augur advances for a theory of quantum gravity  \cite{refKokkotasRev,refBertiCardoso,refKonoplyaZhidenkoReview}. Though Hod’s conjecture gained traction for several years, a possible link to a quantum theory of gravity was quickly proven tenuous when the $\ln 3$ result did not emerge for the 4D Reissner-Nordstr{\"o}m BH in Refs. \cite{refAnderssonHowls,refMotlNeitzke} nor within other multi-horizon BHs inclusive or exclusive of a cosmological constant \cite{refNatarioSchiappa,Das2005,Ghosh2006,Cho2006,Lopez-Ortega2006,Casals2018,refCardosoLemos2003,refCardosoNatarioSchiappa,Daghigh2007eRN,Daghigh2008SAdS,Arnold2013,Arnold2014}. Furthermore, the authors of Refs. \cite{Cho2006,Lopez-Ortega2006,Casals2018} determined that $\ln 3$ was not universal even to Schwarzschild aQNFs, as this result could only be obtained for scalar and gravitational perturbations.

\par Despite this, research continued into aQNFs. For example, a further iteration of Hod's conjecture by Maggiore \cite{refMaggiore2008} caste BH perturbations as a collection of damped harmonic oscillations, with the real frequency defined as $\omega_0 = \sqrt{\omega_I^2 + \omega_R^2}$ and a subsequent result of 
\begin{equation}
\Delta S = 2 \pi 
\end{equation}
\noindent for the Bekenstein-Hawking entropy spectrum. As discussed in Refs. \cite{refKonoplyaZhidenkoReview,Daghigh2008SAdS}, this result is more promising as it holds true for all perturbations and in a variety of BH spacetimes \cite{Medved2008}. Interest in accessing the quantum regime through QNFs has recently led to the study of quantum corrected BHs, where only the highly-damped QNF spectra are altered; QNFs with lesser damping, in contrast, resemble the QNF spectra of classical BHs \cite{Daghigh2011QMBH,Daghigh2020_ScalarPertRegBH}.

\par Spurred initially by these early conjectured insights into quantum gravity, two comprehensive applications of analytical aQNF computational methods were produced: (i) Ref. \cite{Cho2006}, where Cho's implementation of the modified ``phase-integral" method, as put forth by Andersson and Howls \cite{refAnderssonHowls}, yielded the spin $s \in \{ 0,1/2,1,3/2,2 \}$ QNFs in 4D Schwarzschild and Reissner-Nordstr{\"o}m (extremal and non-extremal) BHs; (ii) Ref. \cite{refNatarioSchiappa}, where Nat{\'a}rio and Schiappa obtained aQNF expressions for gravitational perturbations in $d$-dimensional Minkowski and (anti-)de Sitter (AdS) BH spacetimes via the ``monodromy technique" of Ref. \cite{refMotlNeitzke}. Both methods exploit analytic continuity, with the tracing of Stokes and anti-Stokes lines to extract a physical solution from the complex plane. They differ in that the former employs a very careful ``fine-structure" analysis while the latter utilises a more direct approach reliant almost exclusively on the anti-Stokes lines behaviour. These investigations motivate the present work, where we exploit the more flexible methodology used by Nat{\'a}rio and Schiappa to extend the results of Ref. \cite{Cho2006} to higher-dimensional Schwarzschild, Schwarzschild (A)dS, and Reissner-Nordstr{\"o}m BH spacetimes.

\par To do so, we begin with a thorough review of the monodromy technique in section \ref{sec:NatSchiap}. We describe the underlying principles of the method and how it is adjusted to account for BH charge and a non-zero cosmological constant. We also confirm and extend the unified treatment employed in Ref. \cite{Ghosh2006} for the QNFs of Schwarzschild and Schwarzschild dS BHs: although a clear distinction is evident between the behaviour of the quasinormal mode (QNM) potentials within the Schwarzschild ``family" and Reissner-Nordstr{\"o}m ``family" of BH spacetimes near the origin, irrespective of $\Lambda$, we find that the nature of the cosmological constant dictates behaviour near spatial infinity. This observation manifests also in section \ref{sec:fields}, where we study the known effective QNM potentials of various spin in order to extract the field contribution to the aQNF expressions. Finally, in section {\ref{sec:aQNFs}} we demonstrate how the generalised expressions for the aQNFs and the field contributions derived in sections \ref{sec:NatSchiap} and \ref{sec:fields}, respectively, yield the aQNFs of the BHs of interest. Therein, we supply the new aQNFs for fields of half-integer spins within the BHs of interest. Insights, conclusions, and future directions are summarised in section \ref{sec:conc}.

\section{\label{sec:NatSchiap} The monodromy technique: a review}

\par For a perfectly isolated, static, and spherically-symmetric BH of dimension $d \geq 3$, the metric function $f(r)$ is given by
\begin{equation} \label{eq:f(r)}
f(r) = 1 - \frac{2 \mu}{r^{d-3}} + \frac{\vartheta^2}{r^{2d-6}} - \lambda r^2 \;,
\end{equation}
\noindent which parametrises the Arnowitt-Deser-Misner (ADM) BH mass ($M$) and charge ($Q$), as well as the cosmological constant ($\Lambda$) via 
\[ \mu = \frac{8 \pi G_d}{(d-2) \; \Omega_{d-2}} \;M  \;, \hspace{0.5cm} \vartheta^2 = \frac{8 \pi G_d}{(d-2) (d-3)} \; Q^2 \;, \hspace{0.5cm} \text{and} \hspace{0.5cm} \lambda = \frac{2\Lambda}{(d-2)(d-1)} \;,\]
\noindent respectively. $G_d$ indicates the gravitational constant for $d$-dimensional spacetimes, and the area of a unit $(d-2)$-sphere is given by $\Omega_{d-2}$. Minkowski, dS, and AdS spacetimes are characterised by $\lambda = 0$, $\lambda > 0$, and $\lambda <0$, respectively \cite{refIKschwarz1,refIKschwarz2,refIKrn,refIKchap6}.

\par To describe the damped perturbations thereof, we can exploit the stability analyses of Refs. \cite{refRW,ZerMon1,ZerMon2,ZerMon3,ZerMon4,refIKschwarz1,refIKschwarz2,refIKrn,refIKchap6} which encapsulate QNM behaviour. Following the notation of Refs. \cite{refIKschwarz1,refIKschwarz2,refIKrn,refIKchap6}, we can express the QNM in a variable-separable form,
\begin{equation}
\Psi (x^{\mu}) = \sum_{\ell, m} 
\frac{ \Phi(r) e^{+i \omega t}}{r^{(d-2)/2}} \; Y_{\ell m} (\theta_i) \;, 
\end{equation}
\noindent where the angular components are given by the hyper-spherical harmonics for $d-2$ angles $Y_{\ell m} (\theta_i)$, and the radial dependence may be cast into a ``Schr{\"o}dinger-like" ordinary differential equation, 
\begin{equation} \label{eq:ode}
\left[ -\frac{d^2}{dx^2} + V[r(x)] \right] \Phi(x) = \omega^2 \Phi (x)  \;.
\end{equation} 
\noindent Here, $V[r(x)]$ is the effective potential, $\omega$ is the QNF, and $x$ is the ``tortoise coordinate" generally defined as
\begin{equation} \label{eq:xdef}
x = \int \frac{dr}{f(r)} \;.
\end{equation}
\noindent We observe that $x=x[r]$ serves as a bijection from $(r_{_H},+\infty)$ to $(-\infty,+\infty)$ for asymptotically flat BH spacetimes \cite{refDecanini2011}, where $r=r_{_H}$ refers to the BH event horizon. Note that in the case of dS BH spacetimes, inclusive of the cosmological horizon $r_{_C}$, the bijection maps from $(r_{_H},r_{_C})$ to $(-\infty,+\infty)$; for AdS BH spacetimes, the bijection is from $(r_{_H},+\infty)$ to $(-\infty,0)$.

\par Near the horizon, we can express Eq. (\ref{eq:xdef}) as
\begin{equation} \label{eq:xdeflog}
x \thicksim \int \frac{dr}{(r-r_{_H})f'(r_{_H})}= \frac{1}{f'(r_{_H})} \log( r - r_{_H}) \;,
\end{equation}
\noindent such that $x \rightarrow - \infty$ logarithmically when approaching $r_{_H}$. For a non-degenerate horizon, with $r_{_H}$ as a simple zero of $f(r)$, $f'(r_{_H})=2k_{_H}$ where $k_{_H}=2 \pi T_{_H}$ is the surface gravity at the horizon defined in terms of the Hawking temperature $T_{_H}$ \cite{refNatarioSchiappa}. 

\par Near the event horizon, QNMs are purely ingoing:
\begin{equation} \label{eq:BCin}
\Phi (x) \sim e^{+i \omega x} \;, \hspace{1cm} x \rightarrow -\infty \; (r \rightarrow r_{_H}) \;.
\end{equation} 
\noindent At spatial infinity, QNMs are purely outgoing. However, this condition manifests differently based on the nature of the cosmological constant:
\begin{eqnarray}  \label{eq:BCout}
\Phi(x) \sim 
\begin{cases}
e^{-i \omega x} \;, \; \; & \; \; x \rightarrow +\infty \; (r \rightarrow +\infty) \;, \; \; \lambda = 0\;, \\
e^{-i \omega x}\;, \; \; & \; \; x \rightarrow +\infty \; \;\;(r \rightarrow r_{_C}) \;, \; \; \;   \lambda > 0\;,  \\
\; \; 0 \;, & \; \; x \rightarrow +\infty \; (r \rightarrow +\infty) \;, \; \; \lambda < 0\;, \\
\end{cases}
\end{eqnarray}
\noindent where $r_{_C}$ denotes the cosmological horizon, as before. While there are a range of possible boundary conditions applicable to AdS contexts, we remain without a convincing \textit{a priori} argument for a universally applicable set (see Ref. \cite{refnewQNMs2020} for discussion against an uninformed reliance on a singular set of boundary conditions in AdS BH spacetimes). Here, however, we choose to maintain Dirichlet boundary conditions near spatial infinity when considering AdS BH spacetimes, in keeping with Refs. \cite{refNatarioSchiappa,Das2005,Ghosh2006,Lopez-Ortega2006,refCardosoLemos2001,refCardosoNatarioSchiappa,Daghigh2007eRN,Daghigh2008SAdS}. 

\par From the boundary conditions of Eqs. (\ref{eq:BCin}) and (\ref{eq:BCout}), the system is shown to be inherently dissipative: energy is lost at the boundaries and cannot be reintroduced. In a manner reminiscent of normal mode analyses, it is the implementation of these physically-motivated boundary conditions that discretises the QNF spectrum. The discrete QNF can then be decomposed into its real and imaginary part, such that ${\mathbb{R}}e \{ \omega  \}$ represents the physical frequency of oscillation and ${\mathbb{I}}m \{ \omega  \}$  denotes the damping. Its dependencies include $n$ (the ``overtone" number) and $\ell$ (the ``multipolar" or angular momentum number), the asymptotic limits of which represent regimes of interest in QNM studies. We explore the properties of QNFs subjected to $\ell \rightarrow \infty$ within Ref. \cite{refOurLargeL}; here, we focus on the $n \rightarrow \infty$ regime.   

\par As emphasised by Daghigh \textit{et al.} in Ref. \cite{Daghigh2012WKBvalid}, ``aQNFs" are those for which $\vert \omega \vert \rightarrow \infty$ while ``highly damped QNFs" obey the condition $\vert \omega \vert \approx \vert \mathbb{I}m \{ \omega \} \vert \gg \vert \mathbb{R}e \{ \omega \} \vert$ as $n\rightarrow \infty$. This nuanced distinction becomes relevant in the computation of aQNFs within AdS spacetimes, as noted in Ref. \cite{Ghosh2006} and the numerical works referenced therein, where the asymptotic limit is described as the regime in which $\mathbb{R}e \{ \omega \} \approx \mathbb{I}m \{ \omega \}$ \cite{refNatarioSchiappa,Ghosh2006,Daghigh2008SAdS}. Consequently, there exists a slight discrepancy in the decomposition of the QNF within the large overtone limit: in Minkowski and dS BH spacetimes,
\begin{equation} \label{eq:omega}
\omega  = \omega_R + in\omega_I \;, \hspace{1 cm} \omega_R, \omega_I \in {\mathbb{R}} \;,
\end{equation}
whereas in AdS BH spacetimes,
\begin{equation} \label{eq:omegaAdS}
\omega  = n( \omega_R + i\omega_I ) + \omega_0 \;, \hspace{1 cm} \omega_R, \omega_I \in {\mathbb{R}} \;.
\end{equation}
\noindent In setting $n \rightarrow +\infty$, $\omega$ in Eq. (\ref{eq:omega}) for flat and dS BH spacetimes approximates to a purely imaginary number; for $\omega$ in Eq. (\ref{eq:omegaAdS}) in AdS BH spacetimes, real and imaginary parts contribute in equally large magnitudes.

\par As such, the analytical calculation of the aQNF generally involves the tracing of a closed global contour from the origin to infinity in the complex $r$-plane, enclosing regular singular point(s). In the case of the monodromy technique, these singular points are defined by determining the complex roots of $f(r)=0$; they represent the physical and ``fictitious" \cite{refMotlNeitzke,refNatarioSchiappa} horizons of the BH spacetime of interest. Through a comparison of the ``global" monodromy computed from the contour to infinity with the ``local" monodromy around the enclosed singular point(s) \cite{Birmingham2003}, the aQNF may be extracted.

\par In subsection \ref{subsec:principles}, we describe the underlying requirements for this calculation and the mathematical features we exploit. Note that we adhere to the terminology utilised in Ref. \cite{Daghigh2005generic} in our description of these contours: we consider anti-Stokes lines as lines upon which $\omega x$ is purely real (i.e. $\mathbb{I}m \{ \omega x \} = 0$) and Stokes lines as lines upon which $\omega x$ is purely imaginary (i.e. $\mathbb{R}e \{ \omega x \} = 0$). This is to improve comparison with the phase-integral technique, as used in Refs. \cite{refAndersson1993,Cho2006,Daghigh2011QMBH,Daghigh2008SAdS,Daghigh2006RNsmallQ,Daghigh2007eRN,Daghigh2012WKBvalid}.

\subsection{Mathematical background \label{subsec:principles}} 
\par The method put forth in Ref. \cite{refMotlNeitzke} and extended in Refs. \cite{refNatarioSchiappa,Ghosh2006} exploit a form of analytic continuity predicated on the fact that any solution of an ordinary differential equation can be extended from its physical region to a solution on the complex plane through the introduction of a ``Wick" rotation \cite{Wick1954}. From the effect of $n \rightarrow +\infty$ on Eqs. (\ref{eq:omega}) and (\ref{eq:omegaAdS}), it is clear that a continuation from the physical region $r_{_H}<r<\infty$ to the whole complex $r$-plane is necessary. 

\par To accommodate this, the boundary conditions of Eq. (\ref{eq:BCout}) undergo a transformation. For the highly-damped modes of Minkowski and dS spacetimes, the outgoing boundary conditions can be explicitly restructured as ``monodromy boundary conditions" \cite{Ghosh2006} consequent of the extension to the complex plane. These can be clearly stipulated: $x \sim \infty$ is ``Wick rotated" to $\omega x \sim \infty$ \cite{refMotlNeitzke,Ghosh2006}, such that
\begin{eqnarray}  
 \label{eq:monodromyBCs}
\Phi(x) \sim e^{\mp i \omega x} \;, \hspace{0.7cm} 
\begin{cases}
\omega x \rightarrow \pm \infty \;, \; \; & \; \; \mathbb{R} e \{\omega \} > 0 \\
\omega x \rightarrow \mp \infty \;, \; \; & \; \; \mathbb{R} e \{\omega \} < 0 \;.
\end{cases}
\end{eqnarray}
\par Throughout this work, we impose ${\mathbb{R}}e \{ \omega \} > 0$. For ${\mathbb{R}}e \{ \omega \} < 0$, the Wick rotation is taken in the opposite direction; the method remains the same, albeit the contour drawn must be traced in the counterclockwise direction. The subsequent solution would then be the complex conjugate of that obtained under the original ${\mathbb{R}}e \{ \omega \} > 0$ condition \cite{refMotlNeitzke,refNatarioSchiappa}.

\par The imposition of $\vert \omega \vert \rightarrow + \infty$ naturally affects Eq. (\ref{eq:ode}): $V[r(x)]$ becomes negligible across the complex $r$-plane (except near singular points), such that
\begin{equation}
- \frac{d^2 \Phi}{dx^2} \approx \omega^2 \Phi (x) \;.
\end{equation}
\noindent The consequent solution is a superposition of plane waves,
\begin{equation} \label{eq:genQNM}
\Phi(x) \sim A_+ e^{+ i \omega x} + A_- e^{-i \omega x} \;. 
\end{equation}

\par Near the origin, the effective potential has been shown to retain a standard form for static and spherically-symmetric BH spacetimes, irrespective of the nature of the perturbing field \cite{refAndersson1993,refMotlNeitzke,Cho2006,Daghigh2005generic,Daghigh2011QMBH,Daghigh2008SAdS,Daghigh2006RNsmallQ,Daghigh2007eRN,Daghigh2012WKBvalid}:
\begin{equation} \label{eq:genpot}
V[r(x)] \bigg \vert_{r \rightarrow 0} \sim \frac{j^2 - 1 }{4x^2} \;,
\end{equation}
\noindent where $x$ is the complex form of the tortoise coordinate defined in Eq. (\ref{eq:xdef}). As we shall demonstrate in section \ref{sec:fields}, Eq. (\ref{eq:genpot}) is achieved by defining $x$ via Eqs. ({\ref{eq:tortst}})-({\ref{eq:tortads}}) according to the BH context; in sections \ref{sec:fields} and \ref{sec:aQNFs}, we show that this $j$ parameter is characterised by the perturbing field, and serves as the sole field contribution to the aQNF result.

\par Within the neighbourhood of $r \sim 0$,
\begin{equation}  \label{eq:tortst}
x[r]  \sim -\frac{1}{2\mu} \int dr \; r^{d-3} = -\frac{r^{d-2}}{ 2\mu (d-2)}
\end{equation}
\noindent and
\begin{equation} \label{eq:tortrn}
x[r]  \sim  \frac{1}{\vartheta^2} \int dr \; r^{2d-6} = \frac{r^{2d-5}}{ (2d-5)\vartheta^2} 
\end{equation}
\noindent for the Schwarzschild and Reissner-Nordstr{\"o}m BH ``families", respectively, regardless of the nature of the cosmological constant. 

\par Within the region of spatial infinity ($r \sim \infty$), consistency can be found in BH spacetimes inclusive of $\lambda$ rather than within a BH family. For both Schwarzschild and Reissner-Nordstr{\"o}m BHs in asymptotically-flat spacetimes, $f(r) \sim 1$ and
\begin{equation} \label{eq:tortflat}
x[r] \sim \int dr 1 = r \;.
\end{equation}
\noindent This corresponds to a vanishing potential. For dS and AdS BH spacetimes, $f(r) \sim - \lambda r^2$ for $r \sim \infty$, such that 
\begin{equation} \label{eq:tortads}
x[r] \sim 
\begin{cases} 
x_0 + \frac{1}{\lambda r} \; \; \; \; \;\text{for dS} \;\; \; (\lambda >0) \;, \\
x_0 - \frac{1}{\vert \lambda \vert r} \; \; \; \text{for AdS} \; (\lambda <0) \;.
\end{cases}
\end{equation}
\noindent The constant of integration $x_0 \in \mathbb{C}$ is introduced to ensure the choice of $x[r = 0] = 0$ remains fixed, as explained in Appendix C of Ref. \cite{refNatarioSchiappa}. The corresponding potential has a form much like that of Eq. (\ref{eq:genpot}),

\begin{equation} \label{eq:genpotAdS}
V[r(x)] \bigg \vert_{_{r \rightarrow \infty}^{\lambda \neq 0}} \sim \frac{\left(j^{\infty}\right)^2 - 1}{\; \; 4(x-x_0)^2} \;,
\end{equation}
\noindent where we use $j^{\infty}$ to identify the parameter as distinctly associated with the singular point at $r = \infty$. This approximation of the effective potential applies for both dS and AdS BH spacetimes, as $(x-x_0)^2 \sim (\pm \vert \lambda \vert r)^{-2}$ according to Eq. (\ref{eq:tortads}). 

\par With the approximated potentials of Eqs. (\ref{eq:genpot}) and (\ref{eq:genpotAdS}), Eq. (\ref{eq:ode}) within the asymptotic regions of $r \sim 0$ and $r \sim + \infty$, respectively, can then be solved through the introduction of Bessel functions of the first kind \cite{refmathbible}. The general QNM solution becomes a linear combination of these Bessel functions,
\begin{equation} \label{eq:genQNMBessel}
\Phi (x) \sim A_+ \sqrt{2 \pi \omega x} \; J_{+j/2} ( \omega x ) + A_- \sqrt{2 \pi \omega x} \; J_{-j/2} ( \omega x ) \;,
\end{equation} 
\noindent whose asymptotic expansions,
\begin{eqnarray}
J_{\pm j/2} ( \omega x ) & \sim & \sqrt{\frac{2}{\pi \omega x}} \; \cos \left[ \omega x - \frac{\pi}{4} (1 \pm j)\right] \;, \hspace{0.3cm} \omega x \gg +1 \label{eq:Jexpandpos} \;,\\
J_{\pm j/2} ( \omega x ) & \sim & \sqrt{\frac{2}{\pi \omega x}} \; \cos \left[ \omega x + \frac{\pi}{4} (1 \pm j)\right] \;, \hspace{0.3cm} \omega x \ll -1 \;, \label{eq:Jexpandneg}
\end{eqnarray} 
\noindent shall prove useful in the monodromy calculations  (see subsection \ref{subsec:aQNF}). When tracing the path along the contour, rotations can be incorporated into Bessel function solutions using a further asymptotic expansion,
\begin{equation} \label{eq:Jtwist}
J_{\pm j/2} (\omega x) = (\omega x)^{\pm j/2} \phi (\omega x) \; \; \Rightarrow \; \; J_{\pm j/2 } ( e^{i \theta} (\omega x) ) = e^{\pm i \theta}e^{\pm i j/2  } J_{\pm j/2 } ( e^{i \theta} \omega x ),
\end{equation}  
\noindent where $\phi(\omega x)$ is considered an even holomorphic function of $\omega x$, and $e^{i \theta} \omega x $ is real and positive \cite{refNatarioSchiappa,Daghigh2011QMBH}.    

\par Within the complex $r$-plane, we recognise that Eq. (\ref{eq:xdeflog}) holds in close proximity to a specific horizon of choice. This indicates a ``multi-valuedness" in the tortoise coordinate, and subsequently in the QNM solution. To avoid contending directly with this multi-valuedness, branch cuts are placed at singular points such that $x$ ``jumps" the discontinuity when $r=r_{_{H}}$. We describe the corresponding behaviour of $\Phi (x)$ as it subsequently traces a closed circular path around the singular point by introducing the monodromy \cite{refMotlNeitzke,refNatarioSchiappa}. 

\par Near the horizon, $V[r(x)] \sim 0$, which in turn leads to a plane-wave QNM solution. Thus, to determine this local monodromy for $\Phi \sim e^{\pm i \omega x}$, we consider a closed clockwise contour $\gamma \subset \mathbb{C}$ centred on $r=r_{_H}$ (i.e. a rotation of $2 \pi$ in the $r$-plane about the event horizon). From Eq. (\ref{eq:xdeflog}),
\begin{eqnarray}
\log (r-r_{_H}) \; &\rightarrow&  \log (r-r_{_H}) - 2 \pi i \nonumber \\
&\Rightarrow& x \sim \frac{1}{2 k_{_H}} (\log (r-r_{_H}) - 2 \pi i) \sim x - \frac{\pi i}{k_{_H}} \nonumber \\
&\Rightarrow&  e^{\pm i \omega x} \rightarrow e^{\pm i \omega \left( x - \frac{\pi i}{k_{_H}} \right)} = e^{\pm i \omega x} e^{\pm \frac{\pi \omega}{k_{_H}}} \nonumber \;,
\end{eqnarray}
\noindent which yields the local monodromy,
\begin{equation}
\mathfrak{M}_{\gamma, \;r_{_H}} [\Phi (x)]_{\ell} = e^{\pm \frac{\pi \omega}{k_{_H}}} \;. \label{eq:monodromyLocal}
\end{equation}
\par
\begin{figure}[t]
\centering
    \includegraphics[height=8cm]{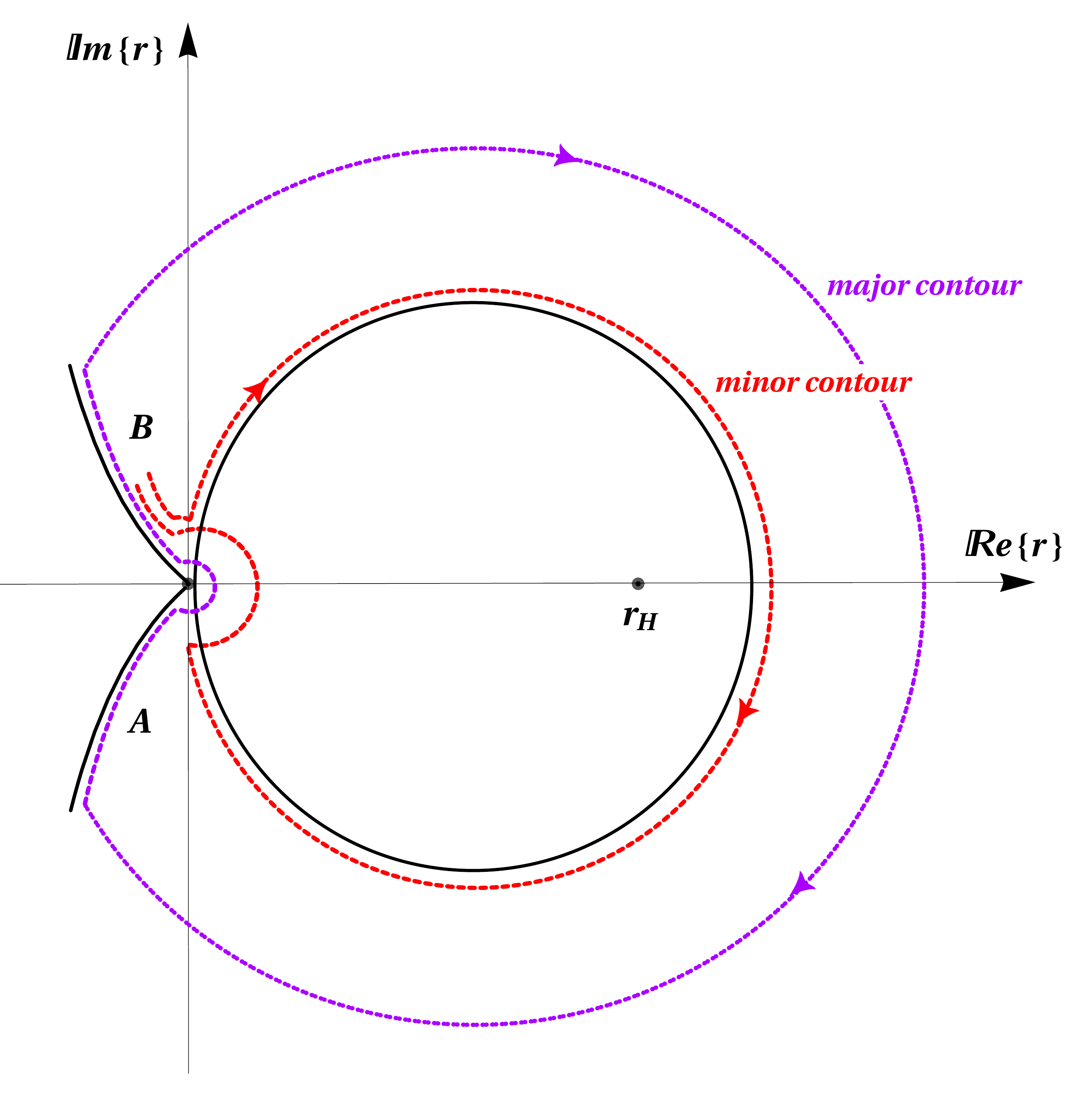}
    \caption{{\textit{Anti-Stokes line contour within the complex $r$-plane for a 4D single-horizon BH. The QNM solution oscillates with neither decay nor growth along the anti-Stokes lines.}}}
    \label{fig:MN}
\end{figure}

 For the global monodromy, 
\begin{equation}
\mathfrak{M}_{\gamma} [\Phi (x)] = \frac{\Phi_B}{\Phi_A} \; \mathfrak{M}_{\gamma, \;r_{_H}} [\Phi (x)]_g \;, \label{eq:monodromyGlobal}
\end{equation}
\noindent where $\Phi_A$ and $\Phi_B$ represent the solutions on branches $A$ and $B$ in the complex $r$-plane, shown in Fig. \ref{fig:MN} \cite{Ghosh2006,Daghigh2005generic}. The expressions for $\Phi_A$ is obtained by applying the appropriate boundary condition associated with branch $A$ to Eq. (\ref{eq:genQNMBessel}); $\Phi_B$ is determined by incorporating the rotation $\Phi_A$ undergoes to reach branch $B$, and then subjecting the resultant expression to outgoing boundary conditions. Note that the monodromy at $r = +\infty$ is zero.

\par Since the ``major" and ``minor" contours include the same singular points, their monodromies are equivalent \cite{refMotlNeitzke,refNatarioSchiappa,Ghosh2006,Daghigh2005generic}. To compute the aQNF, we equate Eqs. (\ref{eq:monodromyLocal}) and (\ref{eq:monodromyGlobal}) and solve for $\omega$.

\par The procedure we follow may be decomposed into two components:  the set-up of the complex $r$-plane and the $\mathbb{I}m \{ \omega x \} = 0$ anti-Stokes lines corresponding to the BH spacetime studied (subsection \ref{subsec:plane}), and the behaviour analysis of the QNM as it is traced along $\mathbb{I}m \{ \omega x \} = 0$ (subsection \ref{subsec:aQNF}). The first component thereby contextualises the problem while the second allows for the extraction of a solution.

\subsection{\label{subsec:plane}The complex \textit{r}-plane set-up}
\par With the extension of the QNM solution space for Eq. (\ref{eq:ode}) to the complex plane comes the need to establish the positions of singular points and the path of the contour traced. These are dictated exclusively by the BH spacetime and are independent of the perturbing field. As illustrated in Refs. \cite{refNatarioSchiappa,Ghosh2006}, that the aQNF in flat and dS spherically-symmetric BH spacetimes is governed by the same relationship between $\mathbb{I}m \{ \omega \}$ and $\mathbb{R}e \{ \omega \}$ dictates that the contour traced within the two complex $r$-planes must be very similar. This is shown explicitly in Figs. \ref{fig:STdS} and \ref{fig:RNdS}. The path traced in the AdS cases, on the other hand, differs noticeably from the flat and dS spacetimes, as exhibited in Fig. \ref{fig:SAdS}.
 
\par
\begin{figure}[t]
\begin{subfigure}{.2\textwidth}
  \centering
 	 \includegraphics[height=3.7cm]{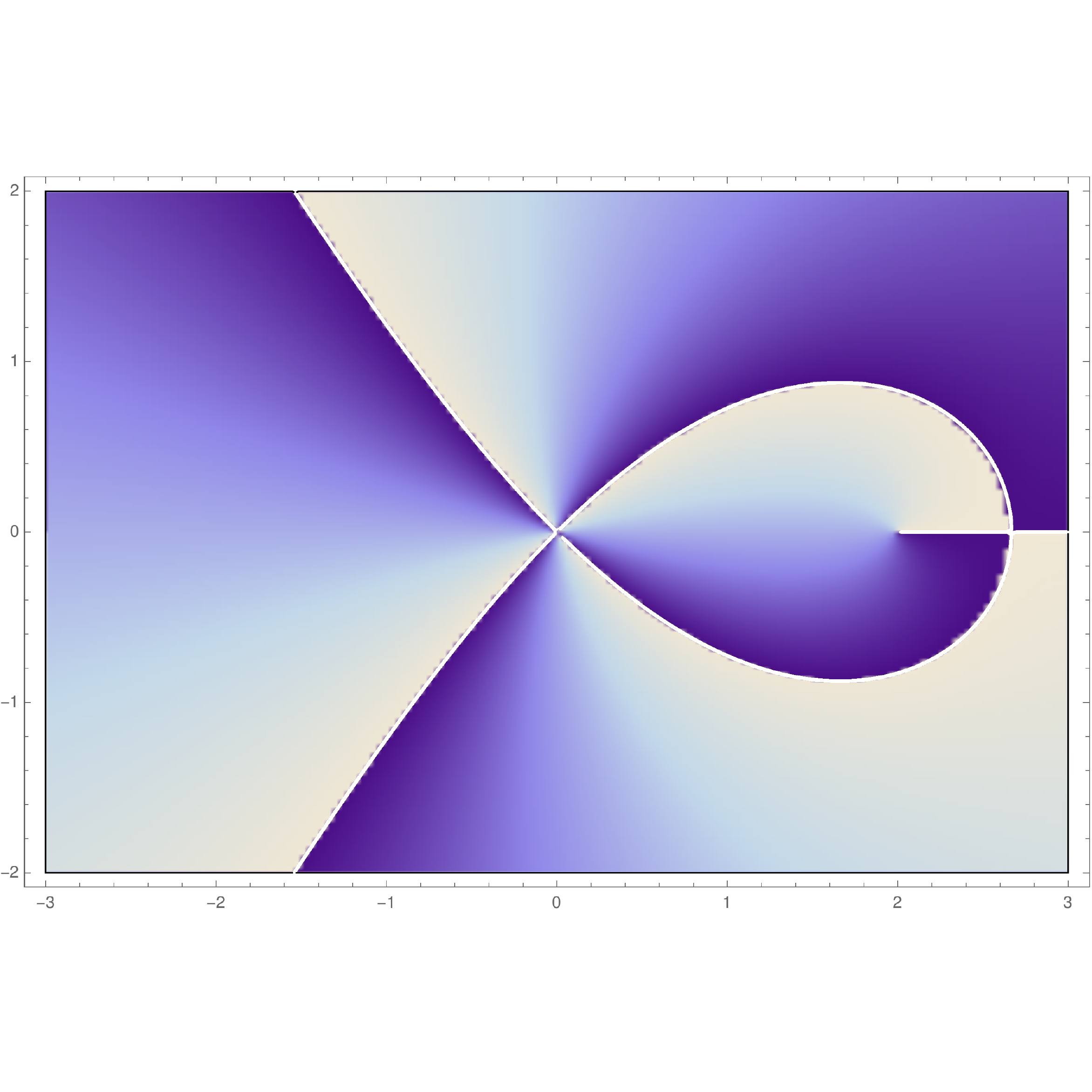}  
  \caption{d=4}
\end{subfigure}
\hfill
\begin{subfigure}{.2\textwidth}
  \centering
  		\includegraphics[height=3.7cm]{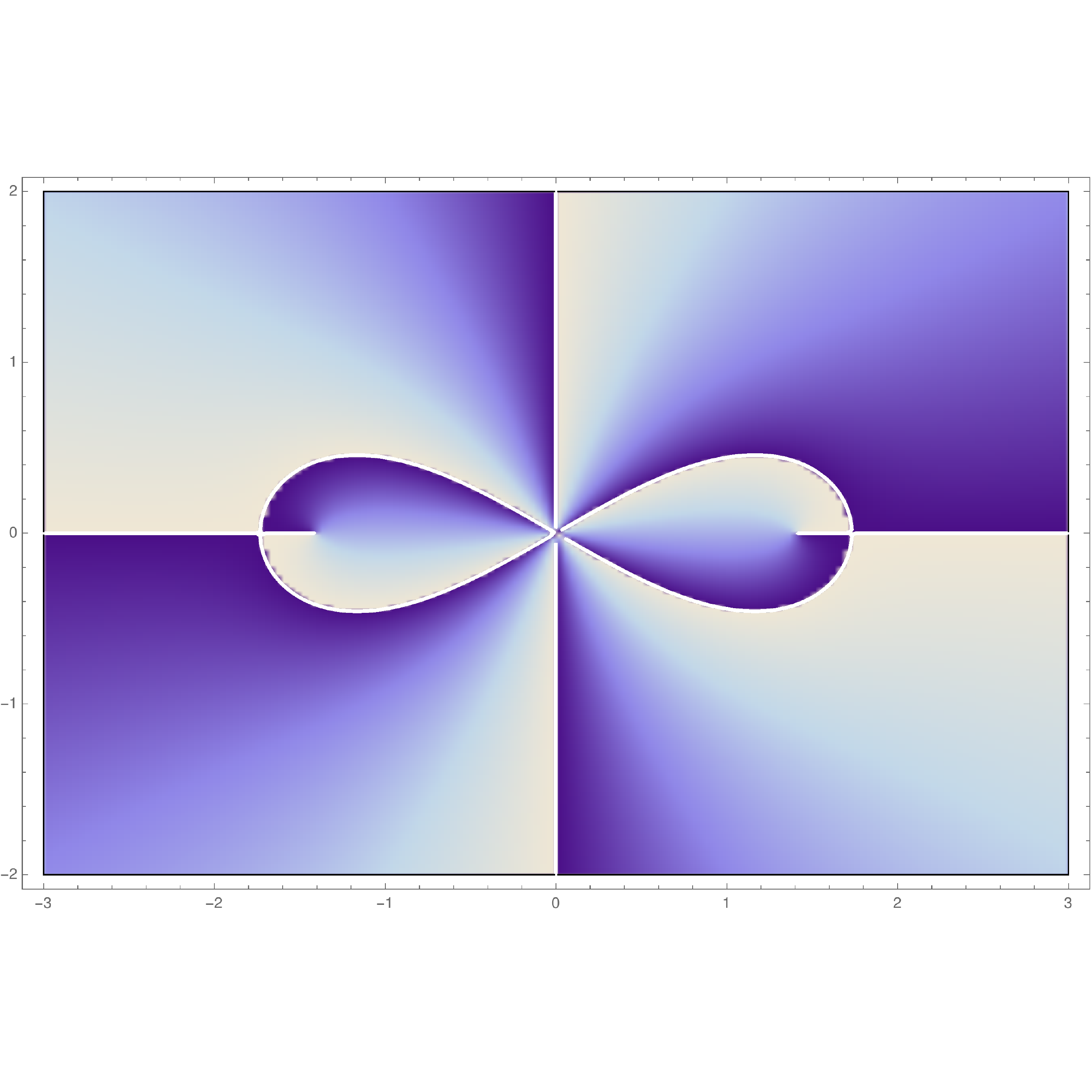}  
  \caption{d=5}
\end{subfigure}
\hfill
\begin{subfigure}{.2\textwidth}
  \centering
  		\includegraphics[height=3.7cm]{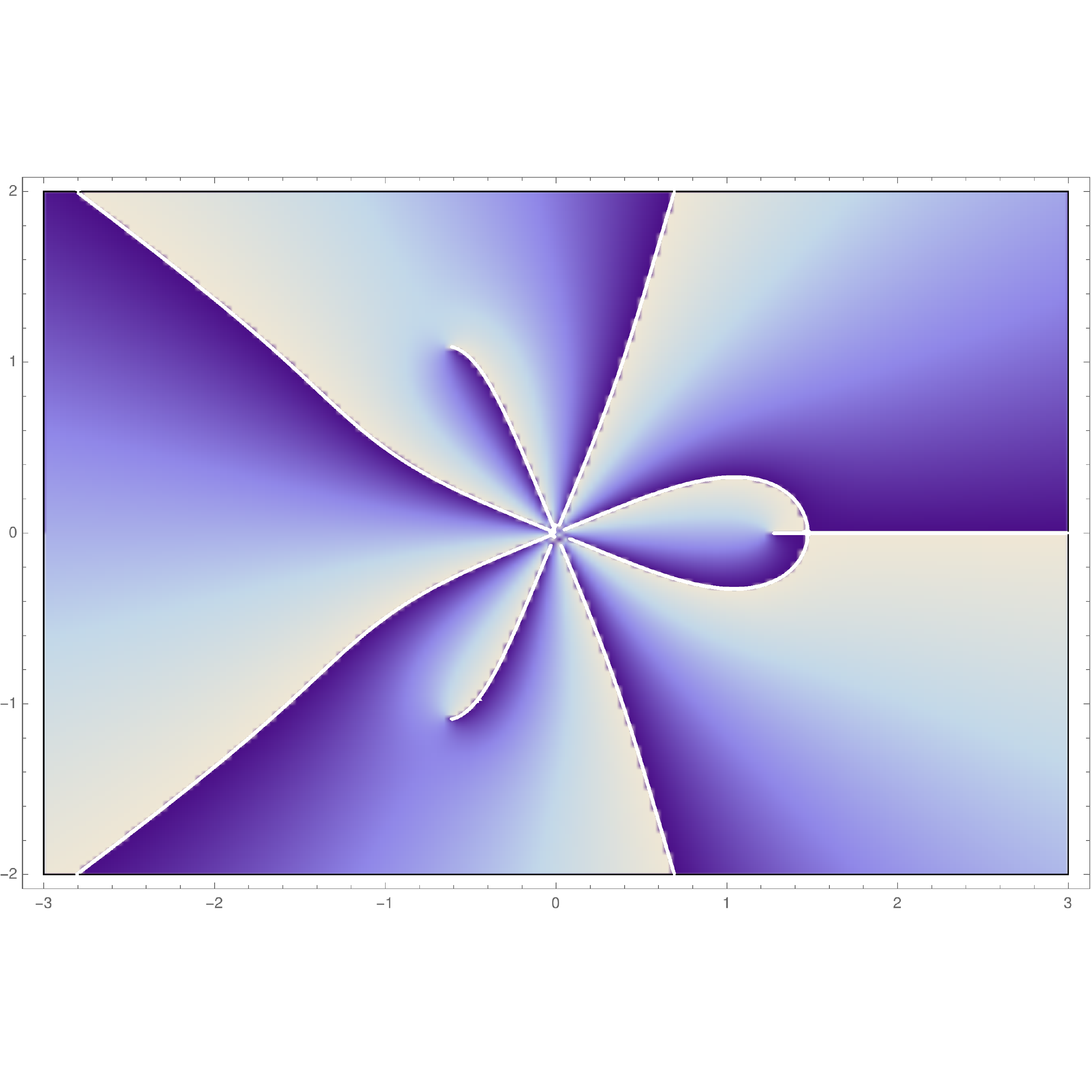}  
  \caption{d=6}
\end{subfigure}
\hfill
\begin{subfigure}{.2\textwidth}
  \centering
  		\includegraphics[height=3.7cm]{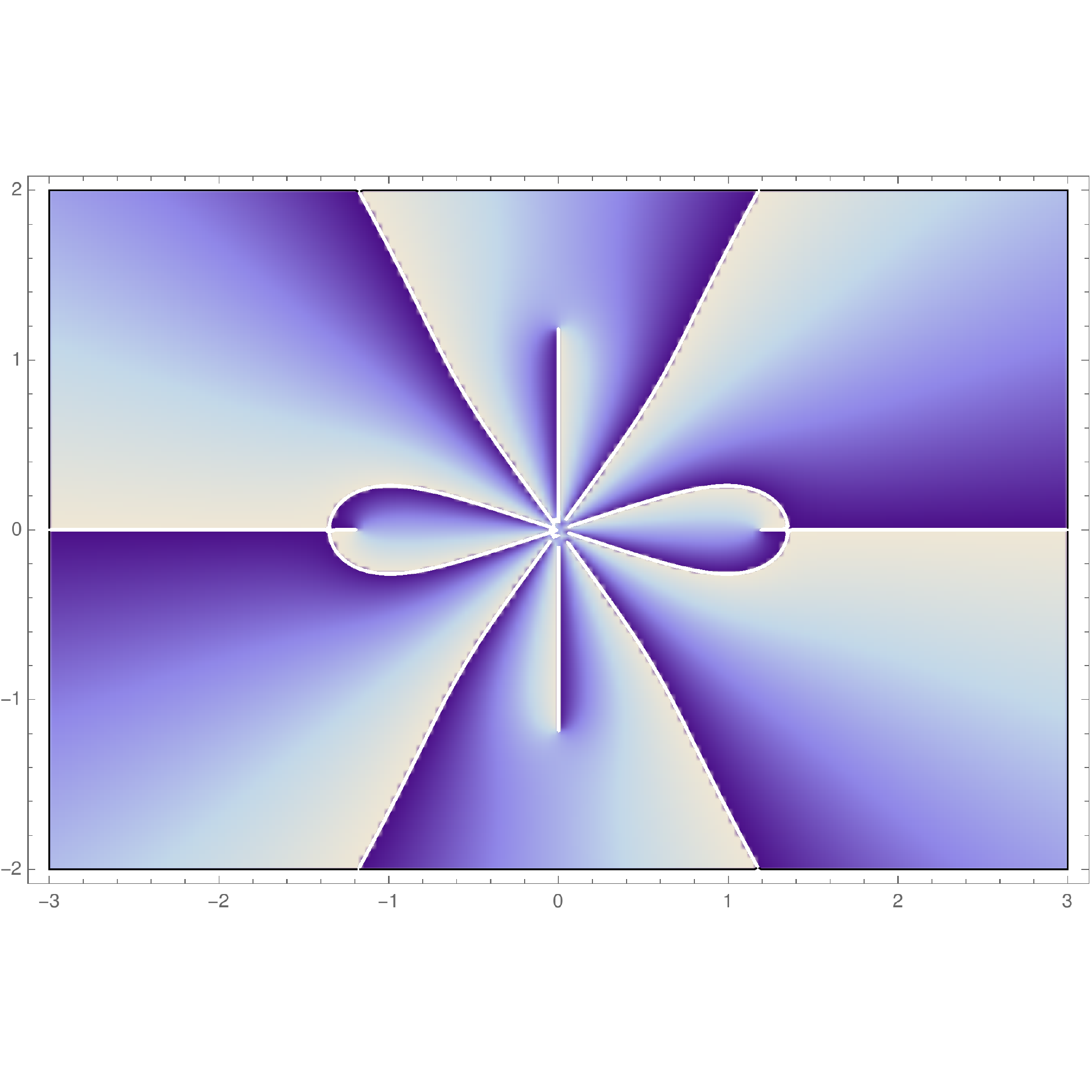}  
  \caption{d=7}
\end{subfigure}
\caption{{\textit{Numerically generated plots illustrating the behaviour of the anti-Stokes lines for Schwarzschild BHs ($ \mu = 1$).}}}
\label{fig:SchwarzNum}

\begin{subfigure}{.2\textwidth}
  \centering
 	 \includegraphics[height=3.7cm]{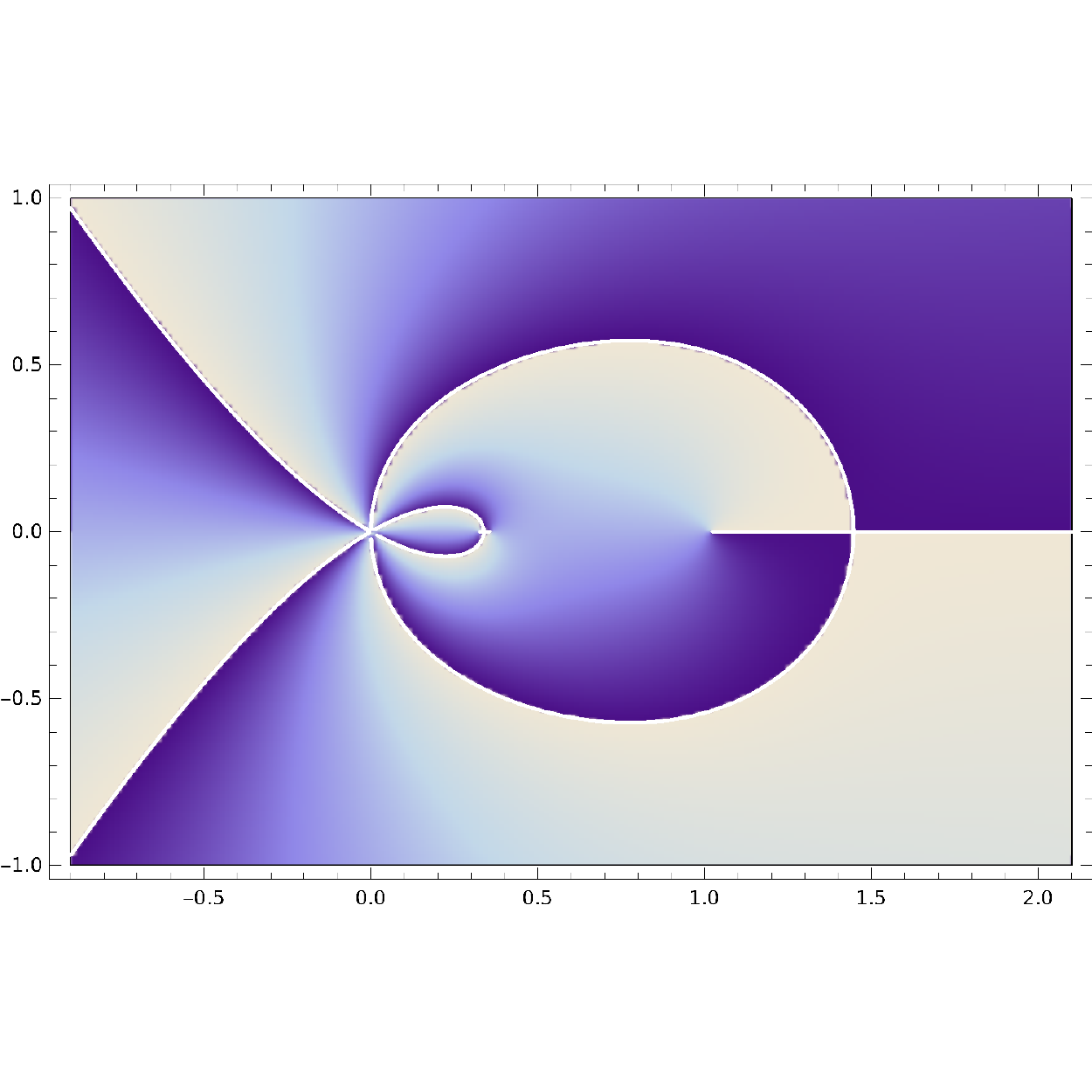}  
  \caption{d=4}
\end{subfigure}
\hfill
\begin{subfigure}{.2\textwidth}
  \centering
  		\includegraphics[height=3.7cm]{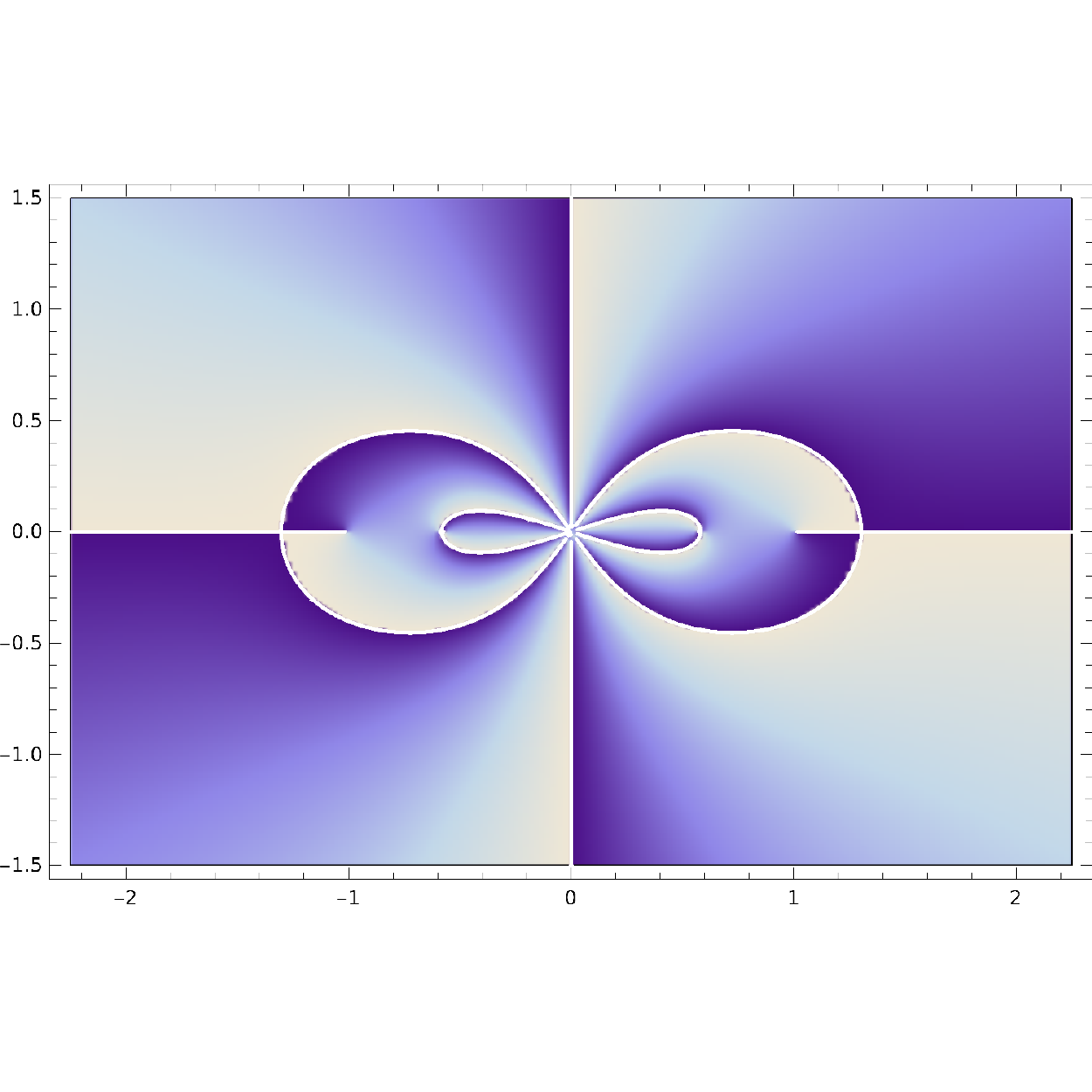}  
  \caption{d=5}
\end{subfigure}
\hfill
\begin{subfigure}{.2\textwidth}
  \centering
  		\includegraphics[height=3.7cm]{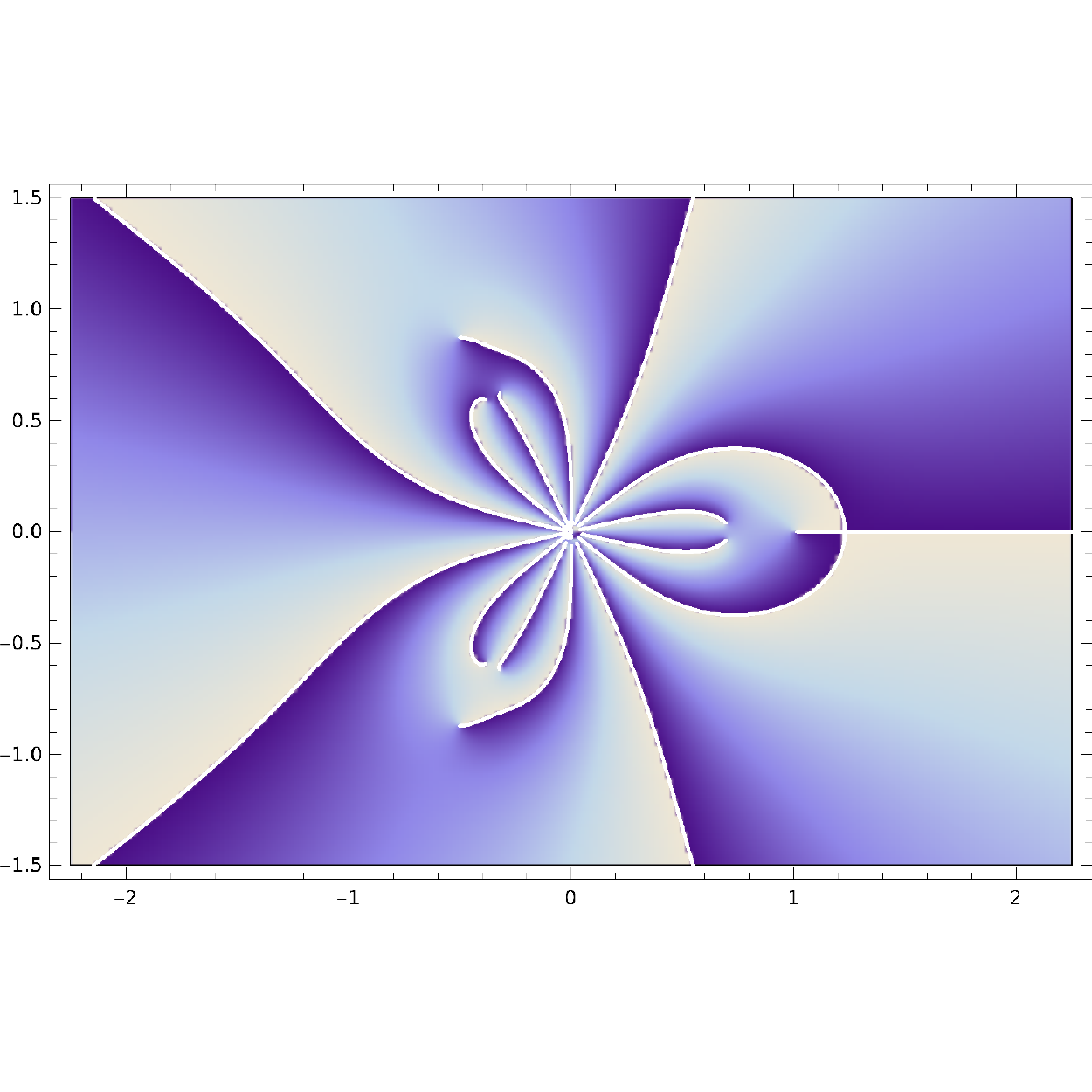}  
  \caption{d=6}
\end{subfigure}
\hfill
\begin{subfigure}{.2\textwidth}
  \centering
  		\includegraphics[height=3.7cm]{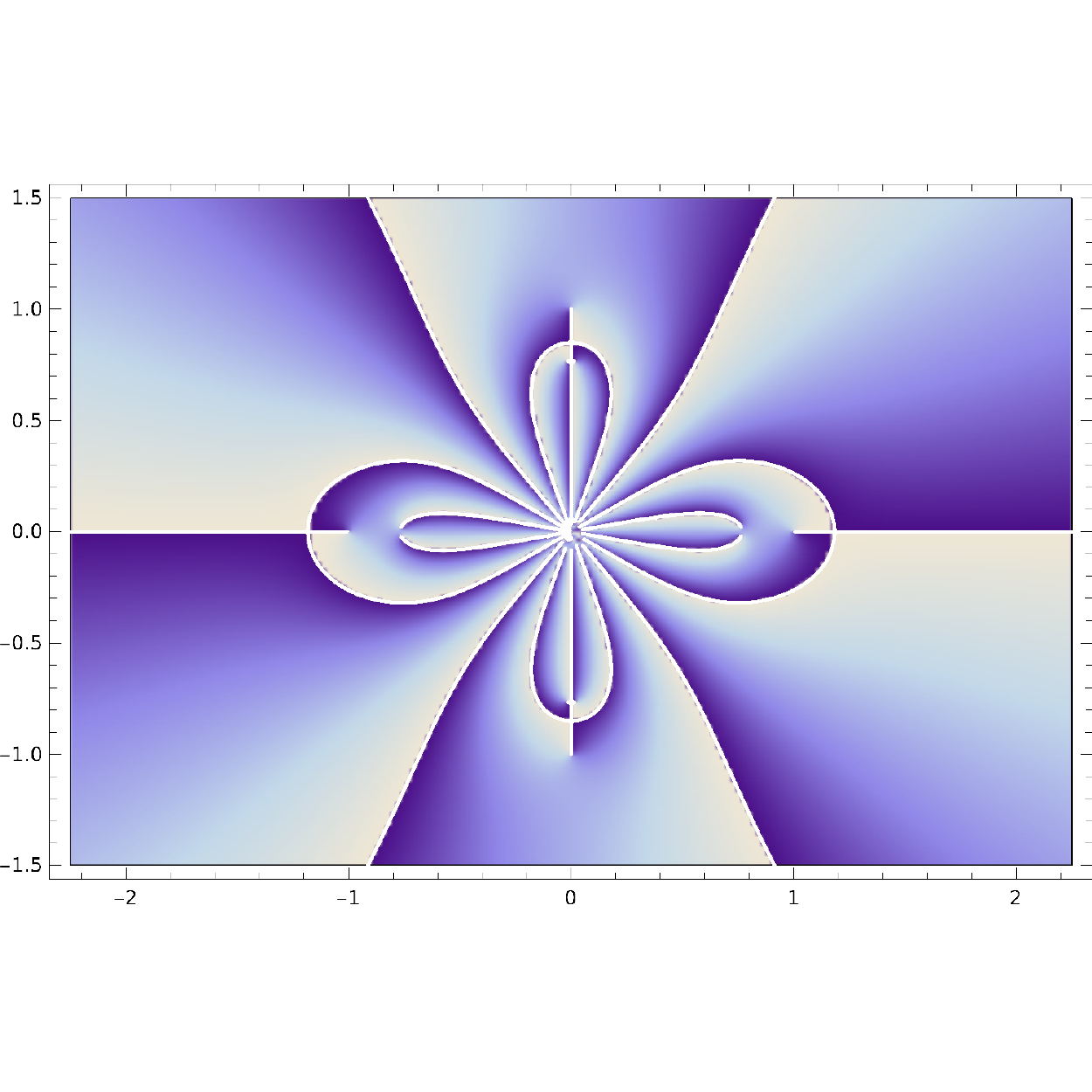}  
  \caption{d=7}
\end{subfigure}
\caption{{\textit{Numerically generated plots illustrating the behaviour of the anti-Stokes lines for Reissner-Nordstr{\"o}m BHs ($ \mu = 1, \; \vartheta = 0.9$).}}}
\label{fig:RNNum}
\end{figure}

 In Ref. \cite{Ghosh2006}, it is recommended that numerical analyses serve as aides in the construction of the contour path; in Refs. \cite{refNatarioSchiappa,Daghigh2007eRN,Daghigh2008SAdS}, numerical plots are used to depict the anti-Stokes lines in various dimensions. We find that we can reproduce these plots by applying the Mathematica function ${\mathtt{ComplexPlot}}$ to expressions of the form $-r^{d-1}/f(r)$, where $f(r)$ represents the metric function for each BH. These are sketched in Figs. \ref{fig:SchwarzNum}, \ref{fig:RNNum} and \ref{fig:SAdSNum}.

\subsubsection{ \label{subsubsec:nonAdSplane}The non-AdS $r$-plane}
\par To determine the position and direction of the anti-Stokes lines, we first identify the location where branch cuts shall be needed within the complex plane. This is achieved by solving for the complex roots of the metric function. For Eq. (\ref{eq:f(r)}) in the dS BH spacetimes, $f(r)=0$ produces
\begin{eqnarray}
\text{Schwarzschild: \hspace{0.5cm}} &\textcolor{white}{.} & - \lambda r^{d-1} + r^{d-3} - 2 \mu = 0 \;, \\
\text{Reissner-Nordstr{\"o}m: \hspace{0.5cm}} &\textcolor{white}{.} & - \lambda r^{2(d-2)} + r^{2(d-3)} - 2 \mu r^{d-3} + \vartheta^2 = 0 \;,
\end{eqnarray}
\noindent for which analytic solutions cannot be obtained. Thus, in order to determine the position of the horizons, roots must be calculated numerically. For even $d$, there are an odd and even number of roots for Schwarzschild and Reissner-Nordstr{\"o}m BHs, respectively. These sum to zero:
\begin{eqnarray}
\text{Schwazrschild: \hspace{0.5cm}} &\textcolor{white}{.} & r_n = r_{_H}, r_{_C}, \gamma_1, \overline{\gamma}_1, ... , \gamma_{\frac{d-4}{2}}, {\overline{\gamma}}_{\frac{d-4}{2}}, {\tilde{r}}  \;,\\
\text{Reissner-Nordstr{\"o}m: \hspace{0.5cm}} &\textcolor{white}{.} &r_n = r^+_{_{H}}, r^-_{_H}, r_{_C}, \gamma_1, \overline{\gamma}_1,... , \gamma_{d-4}, {\overline{\gamma}}_{d-4}, {\tilde{r}} \;,
\end{eqnarray}
\noindent where bars denote complex conjugates, and 
\begin{eqnarray}
\text{Schwarzschild: \hspace{0.5cm}} &\textcolor{white}{.} & {\tilde{r}} = -\left[ r_{_{H}} + r_{_C} + \sum^{\frac{d-4}{2}}_{i=1} (\gamma_i + {\overline{\gamma}}_i ) \right] \;,\\
\text{Reissner-Nordstr{\"o}m: \hspace{0.5cm}} &\textcolor{white}{.} &
{\tilde{r}} = -\left[ r^+_{_{H}} + r^-_{_{H}} + r_{_C} + \sum^{d-4}_{i=1} (\gamma_i + {\overline{\gamma}}_i ) \right] \;.
\end{eqnarray}
\noindent For odd $d$, there are an even number of roots, which sum to zero through pair-wise cancellation:
\begin{eqnarray}
\text{Schwarzschild: \hspace{0.5cm}} &\textcolor{white}{.} & r _n = r_{_{H}}, -r_{_{H}}, r_{_C}, - r_{_C}, \gamma_1, \overline{\gamma}_1, ... , \gamma_{\frac{d-5}{2}}, {\overline{\gamma}}_{\frac{d-5}{2}} \;, \\
\text{Reissner-Nordstr{\"o}m: \hspace{0.5cm}} &\textcolor{white}{.} & r _n = r^+_{_{H}}, r^-_{_{H}}, -r^+_{_{H}}, -r^-_{_{H}}, r_{_C}, - r_{_C},\gamma_1, \overline{\gamma}_1, ... , \gamma_{d-5}, {\overline{\gamma}}_{d-5} \;.
\end{eqnarray}

\par For both even and odd $d$, Schwarzschild dS BHs have $(d-1)$ complex horizons, only two of which are real: the event horizon and the cosmological horizon. For Reissner-Nordstr{\"o}m dS BHs, there are $2(d-2)$ complex horizons, with three real positive roots: the outer and inner BH horizons, as well as the cosmological horizon.  

\par In the absence of a cosmological constant, the positions of the complex horizons become
\begin{eqnarray}
\text{Schwarzschild: \hspace{0.5cm}} &\textcolor{white}{.} & r_{_{H_n}} = \big \vert 2 \mu \big \vert^{\frac{1}{d-3}} \; e^{ \frac{2 \pi i}{d-3}  n } \;, \\
\text{Reissner-Nordstr{\"o}m: \hspace{0.5cm}} &\textcolor{white}{.} &
r^{\pm}_{_{H_n}} = \bigg \vert \left( \mu \pm \sqrt{\mu^2 - \vartheta^2} \right)^{\frac{1}{d-3}} \bigg \vert \; e^{ \frac{2 \pi i }{d-3}  n } \;,
\end{eqnarray} 
\noindent for $n=0,1,...,d-4$. For the Schwarzschild BH with $\lambda = 0$, there are instead $d-3$ complex horizons, one of which is the real event horizon. The Reissner-Nordstr{\"o}m BH has $2(d-3)$ complex horizons and two real positive roots corresponding to the inner Cauchy horizon and outer event horizon.

\par In the case of the extremal Reissner-Nordstr{\"{o}}m BH ($\vartheta \rightarrow \mu$), the two real horizons coalesce. The positions of the complex horizons then become
\begin{equation}
\hspace{-4cm} \text{extremal Reissner-Nordstr{\"o}m: \hspace{0.7cm}} r_{_{H_n}} = \big \vert  \mu \big \vert^{\frac{1}{d-3}} \; e^{ \frac{2 \pi i}{d-3} n } \;. 
\end{equation}
\noindent As observed by Andersson and Howls \cite{refAnderssonHowls}, this presents a topology distinct from both the Schwarzschild and the Reissner-Nordstr{\"{o}}m BHs and therefore requires separate analysis. We shall address this further in section \ref{subsec:aQNF}.

\par To draw the anti-Stokes lines around the singular points, we must adhere to the boundary conditions. As stipulated earlier, $\mathbb{I}m \{ \omega \} \gg \mathbb{R}e \{ \omega \}$ in Minkowski and dS spacetimes within the large overtone limit; since $\mathbb{I}m \{ \omega \} \rightarrow \infty$, the aQNF is effectively a purely imaginary quantity. Consequently, $\omega x \in \mathbb{R}$ for $x \in i \mathbb{R}$ near the origin \cite{refNatarioSchiappa}. $\Phi$ is considered to be ``Wick rotated" 
to the anti-Stokes line of $\mathbb{I}m \{ \omega x \} = 0$ (which approximately corresponds to the  Stokes line of $\mathbb{R}e \{x \} = 0$) \cite{refMotlNeitzke,refNatarioSchiappa,Ghosh2006}.  

\par We may then establish the behaviour of the anti-Stokes lines within the neighbourhood of $r \sim 0$. From the relationship between $x$ and $r$ observed in Eqs. (\ref{eq:tortst}) and (\ref{eq:tortrn}), we set
\begin{equation} \label{eq:Wick}
r(x) = \rho e^{i\eta \left( n + \frac{1}{2} \right)} \;,
\end{equation}
\noindent where $\rho, \eta \in \mathbb{R}$ with $\rho > 0$ as an arbitrary proportionality constant; $\eta$ and $n$ are provided in Table {\ref{table:1}} \cite{refNatarioSchiappa,Ghosh2006}. The above expression represents half-lines extending from the origin, spaced equally from one another by an angle of $\eta$. Since $r$ oscillates between positive and negative values for monotonically increasing values of $n$, so too does the tortoise coordinate $x$. As such, the sign of $\omega x$  alternates between positive and negative on each subsequent half-line, anticlockwise about the origin and starting at $n=0$ \cite{refNatarioSchiappa}. We consider the $n=0$ line to be that which is closest to the imaginary axis in quadrant IV (see Figs. \ref{fig:STdS}, \ref{fig:RNdS}, and \ref{fig:SAdS}). The branches delineating the contour emerge from the origin, superimposed onto a choice of these half-lines.

\subsubsection{\label{subsubsec:AdSplane}The AdS $r$-plane}

\par As specified in Eq. (\ref{eq:BCout}), the boundary conditions for the QNMs in AdS spacetime differ from the standard form near spatial infinity. The aQNF no longer corresponds to the highly damped regime; $\mathbb{I}m \{ \omega \} \sim \mathbb{R}e \{ \omega \}$ implies that a very different anti-Stokes line topology is required. To sketch this, let us determine the positions of the complex horizons. We follow the same procedure: for Eq. (\ref{eq:f(r)}) in the AdS BH spacetimes, $f(r)=0$ produces
\begin{eqnarray}
\text{Schwarzschild: \hspace{0.5cm}} &\textcolor{white}{.} &  \vert \lambda \vert r^{d-1} + r^{d-3} - 2 \mu = 0 \;, \\
\text{Reissner-Nordstr{\"o}m: \hspace{0.5cm}} &\textcolor{white}{.} &  \vert \lambda \vert r^{2(d-2)} + r^{2(d-3)} - 2 \mu r^{d-3} + \vartheta^2 = 0 \;,
\end{eqnarray}
\noindent for which roots must once again be calculated numerically. For even $d$, there are an odd and even number of roots for Schwarzschild and Reissner-Nordstr{\"o}m BHs, respectively. These sum to zero:
\begin{eqnarray}
\text{Schwarzschild: \hspace{0.5cm}} &\textcolor{white}{.} & r_n = r_{_H}, \gamma_1, \overline{\gamma}_1, ... , \gamma_{\frac{d-2}{2}}, {\overline{\gamma}}_{\frac{d-2}{2}}  \;,\\
\text{Reissner-Nordstr{\"o}m: \hspace{0.5cm}} &\textcolor{white}{.} &r_n = r^+_{_{H}}, r^-_{_H}, \gamma_1, \overline{\gamma}_1,... , \gamma_{d-3}, {\overline{\gamma}}_{d-3} \;,
\end{eqnarray}
\noindent where our results are much like the dS cases, albeit with the absence of $r_{_C}$ and ${\tilde{r}} $.
\noindent Similarly, for odd $d$, there are an even number of roots, which sum to zero through pair-wise cancellation:
\begin{eqnarray}
\text{Schwarzschild: \hspace{0.5cm}} &\textcolor{white}{.} & r _n = r_{_{H}}, -r_{_{H}}, \gamma_1, \overline{\gamma}_1, ... , \gamma_{\frac{d-3}{2}}, {\overline{\gamma}}_{\frac{d-3}{2}} \;, \\
\text{Reissner-Nordstr{\"o}m: \hspace{0.5cm}} &\textcolor{white}{.} & r _n = r^+_{_{H}}, r^-_{_{H}}, -r^+_{_{H}}, -r^-_{_{H}},\gamma_1, \overline{\gamma}_1, ... , \gamma_{d-4}, {\overline{\gamma}}_{d-4} \;.
\end{eqnarray}
\par For both even and odd $d$, Schwarzschild AdS BHs have $(d-1)$ complex horizons, only one of which is real: the event horizon. For Reissner-Nordstr{\"o}m AdS BHs, there are $2(d-2)$ complex horizons, with two real positive roots: the outer and inner horizons.

\begin{figure}[t]
\begin{subfigure}{.2\textwidth}
  \centering
 	 \includegraphics[height=3.7cm]{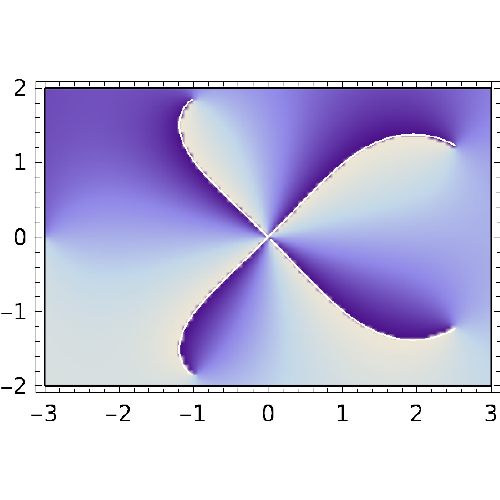}  
  \caption{d=4}
\end{subfigure}
\hfill
\begin{subfigure}{.2\textwidth}
  \centering
  		\includegraphics[height=3.7cm]{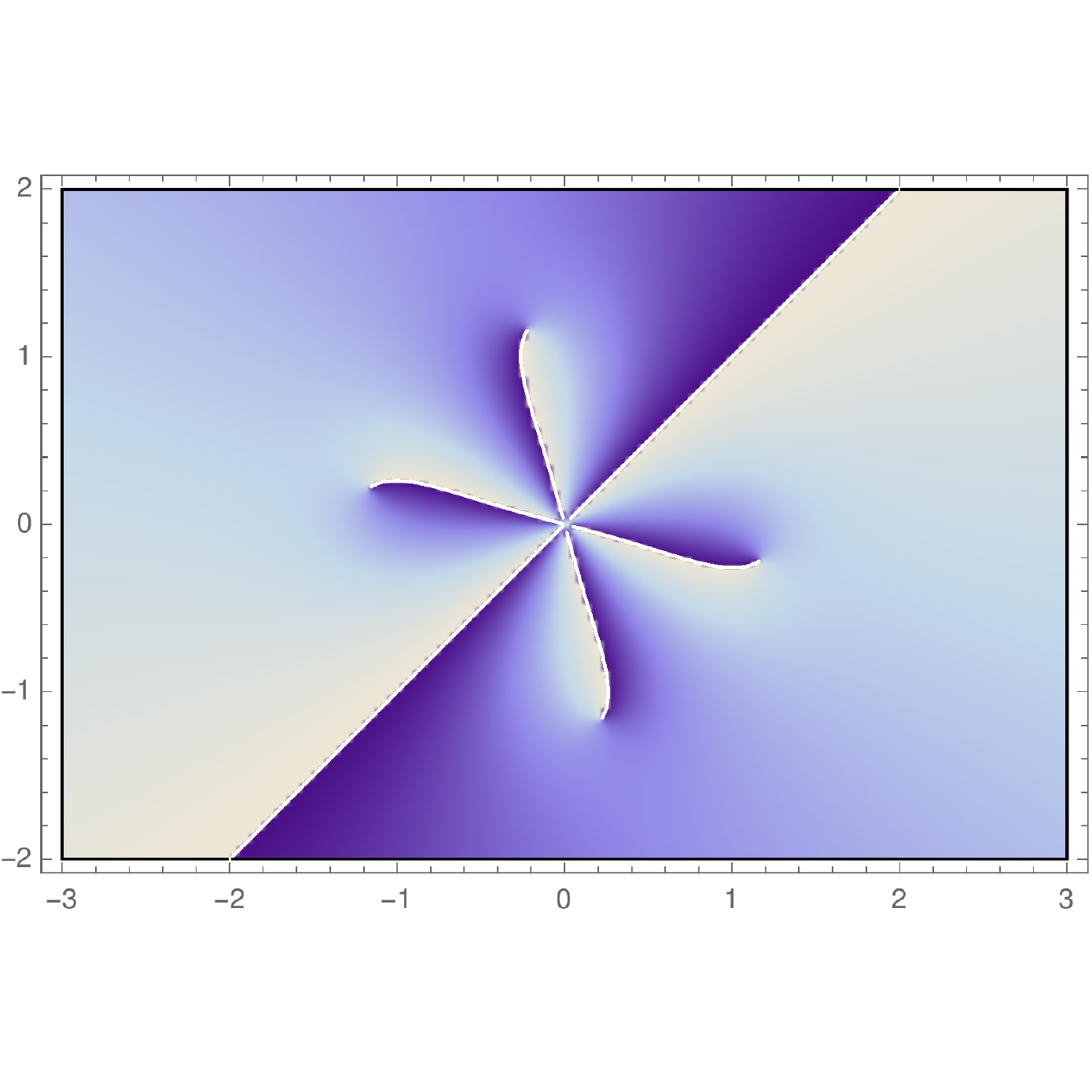}  
  \caption{d=5}
\end{subfigure}
\hfill
\begin{subfigure}{.2\textwidth}
  \centering
  		\includegraphics[height=3.7cm]{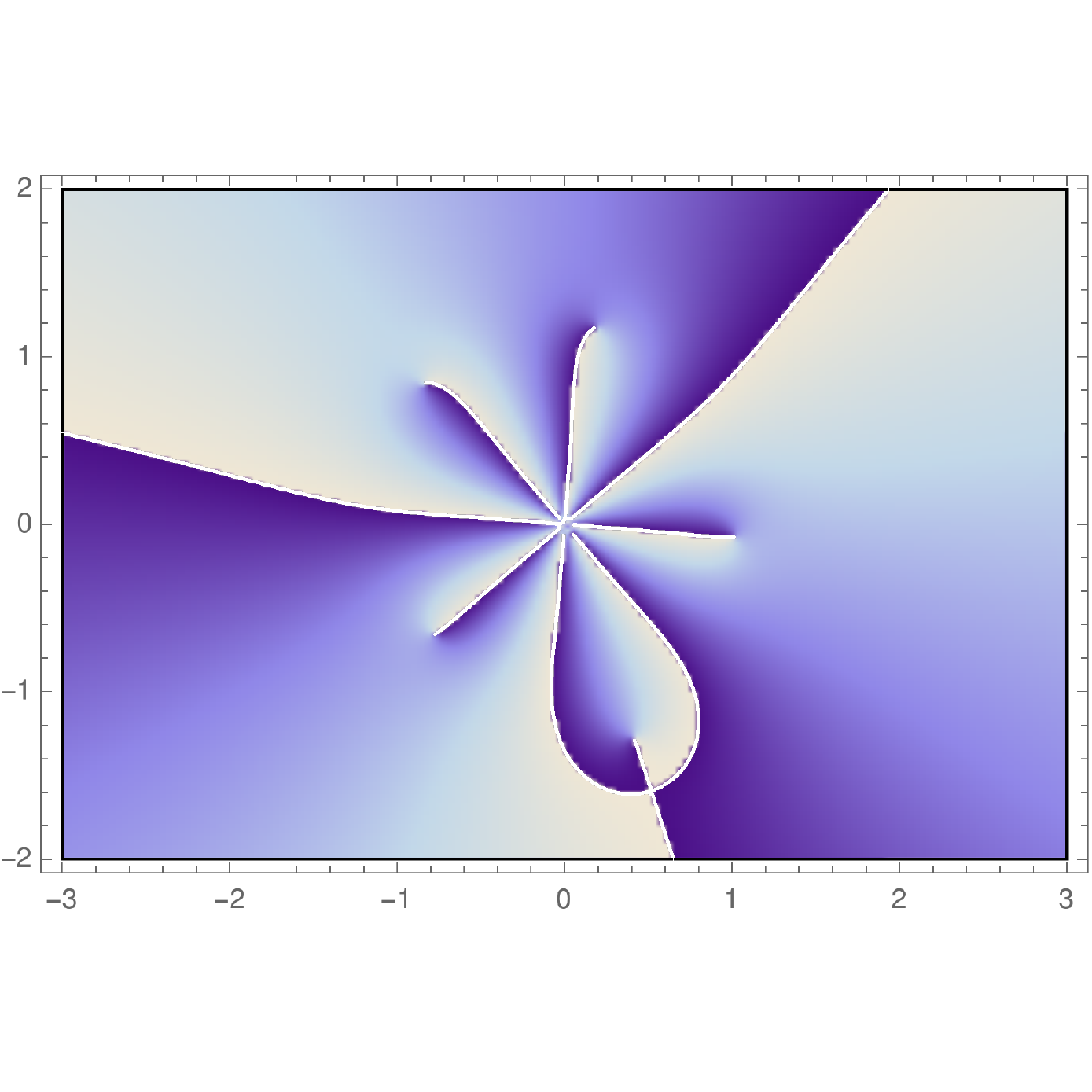}  
  \caption{d=6}
\end{subfigure}
\hfill
\begin{subfigure}{.2\textwidth}
  \centering
  		\includegraphics[height=3.7cm]{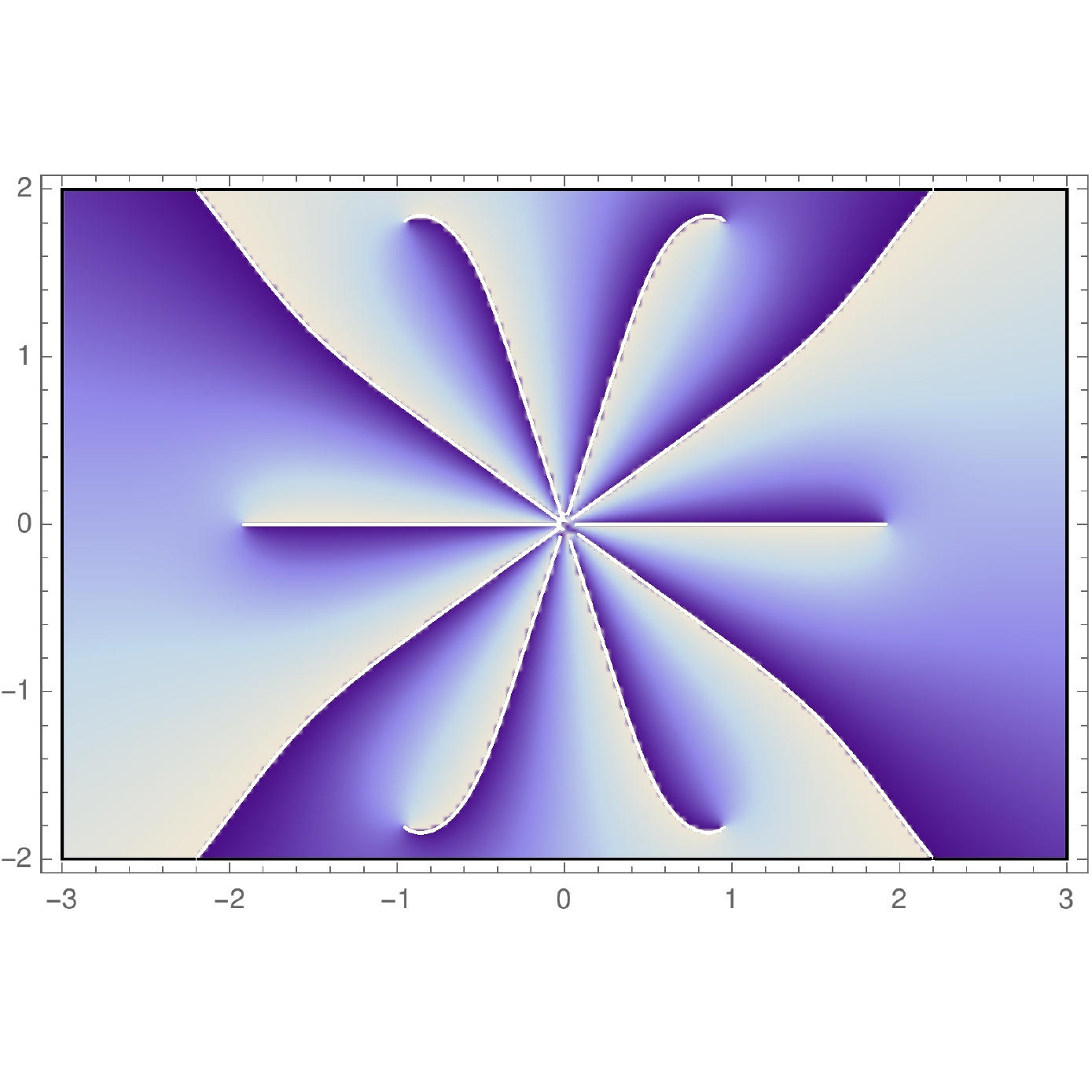}  
  \caption{d=7}
\end{subfigure}
\caption{{\textit{Numerically generated plots illustrating the behaviour of the anti-Stokes lines for Schwarzschild AdS BHs ($ \mu = \lambda = 1$).}}}
\label{fig:SAdSNum}
\end{figure}

\par Near the origin, we once again set
\begin{equation}
r(x) = \rho e^{i \eta \left(n+\varphi_0 \right)} \;,
\end{equation}
\noindent where $\rho, \eta \in \mathbb{R}$ with $\rho > 0$ as an arbitrary proportionality constant; $\eta$ and $n$ are provided in Table {\ref{table:1}} \cite{refNatarioSchiappa,Ghosh2006}. As before, these half-lines extend from the origin and are spaced equally from one another by an angle of $\eta$. The sign of $\omega x$ on each of these half-lines alternates from positive to negative. 

\par However, ${\mathbb{I}}m \{ \omega x \}$ remains multi-valued around all horizons, such that a particular branch must be chosen from which we trace the ${\mathbb{I}}m \{ \omega x \} = 0$ line. The need for a shifting of the branch cuts emerges to ensure that ${\mathbb{I}}m \{ \omega x \} = 0$ does not intersect with the complex horizons established for the AdS BHs. Since at least one branch must extend to spatial infinity, we can select branch cuts in such a way that a branch in quadrant I must correspond to $r \sim \infty$. From Eq. ({\ref{eq:tortads}}),
\begin{equation}
\omega x \sim \omega x_0 - \omega \frac{1}{\vert \lambda \vert r}
\end{equation}
\noindent for $r \sim \infty$. Since ${\mathbb{I}}m \{\omega\} \sim {\mathbb{R}}e \{\omega\}$ for aQNFs in AdS spacetimes, $\omega x_0$ is approximately real in the asymptotic limit. Consequently, we may claim that $\arg \{ \omega \} = - \arg \{ x_0 \} \equiv -\varphi_0$ \cite{refNatarioSchiappa}. The argument of $r$ along the branch extending towards spatial infinity is then equivalent to $- \varphi_0$. The contour informed by these considerations is demonstrated in Fig. {\ref{fig:SAdS}.

\setlength{\extrarowheight}{2.7pt}
\begin{table}[t] 
\centering
\caption{\textit{Details for the sketching of the contour in the complex $r$-plane for flat, dS, and AdS BHs in the Schwarzschild and Reissner-Nordstr{\"o}m (RN) ``families".}}
\begin{tabular}{|c|c|c|c|c|l|l|}
\hline
BH & $\eta $ & $n$ & branches & $x$ for $r \sim 0$ & $\; \omega x$ ($\lambda \geq 0$) & $\; \omega x$ ($\lambda < 0$)   \\
\hline
Schwarz.  & $\; \pi/(d-2) \;$  & $\;0,1, ... ,2d-5 \; $ & $\; 2(d-2) \;$ & $-r^{d-2}/2(d-2)\mu$ & $\; (-1)^{n} \;$ & $\; (-1)^{n+1} \;$ \\
RN  & $\; \pi/(2d-5) \;$  & $\;0,1, ... ,4d-11 \; $ & $\; 2(2d-5) \;$ & $r^{2d-5}/(2d-5)\vartheta^2$ & $\; (-1)^{n+1} \;$ & $\; (-1)^{n} \;$\\
\hline
\end{tabular}
\label{table:1}
\end{table} 

\subsection{\label{subsec:aQNF}The aQNF calculation}
\par While their explicit monodromy calculations are based on contours traced for 6D BH spacetimes, the final gravitational aQNF expressions computed by Nat{\'a}rio and Schiappa in Ref. \cite{refNatarioSchiappa} are claimed to be applicable for spacetimes of dimension $d>3$. Such generalisability of $d=6$ spin-2 aQNF results has also been alluded to in Refs. \cite{Daghigh2007eRN,Daghigh2008SAdS}. Since we find that generalised aQNF expressions can be calculated without consideration of the nature of the perturbing field, these claims of universal applicability for $d>3$ may be extended to the results presented throughout this section.

\subsubsection{Schwarzschild and Schwarzschild dS BHs \label{subsubsec:STnatschiap}}
\par In Fig. \ref{fig:STdS}, the contour drawn remains identical for Schwarzschild and Schwarzschild dS BHs $-$ only the ``contents" enclosed by the contour and the behaviour of the QNM near spatial infinity differ. As such, the solutions on the major contour can be traced uniformly for Schwarzschild and Schwarzschild dS BHs, as first stipulated in Ref. \cite{Ghosh2006}.

\par We begin the path on branch $A$. Near the origin, we know the potential has the standard form depicted in Eq. (\ref{eq:genpot}). The positive $\omega x$ of the $n=0$ branch implies that we can exploit the asymptotic expansion of the Bessel function associated with $\omega x \gg 1$, such that 
\begin{eqnarray}
\Phi(x) & \sim & B_+ \sqrt{2 \pi \omega x } \; J_{+j/2} (\omega x) + B_- \sqrt{2 \pi \omega x } \; J_{-j/2} ( \omega x) \nonumber \\
& \sim & (B_+ e^{-i \alpha_+} + B_- e^{-i\alpha_-} ) \; e^{+i \omega x} + (B_+ e^{+i \alpha_+} + B_- e^{+i\alpha_-} )\; e^{-i \omega x} \;. \label{eq:PhiASTdSGEN} 
\end{eqnarray}
\noindent Furthermore, the appropriate boundary condition on $A$ is purely outgoing, \textit{viz.}
\begin{eqnarray}
\omega x \rightarrow +\infty \hspace{0.5cm} & \Rightarrow & \Phi \sim e^{-i \omega x} \nonumber\\
& \Rightarrow & B_+ e^{-i \alpha_+} + B_- e^{-i\alpha_-} = 0 \;. \label{eq:STBC1}
\end{eqnarray}
\noindent Consequently, the solution on branch $A$ is
\begin{equation}  \label{eq:PhiASTdS}
\Phi_A (x) \sim (B_+ e^{+i \alpha_+} + B_- e^{+i\alpha_-} ) \; e^{-i \omega x}  \;.
\end{equation}

\begin{figure}[t]
  \centering
  \includegraphics[width=.6\textwidth]{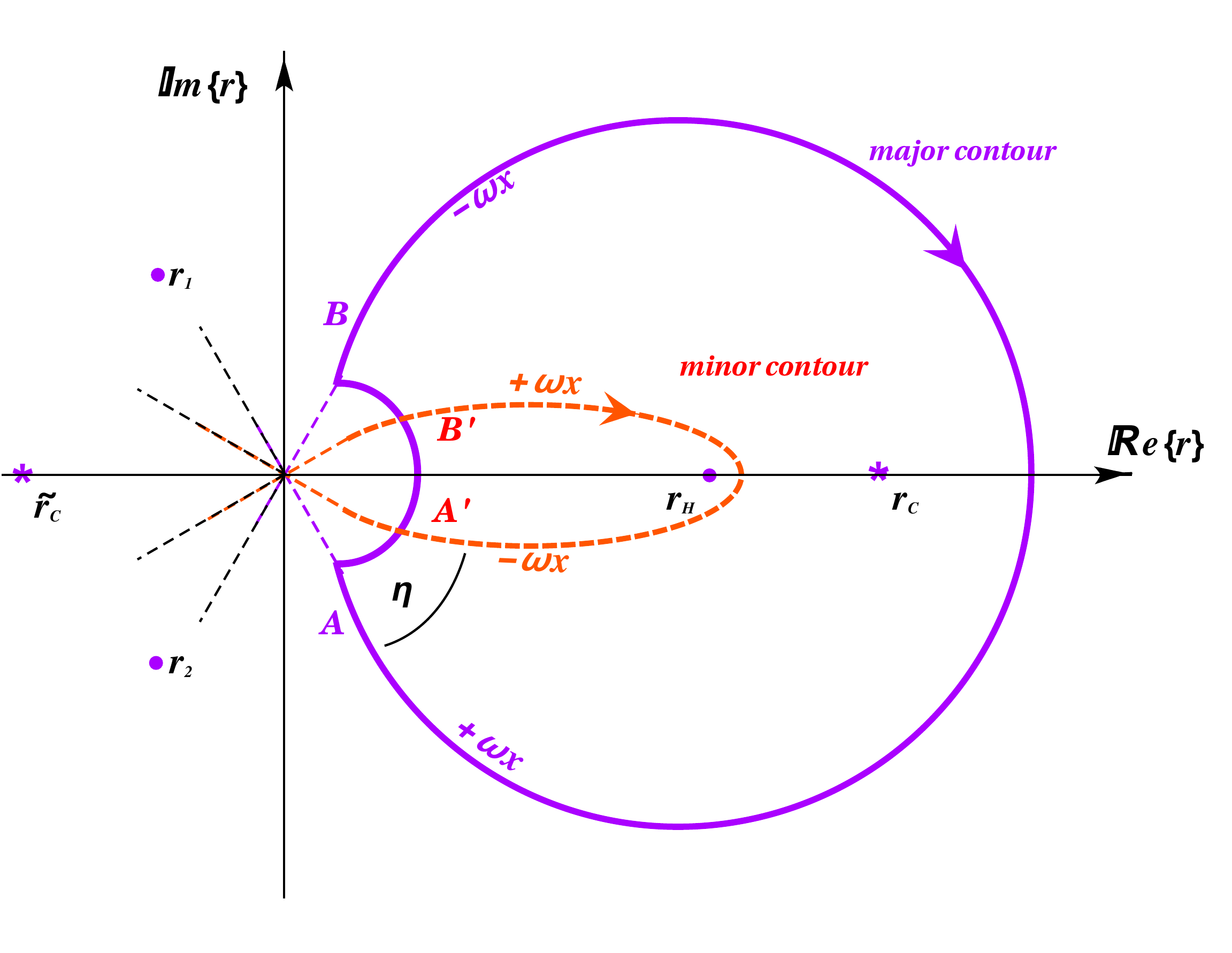}
  \caption{\textit{Anti-Stokes lines for a Schwarzschild dS BH with $d=6$.\\ For $\lambda > 0$, $r_i \rightarrow \gamma_i$ and $r_c$, $\tilde{r}_c$ remain.}}
  \label{fig:STdS}
\end{figure}

\par To reach branch $B$, a rotation of $3 \eta$ in $r$ (corresponding to $3 \pi$ in $x$) is needed. We invoke Eq. (\ref{eq:Jtwist}) to incorporate the rotation into the general solution of Eq. (\ref{eq:genQNMBessel}). To simplify, we utilise the exponential form of $\cos x \equiv (e^{+ix} + e^{-ix})/2$ and $e^{3 \pi i} = -1$; we also define $\alpha_{\pm} \equiv (1 \pm j)\pi/4$:

\begin{eqnarray}
\Phi(x) & \sim & B_+ \sqrt{2 \pi e^{3 \pi i} \; \omega x} \; J_{+j/2} (e^{3 \pi i} \omega x) + B_- \sqrt{2 \pi e^{3 \pi i} \; \omega x} \; J_{-j/2} (e^{3 \pi i} \omega x) \nonumber \\
& \sim & B_+ 2 \cos (-\omega x - \alpha_{+} ) e^{6i \alpha_+ } + B_- 2 \cos (-\omega x - \alpha_{-} ) e^{6i \alpha_- } \nonumber \\
& = & \left( B_+ e^{7 i \alpha_+} + B_- e^{7 i \alpha_-} \right) e^{+ i \omega x} + \left( B_+ e^{5 i \alpha_+} + B_- e^{5 i \alpha_-} \right) e^{- i \omega x} \label{eq:PhiBSTdSGEN} \;.
\end{eqnarray}

\par Though $\omega x < 0$ on branch $B$, the fact that the path is traced towards infinity requires that we maintain the use of outgoing boundary conditions. Thus,
\begin{eqnarray}
\omega x \rightarrow +\infty \hspace{0.5cm} & \Rightarrow & \Phi \sim e^{-i \omega x} \nonumber\\
& \Rightarrow & B_+ e^{7 i \alpha_+} + B_- e^{7 i \alpha_-} = 0 \;,
\end{eqnarray}
\noindent such that the general solution at branch $B$ becomes
\begin{equation}  \label{eq:PhiBSTdS}
\Phi_B (x) \sim \left( B_+ e^{5 i \alpha_+} + B_- e^{5 i \alpha_-} \right) \; e^{-i \omega x}  \;.
\end{equation}

\par For the higher-dimensional Schwarzschild BH, the computation of the local and global monodromies directly follows the discussion outlined in subsection \ref{subsec:principles}. The former requires a clockwise path governed by the ingoing boundary conditions, such that $\Phi \sim e^{+i \omega x}$. Along the path, $x$ increases by $-2 \pi i/2k_{_H}$, such that the local monodromy is given by
\begin{equation} \label{eq:MonoLocalST}
\mathfrak{M}_{\gamma \;, \; r_{_H}} [ \Phi(x) ] = e^{+\frac{\pi \omega}{k_{_H}}} \;.
\end{equation} 
\noindent The global monodromy associated with the major contour of Fig. \ref{fig:STdS} traced to spatial infinity adheres to the outgoing boundary conditions, such that $\Phi \sim e^{-i \omega x}$. We therefore require the coefficients of $e^{-i \omega x}$ from Eqs. (\ref{eq:PhiASTdSGEN}) and (\ref{eq:PhiBSTdSGEN}), as well as the clockwise monodromy of $e^{-i \omega x}$. Since $x$ also increases by $-2 \pi i/2k_{_H}$ on this path, the result is $e^{-\pi \omega/k_{_H}}$. Thus, the global monodromy becomes 
\begin{equation} \label{eq:MonoGlobalST}
\mathfrak{M}_{\gamma } [ \Phi(x) ] = \frac{B_+ e^{5 i \alpha_+} + B_- e^{5 i \alpha_-}}{B_+ e^{+i \alpha_+} + B_- e^{+i\alpha_-}} \; \ e^{-\frac{\pi \omega}{k_{_H}}} \;.
\end{equation}  
\noindent To obtain the final expression for the aQNF within the Schwarzschild BH, we equate the monodromies and solve
\begin{equation} \label{eq:EqaQNFST}
\frac{B_+ e^{5 i \alpha_+} + B_- e^{5 i \alpha_-}}{B_+ e^{+i \alpha_+} + B_- e^{+i\alpha_-}} \; e^{-\frac{\pi \omega}{k_{_H}}} = e^{+\frac{\pi \omega}{k_{_H}}} \;.
\end{equation}
\noindent The aQNF solution is obtained by writing Eq. (\ref{eq:EqaQNFST}) and the additional boundary condition of Eq. (\ref{eq:STBC1}) as a system of linear equations in matrix form $X(B_+,B_-)^T=0$, and then solving for $det(X)=0$. This procedure applies to all spacetimes studied in this work. However, for the Schwarzschild BH family, we solve for the aQNF using the derivative of the determinant $\partial_j det(X)=0$ instead \cite{refNatarioSchiappa}.

\par We can generalise the solution provided in Ref. \cite{refNatarioSchiappa} for the aQNF within the Schwarzschild BH spacetimes, such that
\begin{equation} 
\lim_{n \rightarrow \infty} \omega = T_{_H} \ln (-1 - 2 \cos (\pi j)) + 2 \pi i T_{_H} n \;,
\label{eq:aQNFST}
\end{equation}
\noindent where $T_{_H} = k_{_H}/2 \pi$. This is in agreement with Refs. \cite{refMotlNeitzke,Birmingham2003,CardosoLemosYoshida}, and is in keeping with the aQNF ``structure" given in Eq. (\ref{eq:aQNFoffset}) for the highly-damped QNFs of the 4D Schwarzschild BH.
\par For the $d$-dimensional Schwarzschild dS BH, these considerations are augmented by the presence of the cosmological horizon at $r=r_{_C}$, where $r_{_H} \ll r_{_C}$. We find that this leads to the establishment of two global monodromy expressions: an adapted Eq. (\ref{eq:MonoGlobalST}) incorporating the surface gravity at $r=r_{_C}$ ($k_{_C}<0$) where outgoing boundary conditions dominate the overall behaviour ($\Phi \sim e^{-i \omega x}$), and a second expression where ingoing boundary conditions dominate ($\Phi \sim e^{+i \omega x}$). Thus two ``monodromy equations" emerge that must be solved simultaneously to extract the aQNF. 

\par Let us begin with the local monodromies. Around $r_{_C}$, we trace a clockwise path governed by the outgoing boundary conditions, such that $\Phi \sim e^{-i \omega x}$, along which $x$ increases by $-2 \pi i/2k_{_C}$. Thus, the local monodromy associated with $r_{_C}$ is 
\begin{equation} \label{eq:MonoLocalSdSrc}
\mathfrak{M}_{\gamma \;, \; r_{_C}} [ \Phi(x) ] = e^{-\frac{\pi \omega}{k_{_C}}} \;.
\end{equation} 
\noindent Eq. (\ref{eq:MonoLocalST}) remains valid, such that the tightly wound contour around the regular singular points yields a local monodromy of
\begin{equation} \label{eq:MonoLocalSdS}
\mathfrak{M}_{\gamma \;, \; r_{_i}} [ \Phi(x) ] = e^{+\frac{\pi \omega}{r_{_H}}-\frac{\pi \omega}{k_{_C}}} \;.
\end{equation} 

\noindent The first global monodromy is obtained from Eq. (\ref{eq:MonoGlobalST}), where we now include the increase of $-2 \pi i/2k_{_C}$ in $x$ under the influence of outgoing boundary conditions. Thus,
\begin{equation} \label{eq:MonoGlobalSdS1}
\mathfrak{M}_{\gamma } [ \Phi(x) \sim e^{- i \omega x}] = \frac{B_+ e^{5 i \alpha_+} + B_- e^{5 i \alpha_-}}{B_+ e^{+i \alpha_+} + B_- e^{+i\alpha_-}} \; \ e^{-\frac{\pi \omega}{k_{_H}}-\frac{\pi \omega}{k_{_C}}} \;.
\end{equation}

\noindent For the second global monodromy, we apply ingoing boundary conditions to Eqs. (\ref{eq:PhiASTdSGEN}) and (\ref{eq:PhiBSTdSGEN}) in order to extract the $e^{+i \omega x}$ terms. Since the increase in $x$ by $2\pi i/2k_i$ yields $e^{+i \omega (-\pi i/k_{_H})}e^{+i \omega (-\pi i/k_{_C})}$, the second global monodromy becomes
\begin{equation} \label{eq:MonoGlobalSdS2}
\mathfrak{M}_{\gamma } [ \Phi(x) \sim e^{+ i \omega x}] = \frac{B_+ e^{7 i \alpha_+} + B_- e^{7 i \alpha_-}}{B_+ e^{-i \alpha_+} + B_- e^{-i\alpha_-}} \; \ e^{+\frac{\pi \omega}{k_{_H}}+\frac{\pi \omega}{k_{_C}}} \;.
\end{equation}
\noindent In solving the two simultaneous equations
\begin{eqnarray}
\frac{B_+ e^{5 i \alpha_+} + B_- e^{5 i \alpha_-}}{B_+ e^{+i \alpha_+} + B_- e^{+i\alpha_-}} \; \ e^{-\frac{\pi \omega}{k_{_H}}-\frac{\pi \omega}{k_{_C}}} & = & e^{+\frac{\pi \omega}{r_{_H}}-\frac{\pi \omega}{k_{_C}}} \;,\\
\frac{B_+ e^{7 i \alpha_+} + B_- e^{7 i \alpha_-}}{B_+ e^{-i \alpha_+} + B_- e^{-i\alpha_-}} \; \ e^{+\frac{\pi \omega}{k_{_H}}+\frac{\pi \omega}{k_{_C}}} & = & e^{+\frac{\pi \omega}{r_{_H}}-\frac{\pi \omega}{k_{_C}}} \;,
\end{eqnarray}
\noindent we may extract a fully generalised solution for the aQNF,
\begin{eqnarray}
&&\bigg (-3 e^{\pi i j+2 \left(\frac{\pi  \omega}{k_{_C}}+\frac{\pi  \omega}{k_{H}}\right)} +3 e^{2 \pi i j + 2 \left(\frac{\pi  \omega}{k_{_C}}+\frac{\pi  \omega}{k_{_H}}\right)}+e^{\pi i j + \frac{2 \pi  \omega}{k_{_C}}}  +e^{\pi i j + \frac{2 \pi  \omega}{{k_{_H}}}} -3 e^{ \pi i j}    \hspace{0.6cm}\nonumber \\
& & \hspace{0.6cm} +3 e^{2 \pi i j} +3 e^{2 \left(\frac{\pi \omega}{k_{_C}}+\frac{\pi  \omega}{k_{_H}} \right)} +3 \bigg) \hspace{0.2cm} \times \left(1+e^{ \pi i j}\right) e^{ -\frac{3}{2} \pi i j -\left(\frac{\pi  \omega}{k_{_C}}+\frac{\pi \omega}{k_{_H}} \right)} =0 \;. \label{eq:aQNFSdS}
\end{eqnarray}
\noindent This can be shown to be in agreement with Refs. \cite{refNatarioSchiappa,Lopez-Ortega2006,refCardosoNatarioSchiappa}. As demonstrated in Ref. \cite{Lopez-Ortega2006}, the aQNF solution can instead be extracted in trigonometric form, such that 
\begin{equation} \label{eq:aQNFSdSLopezOrtega}
\sin \left(\frac{3\pi}{2} j \right) \cosh \left( \frac{\pi \omega }{k_{_C}} + \frac{\pi \omega}{k_{_H}} \right) + \sin \left( \frac{\pi}{2} j \right) \cosh \left( \frac{\pi \omega }{k_{_H}} - \frac{\pi \omega}{k_{_C}} \right) = 0 \;.
\end{equation}  
\noindent If differentiated with respect to $j$ and then subjected to $j\rightarrow 0$, Eq. (\ref{eq:aQNFSdSLopezOrtega}) reduces to the Schwarzschild dS aQNF given in Ref. \cite{refNatarioSchiappa}. 

\par Within the investigations of Refs. \cite{refNatarioSchiappa,Lopez-Ortega2006}, it was noted that the Schwarzschild dS aQNF reduces to the aQNF of the Schwarzschild BH. To demonstrate this, we exploit the fact that $r_{_C} \sim \lambda^{-1/2}$ \cite{Daghigh2008SAdS}. Since the Schwarzschild BH spacetime does not possess a cosmological horizon, we set $\lambda \rightarrow 0^+$ such that $r_{_C} \rightarrow + \infty$. Consequently,
\begin{equation} 
k_{_C} = \frac{1}{2}f'(r_{_C}) \thicksim - \lambda r_{_C} \sim -\frac{1}{r_{_C}} \hspace{0.5cm} \Rightarrow \; \; k_{_C} \sim  0^+ \;.
\end{equation}
\noindent For gravitational perturbations, where $j=0,2$ depending on the mode studied, Eq. ({\ref{eq:aQNFSdS}) was shown in Ref. \cite{refNatarioSchiappa} to reduce to
\begin{equation} \label{eq:Ln3}
e^{ \frac{\pi \omega}{k_{_H}}} + 3 e^{- \frac{\pi \omega}{k_{_H}}}=0 \hspace{1cm} \Rightarrow \; e^{\frac{2 \pi \omega}{r_{_H}}} = -3 \;.
\end{equation}  
\noindent This leads precisely to the expression for the gravitational aQNFs of a Schwarzschild BH \cite{refMotlNeitzke,refNatarioSchiappa,Birmingham2003}, which we may obtain from Eq. ({\ref{eq:aQNFST}}) with the appropriate values of $j$. 

\par A further remark from the studies of Refs. \cite{refNatarioSchiappa,Lopez-Ortega2006,refCardosoNatarioSchiappa} concerns the anomalies associated with the $d=4$ and $d=5$ Schwarzschild dS BHs, based on the manner in which the contours are drawn (please see Fig. \ref{fig:SchwarzNum} for comparison, and recall the similarity between Schwarzschild and Schwarzschild dS BH contours). In the case of the 4D Schwarzschild dS BH, the anticlockwise monodromy at ${\tilde{r}}_{_C}$ must be taken into account, in conjunction with the clockwise monodromy around $r_{_H}$ and $r_{_C}$. The first monodromy, however, is equivalent to the latter two, and the final expression for the aQNF is exactly that of Eq. ({\ref{eq:aQNFSdS}}). Since the $d=4$ contour can be deformed into its $d=6$ counterpart \cite{refNatarioSchiappa}, such a result is to be expected. 

\par For the 5D Schwarzschild dS BH, the anti-Stokes line closes near spatial infinity. The solution in this region was provided in Eq. ({\ref{eq:genpotAdS}}), and serves as the QNM expression at point $B$. In Ref. \cite{refNatarioSchiappa}, tensor-, vector-, and scalar-modes were observed to correspond to $j^{\infty}=4,2,0$, respectively. To move from branch $B$ to $A$, a rotation of $\pi/2$ must be introduced, which in turn produces
\begin{equation}
\Phi(x)  \thicksim \left(C_+ e^{3i \beta_+} + C_- e^{3i \beta_-} \right) e^{+i \omega (x-x_0)} + \left( C_+ e^{+ i \beta_+} + C_- e^{+i \beta_-} \right) e^{-i \omega (x - x_0)} \;.
\end{equation}
\noindent With these corrections made, the calculation follows the method outlined above, and yields the aQNF solution
\begin{equation}
\sin \left(\frac{3\pi}{2} j \right) \sinh \left( \frac{\pi \omega}{k_{_H}} + \frac{\pi \omega}{k_{_C}} \right) - \sin \left( \frac{\pi}{2} j \right) \sinh \left( \frac{\pi \omega }{k_{_H}} - \frac{\pi \omega}{k_{_C}} \right) = 0 \;,
\end{equation}  
\noindent as shown in Ref. \cite{Lopez-Ortega2006}. Once again, differentiating with respect to $j$ and then applying $j\rightarrow 0$ produces the corresponding result given in Ref. \cite{refNatarioSchiappa}. These, too, reduce to the Schwarzschild solution within the $r_{_C} \rightarrow + \infty$ limit.

\subsubsection{Reissner-Nordstr{\"o}m and extremal Reissner-Nordstr{\"o}m BHs \label{subsubsec:RNnatschiap}}
\par In Fig. \ref{fig:RNdS}, the shape of the major contour is shown to include a rotation from $A_1$ to $A_2$, a loop around the inner horizon, and a further rotation from $B_2$ to $B_1$. We define the change in $x$ due to this rotation about the Cauchy horizon as $\delta$.

\begin{figure}[t]
  \centering
  \includegraphics[width=.6\textwidth]{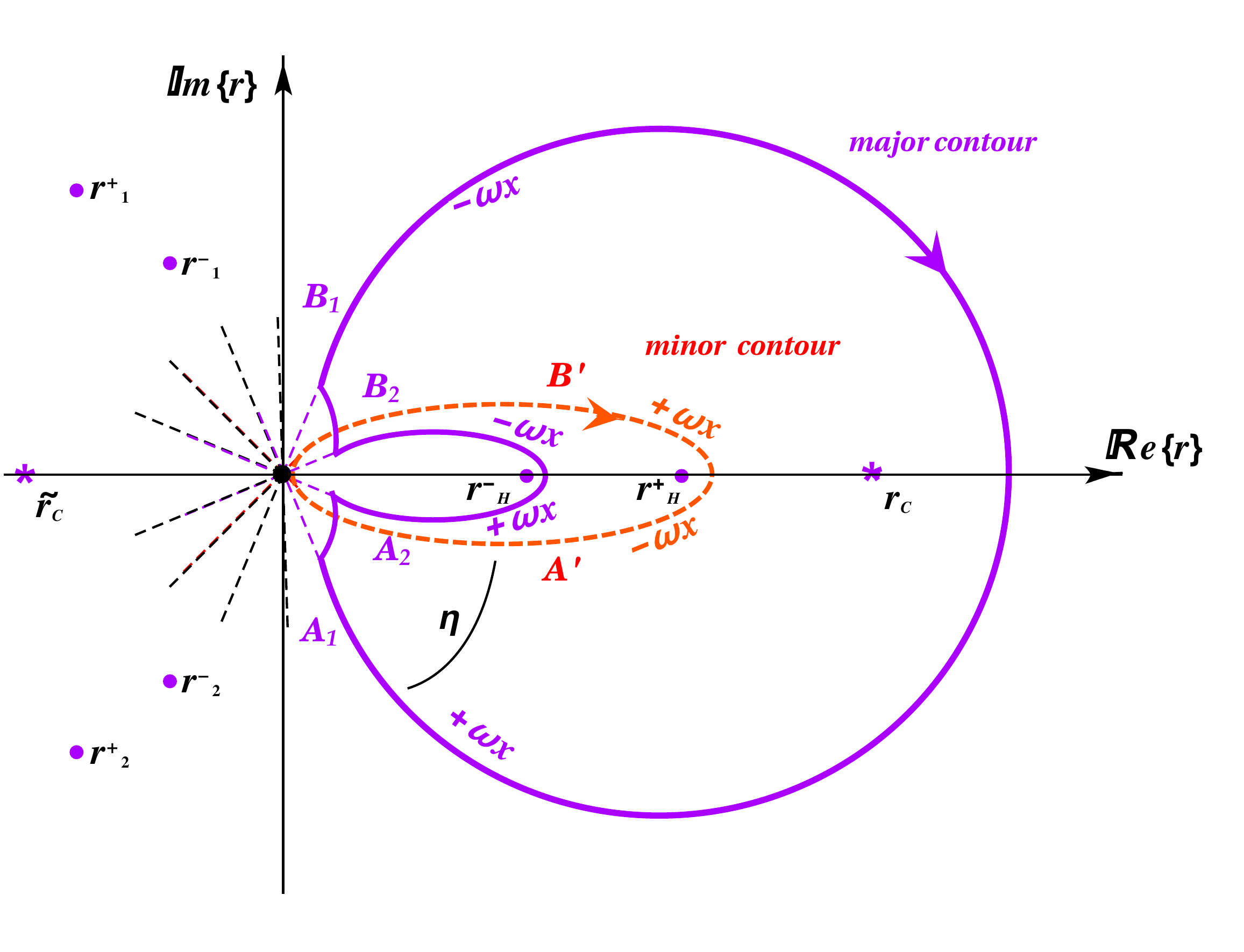}
  \caption{\textit{Anti-Stokes lines for a Reissner-Nordstr{\"o}m dS BH with $d=6$.\\ If we set $\lambda = 0$, $r_c$ and $\tilde{r}_c$ vanish.}}
  \label{fig:RNdS}
\end{figure}

\par Let us begin with the Reissner-Nordstr{\"o}m BH spacetime, within the neighbourhood of the origin. On branch $A_1$, $\omega x \gg 1$. As in the Schwarzschild case, the purely outgoing boundary condition is applied; the general solution of Eq. ({\ref{eq:PhiASTdSGEN}) at $A_1$ becomes
\begin{equation}  \label{eq:PhiA1RN}
\Phi_{A_1} (x) \sim (B_+ e^{+i \alpha_+} + B_- e^{+i\alpha_-} ) \; e^{-i \omega x}  \;,
\end{equation}
\noindent and a boundary condition of
\begin{equation} \label{eq:RNBC1}
B_+ e^{-i \alpha_+} + B_- e^{-i\alpha_-} = 0 
\end{equation}
\noindent emerges.

\par A rotation of $2 \eta$ in $r$, or $2 \pi$ in $x$, is needed to reach branch $A_2$, where Eq. (\ref{eq:Jexpandpos}) applies. We invoke Eq. (\ref{eq:Jtwist}) to incorporate the rotation into the solution of Eq. (\ref{eq:genQNMBessel}), and exploit $e^{2 \pi i} = +1$:
\begin{equation} \label{eq:PhiA2RNGEN}
\Phi_{A_2} (x) \sim \left( B_+ e^{3 i \alpha_+} + B_- e^{3 i \alpha_-} \right) e^{+ i \omega x} + \left( B_+ e^{5 i \alpha_+} + B_- e^{5 i \alpha_-} \right) e^{- i \omega x}  \;.
\end{equation}

\noindent We now encounter the loop around $r=r^{-}_{_H}$. The path from $A_2$ to $B_2$ is anticlockwise, such that $x$ increases by $\delta = + 2\pi i/2k^{-}_{_H}$. Note that on branch $B_2$, $\omega x <0$, such that we expand the Bessel functions according to Eq. (\ref{eq:Jexpandneg}). The general QNM solution therefore becomes
\begin{equation}
\Phi_{B_2} (x) \sim \left( C_+ e^{i (\alpha_+ - \omega \delta)} + C_- e^{i (\alpha_- - \omega \delta)} \right) e^{i \omega x} + \left( C_+ e^{-i (\alpha_+ - \omega \delta)} + C_- e^{-i (\alpha_- - \omega \delta)}\right) e^{- i \omega x} \;. \label{eq:PhiB2RNGEN}
\end{equation}

\par We match the solutions of Eqs. ({\ref{eq:PhiA2RNGEN}}) and ({\ref{eq:PhiB2RNGEN}}) by equating their coefficients of $e^{\pm i \omega x}$:
\begin{eqnarray}
B_+ e^{3i\alpha_+} + B_- e^{3i \alpha_-} & = & C_+ e^{i (\alpha_+ - \omega \delta)} + C_- e^{i (\alpha_- - \omega \delta)} \;, \label{eq:RNBC2}\\
B_+ e^{5i \alpha_+} + B_- e^{5i \alpha_-} & = & C_+ e^{-i (\alpha_+ - \omega \delta)} + C_- e^{-i (\alpha_- - \omega \delta)} \;. \label{eq:RNBC3}
\end{eqnarray} 

\noindent To rotate from $B_2$ to $B_1$, we require a further rotation in $r$ of $2 \eta$ (i.e. $2\pi$ in $(x-\delta)$); with $e^{2 \pi i} = +1$, this introduces no change. On branch $B_1$, $\omega x <0$, such that Eq. (\ref{eq:Jexpandneg}) applies and the QNM solution becomes
\begin{eqnarray}
\Phi_{B_1}  (x) \sim  \left(C_+ e^{5i\alpha_+} + C_- e^{5i \alpha_-} \right) e^{+i \omega (x-\delta)} + \left( C_+ e^{3i \alpha_+} + C_- e^{3i \alpha_-} \right) e^{-i \omega (x-\delta)} \;. \label{eq:PhiB1RNGEN}
\end{eqnarray}

\par For the global and local monodromies, only the outer horizon is taken into account. For the latter, $x$ increases by $-2 \pi i/k^+_{_H}$ along the clockwise path of $\Phi \sim e^{+i \omega x}$, such that
\begin{equation} \label{eq:MonoLocalRN}
\mathfrak{M}_{\gamma \;, \; r^+_{_H}} [ \Phi(x) ] = e^{+ \frac{\pi \omega}{k^+_{_H}}} 
\end{equation} 
\noindent is the local monodromy. Since the global monodromy is associated with the path to spatial infinity, the $e^{- i \omega x}$ coefficients of Eqs. (\ref{eq:PhiASTdSGEN}) and (\ref{eq:PhiB1RNGEN}) combine with $e^{-i \omega (\pi i/k^+_{_H})}$ to produce  
\begin{equation} \label{eq:MonoGlobalRN}
\mathfrak{M}_{\gamma } [ \Phi(x) ] = \frac{\left( C_+ e^{3i \alpha_+} + C_- e^{3i \alpha_-} \right) e^{+i \omega \delta}}{B_+ e^{+i \alpha_+} + B_- e^{+i\alpha_-}} \; e^{- \frac{\pi \omega}{k^+_{_H}}} \;.
\end{equation} 

\par With the aid of the boundary conditions introduced in Eqs. (\ref{eq:RNBC1}), (\ref{eq:RNBC2}), and (\ref{eq:RNBC3}), the aQNF can be extracted from
\begin{eqnarray}
\frac{\left( C_+ e^{3i \alpha_+} + C_- e^{3i \alpha_-} \right) e^{+i \omega \delta}}{B_+ e^{+i \alpha_+} + B_- e^{+i\alpha_-}} \; e^{- \frac{\pi \omega}{k^+_{_H}}} & = & e^{+ \frac{\pi \omega}{k^+_{_H}}}  \\
 \Rightarrow  e^{ \frac{2 \pi \omega}{k^+_{_H}}} & = & - \left( 1 + 2 \cos (\pi j) \right) - \left(2 + 2 \cos (\pi j )\right) e^{- \frac{2 \pi \omega}{k^-_{_H}}} \;. \label{eq:aQNFRN}
\end{eqnarray}  
\noindent Confirmation of this aQNF solution can be found in Refs. \cite{refAnderssonHowls,refMotlNeitzke,refBertiKokkotas}.

\par The Reissner-Nordstr{\"o}m metric lends itself naturally to assessment within different limits. Let us first consider $\vartheta \rightarrow 0$: since the metric reduces to that of the Schwarzschild form under this condition, the Cauchy horizon at $r=r^-_{_H}$ vanishes and only the outer horizon at $r=r^+_{_H}$ remains. The surface gravity defined as $k^{\pm}_{_H} = 1/2 f'(r^{\pm}_{_H})$ at each horizon then becomes
\begin{equation}
k^-_{_H} \sim -\frac{(d-3) \mu}{\; (r^-_{_H})^{d-2}} \; \rightarrow - \infty \hspace{ 0.7 cm} {\text{and}} \hspace{ 0.7 cm} k^+_{_H} \sim \frac{(d-3) \mu}{\; (r^+_{_H})^{d-2}} \;,
\end{equation}
\noindent respectively. Consequently, $e^{-2 \pi \omega/k^-_{_H}} \rightarrow 1$, and Eq. ({\ref{eq:aQNFRN}}) reduces to
\begin{equation}
e^{+\frac{2 \pi \omega}{k^+_{_H}}} = - \left( 3 + 4 \cos (\pi j ) \right) \;.
\end{equation}
\noindent As such, we see that although the Reissner-Nordstr{\"o}m BH metric resembles the Schwarzschild BH metric in the $\vartheta \rightarrow 0$ limit, the expression corresponding to the aQNF does not follow suit. 

\par  This issue was first addressed in Ref. \cite{refAnderssonHowls}. There, the real part of the aQNF of the 4D Reissner-Nordstr{\"o}m BH reduced to $\ln 5$ when $Q \rightarrow 0$, in contradiction to the $\ln 3$ of the 4D Schwarzschild BH for which $Q=0$. To explain this, Andersson and Howls suggested that two separate scales correspond to the Reissner-Nordstr{\"o}m BH problem, demarcated by the non-commuting limits $\vert \omega \vert \approx \vert \omega_I \vert \rightarrow \infty$ and $Q \rightarrow 0$. Applying the former first yields $\ln 5$; applying the latter first yields $\ln 3$. While Andersson and Howls confirm that $\ln 5$ represents the correct expression for the highly-damped aQNFs of the 4D Reissner-Nordstr{\"o}m BH, they consider the existence of an ``intermediate" damping range for which $\mathbb{R}e \{ \omega \} \approx \ln 3$ within the Reissner-Nordstr{\"o}m BH spacetime. By order of magnitude estimates for $d=4$, Refs. \cite{refAnderssonHowls,Daghigh2006RNsmallQ,Daghigh2011QMBH} define this range in terms of Newton's gravitational constant and the ADM mass:
\begin{equation}
1 \ll (GM) \vert \omega \vert \ll \left( \frac{GM}{Q} \right)^4 \;.
\end{equation}

\par Of greater interest is the $\vartheta \rightarrow \mu$ limit that characterises the extremal Reissner-Nordstr{\"{o}}m BH. While it is known that the unique topology of this extremal case requires its own individual analytical treatment \cite{refAnderssonHowls,refNatarioSchiappa,Daghigh2007eRN,Cho2006,Das2005}, there is contention within the literature on the correct way to perform such an analysis. One example lies in Ref. \cite{Das2005}, where interpretation of the anti-Stokes line behaviour at the origin (specifically, a rotation of $\pi$ in the 4D complex plane rather than the expected $5\pi/3$) has been criticised by the authors of Refs. \cite{refNatarioSchiappa,Daghigh2007eRN,Daghigh2008SAdS}. However, we see this inconsistency in the treatment of the extremal BH more clearly when comparing the results of Ref. \cite{refNatarioSchiappa} and \cite{Daghigh2007eRN}. The former produced the aQNF expression for the extremal Reissner-Nordstr{\"{o}}m BH
\begin{equation} \label{eq:aQNFeRN-NatSchiap}
e^{\frac{2 \pi \omega}{k}} = \frac{ \sin \left( \frac{5 \pi j}{2} \right)}{\sin \left( \frac{\pi j}{2} \right) } 
\end{equation} 
\noindent through an application of the monodromy technique utilised throughout this work. The latter applied the more involved phase-integral method and obtained an expression of the form 
\begin{equation} \label{eq:aQNFeRN}
e^{\frac{2 \pi \omega}{k}} = - \left( 2 + 2 \cos (\pi j ) \right) \;.
\end{equation}
\noindent To our knowledge, this is the only known example where the two analytical techniques produce different results for the aQNF expression. If we compare Fig. 5 of Ref. \cite{refNatarioSchiappa} and Fig. 1 of Ref. \cite{Daghigh2007eRN}, it is clear that both groups used the correct BH topology. However, while tracing the contour around the event horizon, the authors of Ref. \cite{refNatarioSchiappa} crossed two anti-Stokes lines connected to the horizon without applying the QNM boundary conditions, which in turn led to an incorrect monodromy. This may invalidate Eq. (\ref{eq:aQNFeRN-NatSchiap}). The more reliable result seems to be that of Eq. (\ref{eq:aQNFeRN}); subsequent discussion on the aQNFs of the extremal Reissner-Nordstr{\"o}m BH shall be based on this result. 

\par Further validation of the Eq. (\ref{eq:aQNFeRN}) is outstanding. We believe this strongly motivates for additional focus on the aQNFs of the extremal Reissner-Nordstr{\"{o}}m BH through numerical approaches, which have been sparse to date. Difficulty in producing a stable numerical method for the computation of the aQNFs of extremal Reissner-Nordstr{\"o}m BHs has also been encountered, as seen in Ref. \cite{Berti2003}. In Ref. \cite{Berti2004}, the authors stated that the direct application of Leaver's continued fraction method $-$ known to be a reliable means by which to calculate aQNFs $-$ fails in the case of the extremal Reissner-Nordstr{\"o}m BH. Through a change in variable introduced by Onozawa \textit{et al.} \cite{refOnozawa1996}, the problem can be reduced to a five-term recurrence relation \cite{Berti2004}.

\par We note with interest that Eq. (\ref{eq:aQNFeRN}) is the result determined in Ref. \cite{refNatarioSchiappa} when $\vartheta \rightarrow \mu$ is applied to Eq. (\ref{eq:aQNFRN}). Although $\vartheta \rightarrow \mu$ and $n \rightarrow \infty$ are also a set of non-commuting limits (just like $\vartheta \rightarrow 0$ and $n \rightarrow \infty$), the extremal limit of the aQNF of the non-extremal Reissner-Nordstr{\"{o}}m BH from Ref. \cite{refNatarioSchiappa} agrees with the extremal limit of the aQNF for the extremal Reissner-Nordstr{\"{o}}m BH calculated in Ref. \cite{Daghigh2007eRN}. We consider this to be a coincidence.

\par In both Eqs. (\ref{eq:aQNFeRN-NatSchiap}) and (\ref{eq:aQNFeRN}), we utilise a parameter $k$ instead of the surface gravity $k_{_H}$, where
\begin{equation}
k = \frac{1}{2} f' (r_{_H}) = \frac{1}{2} \frac{(d-3)^2}{(d-2) \mu^{\frac{1}{d-3}}} \;.
\end{equation} 
\noindent As mentioned in section \ref{sec:intro}, surface gravity is well-defined only for non-extremal BHs; Hawking radiation is therefore not associated with extremal BHs. However, we can define an analogous function $T=k/2 \pi$ \cite{refNatarioSchiappa} where
\begin{equation}
T = \frac{d-3}{d-2} \left( \frac{d-3}{4 \pi \mu^{\frac{1}{d-3}}} \right) \;.
\end{equation}
\noindent Moreover, for certain values of $j$, the real part of the aQNF includes the natural logarithm of an integer. Though provocative, Hod's conjecture cannot apply here: its underlying arguments \cite{refBekenstein1972,refMukhanov1986,refBekenstein1974} are restricted to non-extremal cases, thereby rendering Hod's argument invalid. Moreover, we must recognise that proof of the stability of perturbing fields within the extremal Reissner-Nordstr{\"o}m BH spacetime is still outstanding (see section VIII of Ref. \cite{refKonoplyaZhidenkoReview}), which implies any results within the extremal Reissner-Nordstr{\"o}m BH context must be approached with some scrutiny. Despite this, the extremal Reissner-Nordstr{\"o}m case remains a fascinating BH spacetime with extensive applications ranging from numerical development (see Ref. \cite{refOurLargeL} and references therein) to supersymmetry (SUSY) considerations \cite{Onozawa1996,Kallosh1997}, the latter of which shall be addressed in section \ref{sec:aQNFs}.

\subsubsection{Schwarzschild AdS BHs \label{subsubsec:SAdSnatschiap}}

\par For static and spherically-symmetric AdS BHs, the absence of a closed contour implies that monodromy considerations cannot be incorporated. Consequently, for the Schwarzschild AdS BH Nat{\'a}rio and Schiappa's approach becomes a simple matter of matching solutions across the path traced in Fig. \ref{fig:SAdS}. 

\par We begin in the region near spatial infinity, on branch $B$. The form of the potential is given in Eq. (\ref{eq:genpotAdS}); the corresponding solution is of the form of Eq. (\ref{eq:genQNMBessel}). Since $\omega x <0 $ on this branch, we apply the asymptotic expansion of Eq. (\ref{eq:Jexpandneg}). However, the vanishing energy flux boundary conditions we employ in this work nullify the $C_-$ term. Thus,

\begin{equation} 
\Phi_{B_{r \sim \infty}} (x) \sim  C_+ e^{+i \beta_+} e^{+ i \omega ( x - x_0) } + C_+ e^{-i \beta_+} e^{- i \omega ( x - x_0) } \;, \label{eq:PhiBSAdSinfGEN}
\end{equation}
\noindent where we introduce $\beta_+ = \left( 1 + j^{\infty} \right)\pi/4.$

\par The path is traced from spatial infinity to the neighbourhood of the origin. The potential for $r \sim 0$, Eq. (\ref{eq:genpot}), yields a QNM solution of the form of Eq. (\ref{eq:genQNMBessel}), which decomposes under Eq. ({\ref{eq:Jexpandneg}}) to 
\begin{eqnarray}
\Phi_{B_{r \sim 0}} (x) \sim  \left( B_{+} e^{-i \alpha_{+}} + B_{-} e^{-i \alpha_{-}} \right) e^{i \omega x} + \left( B_{+} e^{i \alpha_{+}} + B_{-} e^{i \alpha_{-}} \right) e^{-i \omega x} \;. \label{eq:PhiBSAdS0GEN}
\end{eqnarray}

\noindent We equate the coefficients of $e^{\pm i \omega x}$ from Eqs. (\ref{eq:PhiBSAdSinfGEN}) and (\ref{eq:PhiBSAdS0GEN}). These resultant expressions should also be equivalent, such that
\begin{equation} \label{eq:SAdSBC1}
 \left( B_{+} e^{-i \alpha_{+}} + B_{-} e^{-i \alpha_{-}} \right) e^{-i \beta_+} e^{+i \omega x_0} = \left( B_{+} e^{i \alpha_{+}} + B_{-} e^{i \alpha_{-}} \right) e^{+i \beta_+} e^{-i \omega x_0}
\end{equation}
\noindent serves as a boundary condition for the aQNF calculation.

\par To move from branch $B$ to $A$, the rotation in $r$ is through $- \eta$ (i.e. $-\pi$ in $x$). Since $e^{-i \pi} = -1$ and $\omega x > 0$ on branch $A$, Eq. (\ref{eq:Jexpandpos}) applies and the general solution on branch $A$ becomes
\begin{equation}
\Phi_A  (x) \sim \left( B_+ e^{-i \alpha_+} + B_- e^{- i \alpha_-} \right) e^{+ i \omega x} + \left( B_+ e^{-3 i \alpha_+} + B_- e^{-3 i \alpha_- } \right) e^{- i \omega x} \;.
\label{eq:PhiASAdS0GEN}
\end{equation}

\noindent We then obtain the final boundary condition by applying Eq. (\ref{eq:BCin}) to Eq. (\ref{eq:PhiASAdS0GEN}), such that $\Phi_A  (x) \sim e^{+ i \omega x}$. This yields
\begin{equation}
B_+ e^{-3 i \alpha_+} + B_- e^{-3 i \alpha_- } = 0 \;. \label{eq:SAdSBC2}
\end{equation}

\par To solve for the aQNF, we treat Eqs. (\ref{eq:SAdSBC1}) and (\ref{eq:SAdSBC2}) as simultaneous equations. The aQNF expression emerges as
\begin{eqnarray} 
&& \Big( i e^{\frac{3 \pi i j}{2} + 2 i \beta }
+3 i e^{2 i (\beta + \pi j)}
+i e^{\frac{i}{2} (4 \beta +\pi  j)}
+3 i e^{2 i \beta }
-3 e^{\frac{3 \pi i  j}{2} + 2 i  \omega x_0 }
+e^{2 i ( \omega x_0 +\pi  j)} \hspace{0.6cm}
 \nonumber \\
&& \hspace{1cm} -3 e^{\frac{i}{2} (4 \omega x_0 +\pi  j)} + e^{2 i \omega x_0}\Big) \hspace{0.2cm}\times e^{-i (\beta + \omega x_0 + \pi  j)}  =0 \label{eq:aQNFSAdS}\;.
\end{eqnarray}
\noindent This is in agreement with Refs. \cite{refNatarioSchiappa,Lopez-Ortega2006,Daghigh2008SAdS,refCardosoNatarioSchiappa}, where the solution for scalar- and tensor-modes of the gravitational aQNFs ($j=0$) were found to be
\begin{equation}
\lim_{n \rightarrow \infty} \omega x_0 = \frac{\pi}{4} + \beta_+ - \arctan \left( \frac{i}{3} \right) + n \pi \;,
\end{equation}
\noindent while the vector-modes ($j=2$) yielded 
\begin{equation}
\lim_{n \rightarrow \infty} \omega x_0 = \frac{3\pi}{4} + \beta_+ - \arctan \left( \frac{i}{3} \right) + n \pi
\end{equation}
\noindent for $n \in \mathbb{N}$. These can be further generalised into a single expression,
\begin{equation}
\lim_{n \rightarrow \infty} \omega x_0 = \frac{\pi}{4}(d+1) - \arctan \left( \frac{i}{3} \right) + n \pi
\end{equation}
\noindent for $n \in \mathbb{N}$, where the phase change of $\beta_+$ between solutions can be absorbed into $\pi (d+1)/4$ \cite{refNatarioSchiappa,Daghigh2008SAdS}.  
\begin{figure}[t]
  \centering
  \includegraphics[width=.55\textwidth]{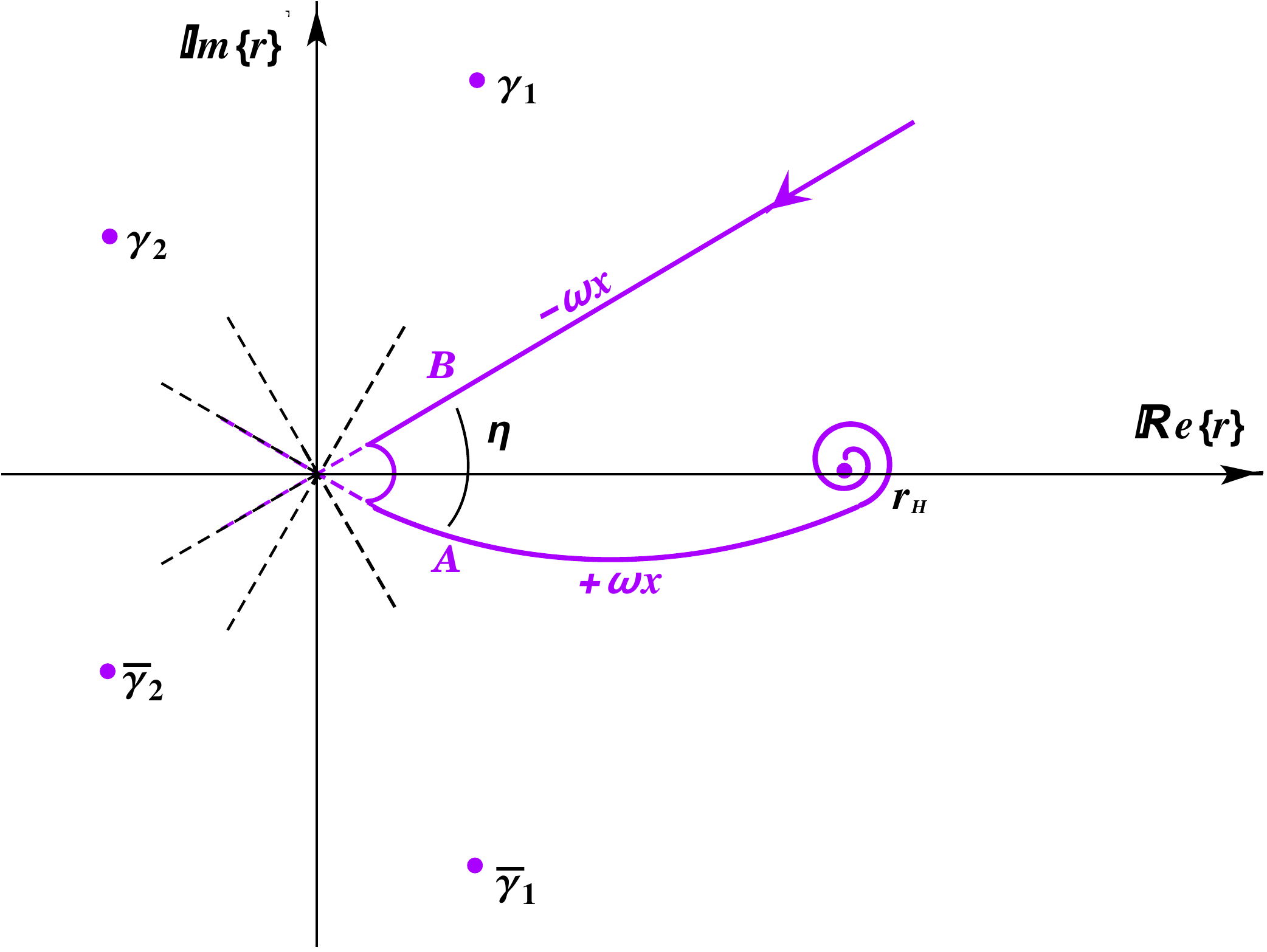}
  \caption{\textit{Anti-Stokes lines for a Schwarzschild AdS BH with $d=6$. \\Note the absence of a closed contour.}}
  \label{fig:SAdS}
\end{figure} 

\par While the aQNF of Eq. ({\ref{eq:aQNFSAdS}}) holds true under the aforementioned circumstances, there are two known exceptions reported in Ref. \cite{refNatarioSchiappa}, $viz.$ the scalar-mode gravitational perturbations of the $d=4$ and $d=5$ Schwarzschild AdS BHs. For the former, the aQNF calculation yields
\begin{equation}
\lim_{n \rightarrow \infty} \omega x_0 = \frac{3\pi }{4} - \arctan \left( \frac{i}{3} \right) + n \pi \; \; \; (n \in \mathbb{N}) \;,
\end{equation}
\noindent as $j^{\infty}$ must be set to $+1$ rather than $-1$ \cite{refNatarioSchiappa}. For $d=5$, we are required to select $j^{\infty} \rightarrow 0^+$. Consequently, $(1 \pm j^{\infty})/2 > 0$, and the solution at $r \thicksim \infty$ therefore vanishes. This leaves but a single constraint, emerging from the BH horizon, which is insufficient to quantise the aQNFs. Thus, a continuous spectrum of aQNFs are produced (where $\omega \in {\mathbb{C}}$) for the 5D Schwarzschild AdS BH in the wake of spin-2 scalar-mode perturbations. In Ref. \cite{Lopez-Ortega2006}, these arguments were shown to apply also to the scalar-mode perturbations of electromagnetic fields.

\section{\label{sec:fields}Behaviour of potentials near singular points}

\par To reduce the effective BH QNM potentials developed in the literature to a generalised form, we extract only the dominant-$r$ term of each potential and incorporate the appropriate tortoise coordinate expression. In so doing, only the definition of the $j$ (and $j^{\infty}$) parameter(s) relays information about the spin of the perturbing field. We find that within the neighbourhood of the origin, the value of $\lambda$ does not influence the outcome of the potential: all potentials take on the form given in Eq. (\ref{eq:genpot}). Near spatial infinity, however, the potentials in asymptotically flat BH spacetimes vanish; integer-spin and spin-3/2 perturbations in BH spacetimes with $\lambda \neq 0$ adopt a common form provided in Eq. (\ref{eq:genpotAdS}).

\subsection{Potentials of integer spin \label{subsec:intpots}}
\par In Refs. \cite{refIKschwarz1,refIKschwarz2,refIKrn,refIKchap6}, Ishibashi and Kodama provide the effective potentials associated with the tensor-, vector-, and scalar-modes of the gravitational perturbations of $d$-dimensional, stationary, spherically-symmetric vacuum BH spacetimes. The vector- and scalar-modes are the higher-dimensional extensions of the axial/odd-parity and polar/even-parity modes, respectively, as described in Refs. \cite{refRW,refZerilli,ZerMon1,ZerMon2,ZerMon3,ZerMon4}. The tensor-mode of these spin-2 perturbations arises to account for the extra degrees of freedom beyond $d =4$ and has been observed to possess the same form as the spin-0 (scalar) perturbations \cite{refBertiCardoso,refKonoplyaZhidenkoReview}. Electromagnetic perturbations similarly require an additional scalar-mode potential, beyond the vector-mode that suffices for their 4D description, in higher dimensions; these are provided in Refs. \cite{Casals2018,refCrispinoHiguchiMatsas}. 

\subsubsection{In the neighbourhood of $r \sim 0$}
\par For the gravitational perturbations of Schwarzschild BHs inclusive and exclusive of a cosmological constant, we observe that the scalar- and tensor-modes collapse identically. The emergent form also describes the spin-0 perturbations. Thus,
\begin{equation}
V^s \sim V^{grav}_{_T} \sim V^{grav}_{_S} \sim \frac{f(r)}{r^2} \left[+ \frac{(d-2)^2 \mu}{2r^{d-3}} \right] \sim -\frac{(d-2)^2 \mu^2}{r^{2(d-2)}} \sim - \frac{1}{4x^2} \;,
\end{equation}
\begin{equation}
V^{grav}_{_V} \sim \frac{f(r)}{r^2} \left[- \frac{3(d-2)^2 \mu}{2r^{d-3}} \right] \sim +\frac{3(d-2)^2 \mu^2}{r^{2(d-2)}} \sim + \frac{3}{4x^2} \;,
\end{equation}
\noindent such that $j_1=0$ and $j_2=2$, respectively \cite{refIKschwarz1,refIKschwarz2,refIKchap6}.

\par From the electromagnetic perturbations of Schwarzschild BHs studied in Refs. \cite{Casals2018,refCrispinoHiguchiMatsas}, we obtain
\begin{equation}
V^{EM}_{_S} \sim \frac{f(r)}{r^2} \left[ \frac{(d-2)(d-4)f(r)}{4} - \frac{(d-4)r}{2} f'(r) \right] \sim \frac{(3d-8)(d-4)}{4(d-2)^2x^2} \;,
\end{equation}

\begin{equation}
V^{EM}_{_V} \sim \frac{f(r)}{r^2} \left[ \frac{(d-4)(d-6)f(r)}{4} + \frac{(d-4)r}{2} f'(r) \right] \sim -\frac{d(d-4)}{4(d-2)^2x^2}  \;,
\end{equation}
\noindent where $j_s=2(d-3)/(d-2)$ and $j_v=2/(d-2)$. As observed in Ref. \cite{Lopez-Ortega2006}, $j_s+j_v=2$. Furthermore, as $d \rightarrow \infty$, $j_s \rightarrow 2$ and $j_v \rightarrow 0$: these are the $j_i$ values corresponding to the scalar- and vector-modes of the gravitational perturbations, indicating that the scalar- and vector-modes of electromagnetic and gravitational fields become increasingly identical for higher dimensional contexts. Note also that $j_s=j_v=1$ for $d=4$, in keeping with the results of Ref. \cite{Cho2006}.

\par For the gravitational perturbations of Reissner-Nordstr{\"o}m BHs inclusive and exclusive of a cosmological constant, spin-0, scalar- and tensor-modes of spin-2 perturbations share a common form; the vector-modes are expressed separately. Thus,
\begin{equation}
V^s \sim V^{grav}_{_T} \sim V^{grav}_{_{\pm S}}  \sim - \frac{(d-2)(3d-8)}{4r^{4d-10}} \; \vartheta^4 \sim - \frac{(d-2)(3d-8)}{4(2d-5)^2 x^2}\;,
\end{equation}
\begin{equation}
V^{grav}_{_{\pm V}} \sim + \frac{(d-2)(5d-12)}{4r^{4d-10}} \; \vartheta^4 \sim + \frac{(d-2)(5d-12)}{4(2d-5)^2 x^2}  \;,
\end{equation}
\noindent for $ j_1 = (d-3)/(2d-5)$ and $j_2 = (3d-7)/(2d-5)$. When $d=4$, $V^s \sim -2\vartheta^4/r^6$ and $V^{grav}_{_{\pm V}}\sim 4\vartheta^4/r^6$, in agreement with Ref. \cite{Cho2006}.

\subsubsection{In the neighbourhood of $r \sim \infty$}
\par From the potentials cited above, a universal term of $f(r)/r^2$ prefixes each expression. Since $f(r) \sim 1$ for both Schwarzschild and Reissner-Nordstr{\"o}m BHs in asymptotically-flat space under the $r \sim +\infty$ condition, $V[r(x)] \sim 0$ near spatial infinity. 

\par For Schwarzschild (A)dS BH spacetimes, we exploit $f(r) \sim - \lambda r^2$ and the subsequent expression for $x$ given in Eq. ({\ref{eq:tortads}}) to claim that $\pm \vert \lambda \vert r \sim (x-x_0)^{-1}$, such that the potentials for the tensor-, vector-, and scalar-modes of the gravitational perturbations become

\begin{equation}
V^s \sim V^{grav}_{_T} \sim \frac{f(r)}{r^2} \left[ - \frac{d(d-2)}{4} \lambda r^2 \right] \sim \frac{-\lambda r^2}{r^2}\left[ - \frac{d(d-2)}{4} \lambda r^2 \right]  \sim \frac{(d-1)^2-1}{4 (x-x_0)^2}\;,
\end{equation}
\begin{equation}
V^{grav}_{_{S}} = \frac{f(r)}{16r^2} \frac{U(r)}{H(r)^2} \sim -\frac{\lambda r^2}{16r^2} \left[- 4\lambda r^2 (d-4)(d-6) \right]  \sim  \frac{(d-5)^2-1}{4(x-x_0)^2} \;,
\end{equation}
\begin{equation}
V^{grav}_{_{V}} \sim \frac{f(r)}{r^2} \left[ - \frac{(d-2)(d-4)}{4} \lambda r^2 \right] \sim  \frac{(d-3)^2-1}{4(x-x_0)^2} \;. 
\end{equation}
\noindent Note that although the spin-0 and the tensor-mode of the spin-2 perturbations remain identical, we lose the commonality between gravitational scalar- and tensor-modes. 

\par For the electromagnetic perturbations in Schwarzschild (A)dS BH spacetimes, we return to the expressions given in Ref. \cite{refCrispinoHiguchiMatsas}. However, the dominant-$r$ terms are now those containing the cosmological constant, such that

\begin{equation}
V^{EM}_{_S} \sim \frac{f(r)}{r^2} \left[ \frac{(d-2)(d-4)f(r)}{4} - \frac{(d-4)r}{2} f'(r) \right]  \sim \frac{(d-5)^2-1}{4(x-x_0)^2} \;,
\end{equation}

\begin{equation}
V^{EM}_{_V} \sim \frac{f(r)}{r^2} \left[ \frac{(d-4)(d-6)f(r)}{4} + \frac{(d-4)r}{2} f'(r) \right]  \sim \frac{(d-3)^2-1}{4(x-x_0)^2} \;.
\end{equation}
\noindent Here, we see that the scalar- and vector-modes of gravitational and electromagnetic perturbations have the same behaviour within the neighbourhood of spatial infinity, $viz.$ $j^{\infty}_s=d-5$ and $j^{\infty}_v=d-3$, respectively. For scalar- and vector-modes of the electromagnetic perturbations, $(j_i^{\infty})^2=1$ for $d=4$ such that the perturbation vanishes for $r \sim \infty$.

\subsection{Potentials of half-integer spin \label{subsec:halfintpots}}
\par As observed in Ref. \cite{refSUSYpot}, the effective potentials corresponding to perturbing fields of half-integer spin can be expressed through
\begin{equation} \label{eq:SUSYpot}
V_{\pm} =\pm F(r) \frac{d}{dr}W + W^2 \;,
\end{equation}
\noindent where $W$ represents the superpotential and $F(r)$ is a function of $f(r)$. This expression applies to the Dirac fields, as well as to the Rarita-Schwinger fields studied in the gauge-invariant formalism of Refs. \cite{refRS15,refRSSchwarz16,refRSRN18,refRSSAdS19}. To compute the aQNFs via the monodromy technique, we approximate these QNM effective potentials into the form of Eqs. ({\ref{eq:genpot}}) and ({\ref{eq:genpotAdS}}), and thereby extract the characteristic $j$ behaviour. This can be achieved through a simple asymptotic analysis applied to spin-1/2 and spin-3/2 fields, where we consider both the transverse-traceless (TT) and non transverse-traceless (nonTT) eigenmodes of the latter.

\subsubsection{In the neighbourhood of $r \sim 0$}
\par From our study of integer perturbations within Schwarzschild BH spacetimes, we expect the potentials to approximate to $V[r(x)] \propto r^{-2(d-2)} \sim x^{-2}$. For spin-1/2 perturbations \cite{refDirac07}, the superpotential yields expressions of the order of
\begin{eqnarray}
W \propto \frac{r^{-(d-3)/2}}{r} = \frac{1}{r^{(d-1)/2}} 
\hspace{0.5cm} & \Rightarrow & \hspace{0.2cm} f(r)\frac{dW}{dr} \propto \frac{1}{r^{d-3}} \frac{1}{r^{(d+1)/2}} = \frac{1}{r^{(3d-5)/2}} \;,
\end{eqnarray}  
\noindent such that 
\begin{equation}
V^D_{\pm} \sim \pm f(r) \frac{dW}{dr} \propto \frac{1}{x^{(3d-5)/(2(d-2))}} \;,
\end{equation}
\noindent with the incorporation of $r \propto x^{-(d-2)}$ from Eq. (\ref{eq:tortst}). By virtue of the fact that 
\begin{equation}
\frac{3d-5}{2(d-2)} = \frac{3}{2} + \frac{1}{2(d-2)} \; < \; 2 \hspace{0.3cm} \forall \hspace{0.2cm} d \geq 4 \;,
\end{equation} 
\noindent we surmise that $j^2-1$ of Eq. (\ref{eq:genpot}) must vanish, such that $j = \pm 1$.

\par For spin-3/2 perturbations in Schwarzschild BH spacetimes, the superpotential of the TT eigenmodes is given by $W = \zeta \sqrt{f(r)}/r^2$, where $\zeta$ is the spinor-vector eigenvalue \cite{refRSSchwarz16}. The Dirac analysis therefore applies to these potentials, such that $j = \pm 1$.

\par The superpotential of the nonTT eigenmodes \cite{refRSSchwarz16}, however, is 
\begin{equation}
W = \frac{\sqrt{f(r)}}{r} \; \kappa \; \left[ \frac{\kappa^2 - \frac{(d-2)^2}{4} \left( 1 + \frac{d-4}{d-2} \frac{2 \mu}{r^{d-3}}\right)}{\kappa^2 - \frac{(d-2)^2}{4} \left( 1 - \frac{2 \mu}{r^{d-3}}\right)} \right] \propto \frac{1}{r^{(d-1)/2}} \; [\; r^0 \;] \;,
\end{equation} 
\noindent where $\kappa$ is related to the spinor eigenvalue on $S^{d-2}$ (see Refs. \cite{refRSSchwarz16} for details). The $r$-dependence of the superpotential is identical to that of the Dirac case. Consequently, $j = \pm 1$. Though the expressions for both the TT and nonTT eigenmodes of the spin-3/2 fields in Schwarzschild (A)dS BH spacetimes appear more complicated (see Refs. \cite{refRSSAdS19,refnewQNMs2020}), they reduce in a similar fashion such that $j = \pm 1$ holds true.

\par From our study of integer fields in Reissner-Nordstr{\"o}m BH spacetimes, we find that $V[r(x)] \propto r^{-2(2d-5)} \sim x^{-2}$. For the spin-1/2 perturbations \cite{refChakrabarti}, the superpotential yields expressions of the order of
\begin{eqnarray}
W \propto \frac{r^{-(d-3)}}{r} = \frac{1}{r^{d-2}} 
\hspace{0.5cm} & \Rightarrow & \hspace{0.2cm} f(r)\frac{dW}{dr} \propto \frac{1}{r^{2(d-3)}} \frac{1}{r^{d-1}} = \frac{1}{r^{3d-7}} \;, 
\end{eqnarray}  
\noindent such that 
\begin{equation}
V^{D}_{\pm} \sim \pm f(r) \frac{dW}{dr} \propto \frac{1}{x^{(3d-7)/(2d-5)}} \;,
\end{equation}
\noindent with the incorporation of $r \propto x^{-(2d-5)}$ from Eq. (\ref{eq:tortrn}). By virtue of the fact that 
\begin{equation}
\frac{3d-7}{2d-5} = \frac{3}{2} + \frac{1}{2(2d-5)} \; < \; 2 \hspace{0.3cm} \forall \hspace{0.2cm} d \geq 4 \;,
\end{equation} 
\noindent we surmise that $j^2-1$ of Eq. (\ref{eq:genpot}) must vanish, such that $j = \pm 1$.

\par For spin-3/2 perturbations in Reissner-Nordstr{\"o}m BH spacetimes, as studied in Ref. \cite{refRSRN18}, the superpotential associated with the nonTT eigenmodes becomes
\begin{eqnarray}
 W & = & \frac{\sqrt{f(r)}}{r} \; \left(\kappa + \frac{d-2}{2} \frac{\vartheta}{r^{d-3}}\right) \left[1 + \frac{(d-2)(d-3)}{2 \left(\kappa + \frac{d-2}{2} \frac{\vartheta}{r^{d-3}} \right)} \left( \frac{(1-f(r))\kappa + \frac{d-2}{2} \frac{\vartheta}{r^{d-3}}}{\frac{(d-2)^2}{4}f(r) - \left( \kappa + \frac{d-2}{2} \frac{\vartheta}{r^{d-3}}\right)^2} \right) \right] \nonumber \\
 & \sim & \frac{\vartheta^2}{r^{2d-5}} \left( \frac{d-2}{2} \right) \left(  \frac{(d-2)^2 \mu + 2 (2 d-5) \vartheta  \kappa }{ (d-2)^2 \mu + 2 (d-2) \vartheta  \kappa } \right) 
\end{eqnarray}
\noindent when $r \sim 0.$ Here, both $f(r)dW/dr$ and $W^2$ contribute equally to the nonTT spin-3/2 potential. With the incorporation of $\vartheta^4/r^{4d-10} \sim (2d-5)^2 x^2$ (see eq. (\ref{eq:tortrn})),
\begin{eqnarray}
V^{RS_{non}}_{+} &\sim & -\frac{\left[ (d-2)^2 \mu + 2 (2 d-5) \vartheta  \kappa \right] 
\left[ (3d-8)(d-2) \mu + 2 (2 d-5) \vartheta  \kappa \right] }{4 \left( (d-2)^2 \mu + 2\vartheta  \kappa \right)^2 (2d-5)^2 x^2} \;, \\
V^{RS_{non}}_{-} & \sim & +\frac{\left[ (d-2)^2 \mu + 2 (2 d-5) \vartheta  \kappa \right] 
\left[ (5d-12)(d-2) \mu + 6 (2 d-5) \vartheta  \kappa \right] }{4 \left( (d-2)^2 \mu + 2\vartheta  \kappa \right)^2 (2d-5)^2 x^2}  \;.
\end{eqnarray}
\noindent Throughout this work, the superscript $RS_{non}$ denotes non-TT Rarita Schwinger perturbations. Despite the complicated nature of the expressions, these potentials may still be reduced to the form of eq. (\ref{eq:genpot}), where \begin{eqnarray}
j_+ & = & \frac{(d-3)(d-2)\mu}{\left((d-2)\mu + 2\vartheta  \kappa \right) (2d-5)} \;, \label{eq:jRNplus}\\
j_- & = & \frac{(3d-7)(d-2)\mu + 4 (2 d-5) \vartheta  \kappa}{\left((d-2)\mu + 2\vartheta  \kappa \right) (2d-5)} \label{eq:jRNminus}\;,
\end{eqnarray}
\noindent for $V^{RS_{non}}_{+}$ and $V^{RS_{non}}_{-}$, respectively.

\par We note with interest that in the $\mu \gg \vartheta \kappa$ limit,
\begin{eqnarray}
V^{RS_{non}}_{+} \Big \vert_{\mu \gg \; \vartheta \kappa} \equiv \widetilde{V}^{RS_{non}}_{+} &\sim & -\frac{ (3d-8)(d-2) }{4(2d-5)^2 x^2} \;, \\
V^{RS_{non}}_{-} \Big \vert_{\mu \gg \; \vartheta \kappa} \equiv \widetilde{V}^{RS_{non}}_{-} & \sim & +\frac{(5d-12)(d-2)}{4 (2d-5)^2 x^2}  \;.
\end{eqnarray}
\noindent We shall use a tilde to denote expressions in this limit. These potentials are identical to the scalar-/tensor- and vector-modes of the gravitational perturbations of the Reissner-Nordstr{\"o}m BH spacetime, respectively. As such, we find that $\widetilde{j}_+ = (d-3)/(2d-5)$ and $\widetilde{j}_- = (3d-7)/(2d-5)$. Furthermore, we find that $\widetilde{V}^{RS_{non}}_{+}=-2\vartheta^4/r^6$ for $d=4$ in this limit, which is equivalent to the 4D spin-3/2 result of Ref. \cite{Cho2006}.

\par For the TT spin-3/2 eigenmodes in Reissner-Nordstr{\"o}m BH spacetimes,
\begin{equation}
W = \frac{\sqrt{f(r)}}{r} \; \left(\zeta - \frac{d-2}{2} \frac{\vartheta}{r^{d-3}}\right) \sim - \frac{d-2}{2} \frac{\vartheta^2}{r^{2d-5}} \;.
\end{equation}
\noindent Thus,
\begin{eqnarray}
V^{RS_{TT}}_{+} &\sim & + \left(\frac{\vartheta^2}{r^{2d-6}}\right) \frac{d}{dr} \left( -\frac{d-2}{2} \frac{\vartheta^2}{r^{2d-5}} \right) + \frac{(d-2)^2}{4} \frac{\vartheta^4}{r^{4d-10}} \sim +\frac{(d-2)(5d-12)}{4(2d-5)^2 x^2} \;, \\
V^{RS_{TT}}_{-} & \sim & - \left(\frac{\vartheta^2}{r^{2d-6}}\right) \frac{d}{dr} \left( - \frac{d-2}{2} \frac{\vartheta^2}{r^{2d-5}} \right) + \frac{(d-2)^2}{4} \frac{\vartheta^4}{r^{4d-10}} \sim - \frac{(d-2)(3d-8)}{4(2d-5)^2 x^2}\;.
\end{eqnarray}
\noindent Here, $j_+ = (3d-7)/(2d-5)$ for $V^{RS_{TT}}_{+}$ and $j_- = (d-3)/(2d-5)$ for $V^{RS_{TT}}_{-}$, which also correspond to the vector- and scalar-/tensor modes of the gravitational perturbations, respectively. As such, $V^{RS_{TT}}_{\pm}$ matches $\widetilde{V}^{RS_{non}}_{\mp}$ within the neighbourhood of the origin.

\subsubsection{In the neighbourhood of $r \sim \infty$}
\par As in the case of integer-spin fields, $f(r) \sim 1$ for both Schwarzschild and Reissner-Nordstr{\"o}m BHs in asymptotically-flat space under the $r \sim +\infty$ condition. As such, $V[r(x)] \sim 0$ near spatial infinity. 
\par In (A)dS BH spacetimes, $f(r) \sim - \lambda r^2$ near spatial infinity. Thus, for spin-1/2 perturbations in Schwarzschild and Reissner-Nordstr{\"o}m BHs inclusive of a cosmological constant,
\begin{eqnarray}
W &\sim & \frac{\sqrt{-\lambda r^2}}{r} \; \kappa = (-\lambda)^{1/2} \; \kappa \nonumber \\
& \Rightarrow & V^{D}_{\pm} \sim \lambda \; \kappa^2 \;.
\end{eqnarray}
\noindent With $V^D_{\pm} \propto r^0$ for both Schwarzschild (A)dS and Reissner-Nordstr{\"o}m (A)dS BH spacetimes, we consider $j^{\infty}= \pm 1$. 

\par For spin-3/2 perturbations in Schwarzschild (A)dS BH spacetimes \cite{refRSSAdS19}, we consider
\begin{equation} \label{eq:RSSdSnontt}
V^{RS_{non}}_{\mp} =\mp \partial_x W + W^2 \;, \hspace{0.5cm}  W = \left[ {\cal D}^2 - {\cal B}^2 \right]^{1/2} f^{-1} {\cal F} \;,
\end{equation}
\noindent for the nonTT eigenmodes. Here, $\partial_x = {\cal F} \partial_r \;,$ for which
\begin{equation}
{\cal F} = f(r) \left[ 1 + \frac{f(r)}{2 \omega} \left( \frac{\partial}{\partial r} \frac{{\cal D}}{i {\cal B}} \right) \left(\frac{ {\cal B}^2}{{\cal B}^2 - {\cal D}^2} \right) \right]^{-1} \;.
\end{equation} 
\noindent Since the asymptotic limit demands that $\vert \omega \vert \rightarrow + \infty$, we may approximate ${\cal F} \approx f(r)$. Thus, $W = \left[ {\cal D}^2 - {\cal B}^2 \right]^{1/2}$.

\par With the use of
\begin{equation} 
z = \frac{- \frac{(d-3)(d-2)}{2} \frac{2 \mu}{r^{d-3}}}{\kappa^2 - \frac{(d-2)^2}{4} \left(1 - \frac{2 \mu}{r^{d-3}} \right)} \propto 0 \;,
\end{equation}
\noindent we may reduce ${\cal D}$ and ${\cal B}$ in the following manner:
\begin{eqnarray}
{\cal B} & = & i \kappa \frac{\sqrt{f(r)}}{r} \; (z+1) \hspace{0.2cm} \sim i (-\lambda)^{1/2} \; \kappa \;, \\ 
{\cal D} & = & -i \sqrt{\lambda f(r)} \;\frac{(d-2)}{2} \; \left( z+ \frac{d-4}{d-2} \right) \sim -i \frac{(d-4)}{2} (-\lambda^2)^{1/2} r = \frac{(d-4)}{2} \vert \lambda \vert r \; .
\end{eqnarray}
\noindent Therefore,
\begin{equation}
W \sim \sqrt{ \frac{(d-4)^2}{4} \lambda^2  r^2- \lambda \kappa ^2 } \sim \frac{(d-4)}{2} \vert \lambda \vert r \;,
\end{equation}
\noindent such that
\begin{eqnarray}
V^{RS_{non}}_{-} & = & - \partial_x W + W^2 \sim -(-\lambda r^2)\frac{(d-4)}{2} \vert \lambda \vert + \frac{(d-4)^2}{4} \vert \lambda \vert^2 r^2  \sim  \frac{(d-4)(d-6)}{4(x-x_0)^2} \;,\\
V^{RS_{non}}_{+} & = & + \partial_x W + W^2 \sim +(-\lambda r^2)\frac{(d-4)}{2} \vert \lambda \vert + \frac{(d-4)^2}{4} \vert \lambda \vert^2 r^2  \sim \frac{(d-2)(d-4)}{4(x-x_0)^2}\;.
\end{eqnarray}
\noindent Note that have imposed $-\lambda = \vert \lambda \vert$ within the potentials, in keeping with the arguments of Ref. \cite{refnewQNMs2020}. From $V^{RS_{non}}_{-}$ and $V^{RS_{non}}_{+}$, respectively, we determine that $j^{\infty}_- =  d-5$ and $j^{\infty}_+ =  d-3$. The behaviour of $V^{RS_{non}}_{-}$ corresponds to $V^{grav}_{_S}$ and $V^{EM}_{_S}$ while $V^{RS_{non}}_{+}$ matches that of $V^{grav}_{_V}$ and $V^{EM}_{_V}$ near spatial infinity.

\par Similarly, for the TT eigenmodes,
\begin{equation}
V^{RS_{TT}}_{\mp} =\mp \partial_x \mathbb{W} + \mathbb{W}^2  \;, \hspace{0.5cm} \mathbb{W} = \left[ \mathbb{D}^2 - \mathbb{B}^2 \right]^{1/2} f^{-1} \mathbb{F} \;,
\end{equation}
where $\partial_x = \mathbb{F} \partial_r$, with
\begin{equation}
\mathbb{F} = f(r) \left[ 1 + \frac{f(r)}{2 \omega} \left( \frac{\partial}{\partial r} \frac{{\mathbb D}}{i {\mathbb B}} \right) \left(\frac{ {\mathbb B}^2}{{\mathbb B}^2 - {\mathbb D}^2} \right) \right]^{-1} \;.
\end{equation}
\noindent We once again claim that $\vert \omega \vert \rightarrow + \infty$ implies ${\mathbb F} \sim f(r)$. Thus, $W \sim \left[ {\mathbb D}^2 - {\mathbb B}^2 \right]^{1/2}$.

\par We may reduce ${\mathbb D}$ and ${\mathbb B}$ in the following manner:
\begin{eqnarray}
{\mathbb B} & = & i \zeta \frac{\sqrt{f(r)}}{r} \sim i(-\lambda)^{1/2} \; \zeta  \;, \\ 
{\mathbb D} & = & -i \sqrt{\lambda f(r)} \;\frac{(d-2)}{2} 
\sim -i \frac{(d-2)}{2} (-\lambda^2)^{1/2} r = \frac{(d-2)}{2} \vert \lambda \vert r \; 
\end{eqnarray}
\noindent Then
\begin{equation}
{\mathbb{W}} \sim \sqrt{\frac{(d-2)^2}{4} \lambda^2  r^2- \lambda \zeta^2} \sim \frac{(d-2)}{2} \vert \lambda \vert r \;,
\end{equation}
\noindent such that
\begin{eqnarray}
V^{RS_{TT}}_{-} & = & - \partial_x W + W^2 \sim -(-\lambda r^2)\frac{(d-2)}{2} \vert \lambda \vert + \frac{(d-2)^2}{4} \vert \lambda \vert^2 r^2 
\sim  \frac{(d-2)(d-4)}{4(x-x_0)^2}\;,\\
V^{RS_{TT}}_{+} & = & + \partial_x W + W^2 \sim +(-\lambda r^2)\frac{(d-2)}{2} \vert \lambda \vert + \frac{(d-2)^2}{4} \vert \lambda \vert^2 r^2 \sim  \frac{ d(d-2)}{4(x-x_0)^2} \;.
\end{eqnarray}
\noindent Note that have once again imposed $-\lambda = \vert \lambda \vert$ within the potentials, in keeping with the arguments of Ref. \cite{refnewQNMs2020}. From $V^{RS_{TT}}_{-}$ and $V^{RS_{TT}}_{+}$, respectively, we determine that $j^{\infty}_- =  d-3$ and $j^{\infty}_+ =  d-1$. The behaviour of $V^{RS_{TT}}_{+}$ matches that of $V^s$ and $V^{grav}_{_T}$ while $V^{RS_{TT}}_{-}$ corresponds to $V^{grav}_{_V}$, $V^{EM}_{_V}$, and $V^{RS_{non}}_{+}$ near spatial infinity.

\section{AQNF expressions of spherically-symmetric BH spacetimes \label{sec:aQNFs}}
\par The final results for QNFs corresponding to perturbing fields of spin $s \in \{0,1/2,1,3/2,2\}$ within the large-$n$ limit are produced by incorporating the appropriate expressions of $j$ (and $j^{\infty}$) into the generalised aQNF expressions of subsection \ref{subsec:aQNF}. Where $j$ is a dimensionally-independent parameter, the aQNF expressions may be considered applicable for $d>3$ \cite{refNatarioSchiappa}. However, since most aQNF expressions feature dimensionally-dependent $j$ and $j^{\infty}$ parameters, we specify that such results hold provided their associated perturbing fields are stable within the BH spacetime of interest (see Table I of Ref. \cite{refIKchap6} and Table IV of Ref. \cite{refKonoplyaZhidenkoReview} for known stable contexts). As far as possible, we validate our new half-integer results against extant aQNF expressions available in the literature.
\par Note that in the following tables (Tables \ref{table:aQNFST}-\ref{table:aQNFSAdS}), horizontal lines demarcate common final aQNF solutions: when the specified $j$ and/or $j^{\infty}$ value(s) are substituted into the equation provided for some specified $d$, the aQNF output is identical for grouped perturbations.

\subsection{aQNFs of the Schwarzschild BH spacetimes}

\begin{table}[t]
\centering
\caption{\textit{For the highly-damped QNFs within the Schwarzschild BH spacetime, only the electromagnetic perturbations exhibit dimensional dependence. Irrespective of the perturbing mode or chirality of the QNM potential, the aQNFs of each spin reduce to a common form.}}
\def\arraystretch{1.22}
\begin{tabular}{|C|C|C|}
\hline
potential & \pm j  & \displaystyle  \lim_{n \rightarrow \infty} \omega \; \; (n \in \mathbb{N}) \\[1ex]
\hline
V^s \; , \; V^{grav}_{_T} \; , \; V^{grav}_{_S} &  0 &  \multirow{2}{4.2cm}{$ \;T_{_H} \ln 3 + 2 \pi i T_{_H} \left( n + \frac{1}{2} \right)$} \\
V^{grav}_V & 2 & \\[1ex] 
\hline
V^{EM}_{_S} & \displaystyle \frac{2(d-3)}{d-2} & \multirow{2}{5.8cm}{$ \;\;T_{_H} \ln (-1 - 2 \cos (\pi j)) + 2 \pi i T_{_H} n $} \\[2.5ex]
V^{EM}_{_V} & \displaystyle \frac{2}{d-2} & \\[2.5ex]
\hline
V^D_{\pm} \;, \; V^{RS_{non}}_{\pm} \;, \; V^{RS_{TT}}_{\pm}  &  1 & 2 \pi i T_{_H} n \\[1ex]
\hline
\end{tabular}
\label{table:aQNFST}
\end{table}

\par To calculate the highly-damped QNFs of the Schwarzschild BH spacetime, we use Eq. (\ref{eq:aQNFST}). For the spin-0, spin-2, spin-1/2, and spin-3/2 perturbations, the $j$ values are dimensionally-independent; if Eq. (\ref{eq:aQNFST}) holds, then these solutions of Table \ref{table:aQNFST}  remain consistent for all $d>3$. 
\par While the scalar- and vector-modes of the spin-1 perturbations consistently yield aQNF expressions equivalent to one another, these results vary from dimension to dimension. For $d=6$, we observe that the real part vanishes and only a $2 \pi i T_H(n+1/2)$ term remains. When $d=5$, an ill-defined $\ln (0)$ emerges. This was observed in Ref. \cite{Lopez-Ortega2006}, where L{\'o}pez-Ortega suggested that the $\ln (0)$ term was a consequence of the real part of the QNF rapidly reducing to zero, citing the numerical work of Ref. \cite{CardosoLemosYoshida}. For the 4D case, the electromagnetic $j$ reduces to 1 and the aQNF becomes $2 \pi i T_Hn$, in agreement with Ref. \cite{Cho2006}. Once $d>6$, the argument of the natural logarithm remains greater than 1 such that the real part of the aQNF cannot be dismissed. 
\par Moreover, we note that Eq. (\ref{eq:aQNFoffset}) holds for the Schwarzschild BH, irrespective of the spin of the perturbing field. The offset for the scalar and gravitational aQNFs remains $\ln 3$ while the gap is given by $2 \pi i T_{_H}=ik_{_H}$. The newly-computed half-integer results (and the spin-1 aQNFs in the $d=4,6$ contexts), however, have an offset of zero. These are in agreement with the 4D results of Ref. \cite{Cho2006}. The oscillation frequency of the emitted radiation therefore cannot be extracted from these highly-damped QNFs, as they are purely imaginary.

\subsection{aQNFs of the Reissner-Nordstr{\"o}m and extremal Reissner-Nordstr{\"o}m BHs}
\begin{table}[t]
\centering
\caption{\textit{For the highly-damped QNFs within the non-extremal Reissner-Nordstr{\"o}m BH spacetime, only the spin-1/2 aQNFs are dimensionally-independent. The aQNFs of spin $ s \in \{0,3/2,2 \}$ reduce to a common form for all but the unmodified nonTT spin-3/2 modes, irrespective of the perturbing mode or chirality. Demarcated rows represent common results.}}
\def\arraystretch{1.3}
\begin{tabular}{|C|C|C|}
\hline
potential & \pm j  &  e^{+ \frac{2 \pi \omega}{k^+_{_H}}} \\[1ex] 
\hline
V^s \;, \; V^{grav}_{_{\pm S}}, \; V^{grav}_{_T} \;, \; \widetilde{V}^{RS_{non}}_{+} \;, \; V^{RS_{TT}}_{-} & \displaystyle \frac{d-3}{2d-5} &  \multirow{2}{1.8cm}{\text{Eq. (\ref{eq:aQNFRN})}} \\[2.5ex] 
V^{grav}_{\pm V}\;, \; \widetilde{V}^{RS_{non}}_{-}  \;, \; V^{RS_{TT}}_{+} & \displaystyle \frac{3d-7}{2d-5} & \\[2.5ex] 
\hline
V^{RS_{non}}_+ & \displaystyle \frac{(d-3)(d-2)\mu}{\left((d-2)\mu + 2\vartheta  \kappa \right) (2d-5)} & \text{Eq. (\ref{eq:aQNFRN})}  \\[2.5ex]
\hline
V^{RS_{non}}_- & \displaystyle \frac{(3d-7)(d-2)\mu + 4 (2 d-5) \vartheta  \kappa}{ \left((d-2)\mu + 2\vartheta  \kappa \right) (2d-5) }
& \text{Eq. (\ref{eq:aQNFRN})} \\[2.5ex] 
\hline
V^D_{\pm} & 1 &  1 \\[1ex] 
\hline
\end{tabular}
\label{table:aQNFRN}
\end{table}

\par For the non-extremal Reissner-Nordstr{\"o}m BH spacetimes, we compute the aQNFs of spin $s \in \{ 0,1/2,3/2,2\}$ via Eq. (\ref{eq:aQNFRN}). In Table \ref{table:aQNFRN} we then demonstrate that our Rarita-Schwinger results match those of the gravitational aQNFs. This can be inferred from section \ref{sec:fields}, where the $\widetilde{j}$ parameters extracted from the spin-3/2 nonTT-modes subjected to the $\mu \gg \vartheta \kappa$ limit ($\widetilde{V}^{RS_{non}}_{\pm}$) were shown to be equivalent to the $j$ parameters of the TT-modes with opposing chirality ($V^{RS_{TT}}_{\mp}$). As such, $\widetilde{V}^{RS_{non}}_{+}$ ($V^{RS_{TT}}_{-}$) corresponds to spin-0 and the scalar- and tensor-modes of the spin-2 perturbations while $\widetilde{V}^{RS_{non}}_{-}$ ($V^{RS_{TT}}_{+}$) corresponds to the spin-2 vector-modes. The aQNFs follow suit.
\par Irrespective of the mode of the perturbing field, the aQNFs of spin $s \in \{0,3/2,2 \}$ reduce to a common aQNF expression for all $d>3$, which differs from dimension to dimension. For $d \geq 5$, the aQNFs of spin $s \in \{0,3/2,2 \}$ cannot be solved analytically. For the nonTT modes of the spin-3/2 aQNFs, these observations apply only if $\mu \gg \vartheta \kappa$. 

\par When $d=4$, the aQNFs of spin $s \in \{0,3/2,2 \}$ within the Reissner-Nordstr{\"o}m BH spacetime reduce to
\begin{equation}
e^{\frac{2 \pi  \omega}{k^+_{_H}}} = -2 -3 e^{-\frac{2 \pi  \omega}{k^-_{_H}}} \;,
\end{equation}
\noindent which corresponds to the results of Ref. \cite{Cho2006}. Note that for this 4D result, the spin-3/2 aQNF is calculated from ${\widetilde{V}}^{non}_{+}$.
\par For the spin-1/2 aQNFs,
\begin{equation} \label{eq:aQNFspinHalfRN}
e^{\pi \omega/k^+_{_H}} = 1 \; \; \; \; \Rightarrow \; \; \; \; \lim_{n\rightarrow + \infty} \omega = 2 \pi i T_{_H} n 
\end{equation} 
\noindent holds, irrespective of $d$. This is in agreement with the 4D results of Ref. \cite{Cho2006}, since only $k^+_{_H}$ contributes to $T_{_H}$. As in the Schwarzschild case, we find Dirac aQNFs that are purely imaginary, with a gap given by a multiple of the Hawking temperature. Note that this result therefore ignores the surface gravity of the Cauchy horizon, such that the spin-1/2 aQNFs allow for an isolated study of the surface gravity of the outer horizon.

\par However, we note that if the $\mu \gg \vartheta \kappa$ limit is not imposed, the nonTT spin-3/2 results in non-extremal Reissner-Nordstr{\"o}m BHs showcase discrepancies from the other aQNFs demonstrated here. Though they differ from dimension to dimension, these spin-3/2 aQNFs of opposing chiralities do not match one another. Furthermore, these retain their dependence on the BH mass and charge, as well as on the spinor eigenvalue on $S^{d-2}$. Such features in the aQNF are to be expected from the form of the $\widetilde{j}$ parameters of Eqs. (\ref{eq:jRNplus}) and (\ref{eq:jRNminus}), which are distinct from one another and dependent on $\vartheta$, $\mu$, and $\kappa$.
 
\par In the case of the extremal Reissner-Nordstr{\"o}m BH spacetimes, the aQNFs can be computed using Eq. (\ref{eq:aQNFeRN}). These results are compiled in Table \ref{table:aQNFeRN}, where identical aQNF outputs are indicated by shared rows. The aQNFs of spin $s \in \{0,3/2,2\}$ manifest in the form of
\[ \lim_{n \rightarrow \infty} \omega = T \ln [\text{trigonometric function}] + 2 \pi i T [\text{gap}] \;.\]  
\noindent Spin-0,2 and TT spin-3/2 aQNFs have a common real part. Specifically,
\[ \ln (2) < \mathbb{R}e \{ \lim_{n \rightarrow \infty} \Big \{  \lim_{n \rightarrow \infty} \omega \Big \} \leq \ln (3) \;,\]
\noindent where the natural logarithm decreases from $\ln (3)$ for $d=4$ with increasing $d$. For $\kappa = 0$, the nonTT spin-3/2 aQNFs are equivalent to their TT counterparts for each $d\geq 4$. For a fixed $d$ but increasing $\kappa$, the magnitude of the real part of the nonTT spin-3/2 aQNF increases from $\ln (3.7321 )$ for $\kappa = 1$ (and $d=4$). As such, for $\kappa \neq 0$, the aQNFs of the nonTT spin-3/2 fields decrease (increase) with increasing $d$ ($\kappa$) with other parameters fixed in the range
\[ \ln (2) < \mathbb{R}e \Big \{  \lim_{n \rightarrow \infty} \omega \Big \} < \ln (4) \;.\] 
\noindent $V^{RS_{non}}_{-}$ and $V^{RS_{non}}_{+}$ are isospectral. 

\par The gap for spin-0,2 and TT spin-3/2 aQNFs is given by $2 \pi i T (n + 1/2)$; for nonTT spin-3/2 aQNFs, it is $2 \pi i T n$ for all $\kappa$, $d$.

\par Thus, these spin $s \in \{0,3/2,2\}$ aQNFs of the extremal Reissner-Nordstr{\"o}m BH spacetime resemble the Schwarzschild form more so than the aQNFs of any other spacetime, albeit with a trigonometric function as the argument of the natural logarithm rather than an integer. This applies for all $d\geq 4$ cases computed. For the special case of $d=4$ for the spin-0,2 aQNFs, we find that the $\mathbb{R}e \{\omega \} \sim \ln 3$ result is recovered, in agreement with Refs. \cite{Berti2004,refAnderssonHowls}. However, as explained in subsection \ref{subsubsec:RNnatschiap}, an attempt to link these results to Hod's conjecture would be invalid since the underlying arguments thereof are based on Bekenstein and Mukhanov's BH area quantisation \cite{refBekenstein1972,refMukhanov1986,refBekenstein1995}, which itself is predicated on non-extremal BH spacetimes. In subsection \ref{subsubsec:RNnatschiap} we also discussed the $T=k/2 \pi$ function that is analogous $-$ but not equivalent $-$ to the Hawking temperature. 

\par For the spin-1/2 aQNFs, we once again observe dimensional independence. However, since Eq. (\ref{eq:aQNFeRN}) reduces to $e^{2 \pi \omega / k} = 0$, the aQNF spectrum is ill-defined. The reasoning for this is unclear and demands further investigation.
\begin{table}[t]
\centering
\caption{\textit{For the highly-damped QNFs within the extremal Reissner-Nordstr{\"o}m BH spacetime, the aQNFs of spin $s \in \{0,3/2,2 \}$ reduce to a common form. The spin-1/2 aQNFs are ill-defined. Demarcated rows represent shared results.}}
\def\arraystretch{1.3}
\begin{tabular}{|C|C|C|}
\hline
 potential & \pm j  &  e^{\frac{2 \pi \omega}{k}}  \\[1ex]
\hline
V^s \;, \; V^{grav}_{_{\pm S}}, \; V^{grav}_{_T} \;, \; V^{RS_{TT}}_{-}  & \displaystyle \frac{d-3}{2d-5} &  \multirow{2}{1.8cm}{\text{Eq. (\ref{eq:aQNFeRN})}} \\[2.5ex] 
V^{grav}_{\pm V}  \;, \; V^{RS_{TT}}_{+} &\displaystyle  \frac{3d-7}{2d-5} & \\[2.5ex] 
\hline
V^{RS_{non}}_{+} & \displaystyle \frac{(d-3)(d-2)}{(2d-5)(d-2+2\kappa)} & \text{Eq. (\ref{eq:aQNFeRN})} \\[2.5ex] 
V^{RS_{non}}_{-} & \displaystyle \frac{(3d-7)(d-2)+4(2d-5)\kappa}{(2d-5)(d-2+2\kappa)} & \text{Eq. (\ref{eq:aQNFeRN})} \\[2.5ex] 
\hline 
V^D_{\pm} & 1 &  0  \\[1ex] 
\hline
\end{tabular}
\label{table:aQNFeRN}
\end{table}

\par As indicated by Cho in Ref. \cite{Cho2006}, the shared forms recorded in Table \ref{table:aQNFeRN} for aQNFs of spin $s \in \{0, 3/2, 2 \}$ for the extremal Reissner-Nordstr{\"o}m BH spacetimes may indicate a connection to SUSY frameworks.

\par On the basis of the inherent characteristics of the extremal Reissner-Nordstr{\"o}m BH related to supergravity conjectures \cite{Kallosh1992SUSYcosmic,Kallosh1992SUSYBH,Kallosh1997,GibbonsextRN}, Onozawa {\textit{et al.}} in Ref. \cite{Onozawa1996} associated the isospectrality between photon, graviton, and gravitino QNFs observed in the 4D extremal Reissner-Nordstr{\"o}m BHs \cite{refOnozawa1996} with a manifestation of a hidden $\mathcal{N}=2$ SUSY. In particular, they considered that since the extremal Reissner-Nordstr{\"o}m BH admits a Killing spinor field \cite{GibbonsextRN}, the background solution is expected to be invariant under SUSY transformations with respect to this Killing spinor field. Consequently, all perturbed fields would be expected to be related via SUSY transformations that conserve the S-matrix. In the case  of Refs. \cite{refOnozawa1996,Onozawa1996}, the SUSY transformation was an increase of $1/2$ in the multipolar index, corresponding to an increase of $1/2$ in the spin of the field. Cho in Ref. \cite{Cho2006} considered that this could relate to the commonality observed between QNFs of $s \in \{1,3/2,2\}$ within the large overtone limit, but could not find a clear way to demonstrate that between the scalar and Dirac aQNFs.

\par Although our results concern higher-dimensional QNFs within the large overtone limit, the shared form between spin-2 and spin-3/2 results observed in Table \ref{table:aQNFeRN} may indicate some manifestation of SUSY beyond the known $d=4$ results (particularly in the wake of the extension of SUSY qualities to extremal Reissner-Nordstr{\"o}m BHs in higher dimensions \cite{HartleHawkingextRN}). Whether this commonality also extends to the electromagnetic aQNFs in these BH spacetimes is an open question and motivates for the development of effective QNM potentials for electromagnetic perturbations within higher dimensional Reissner-Nordstr{\"o}m BH spacetimes.

\par By virtue of the consistent $j$ parameter in most cases, we observe that Tables \ref{table:aQNFRN} and \ref{table:aQNFeRN} showcase the same shared behaviour between potentials. However, since we expect only extreme
BHs to preserve SUSY \cite{Onozawa1996}, the connection between supergravity considerations and uniform aQNFs applies only to the extremal Reissner-Nordstr{\"o}m BH context.

\subsection{aQNFs of the Schwarzschild dS and Schwarzschild AdS BH spacetimes}

\begin{table}[t]
\centering
\caption{\textit{For the highly-damped QNFs within the Schwarzschild dS BH spacetime, the spin-0 and spin-2 aQNFs are dimensionally-independent and reduce to an identical form. This is also the case for all half-integer aQNFs. A common aQNF expression emerges for the scalar- and vector-modes of the spin-1 aQNFs, varying from dimension to dimension.}}
\def\arraystretch{1.22}
\begin{tabular}{|C|C|C|}
\hline
potential & \pm j  & aQNF \\[1ex]
\hline
V^s \; , \; V^{grav}_{_T} \; , \; V^{grav}_{_S} &  0 &  \multirow{2}{7cm}{$ \; \; \cosh\left( \frac{\pi \omega }{k_{_H}} - \frac{\pi \omega}{k_{_C}} \right) + 3 \cosh \left( \frac{\pi \omega }{k_{_H}} + \frac{\pi \omega}{k_{_C}} \right) = 0 $} \\
V^{grav}_V & 2 & \\[2.5ex] 
\hline
V^{EM}_{_S} & \displaystyle \frac{2(d-3)}{d-2} &  \multirow{2}{6cm}{\hspace{2cm} Eq. (\ref{eq:aQNFSdSLopezOrtega})}  \\[2.5ex]
V^{EM}_{_V} & \displaystyle \frac{2}{d-2} &  \\[2.5ex] 
\hline
V^D_{\pm} \;, \; V^{RS_{non}}_{\pm} \;, \; V^{RS_{TT}}_{\pm}  &  1 & 2 \pi i T_{_H} n \hspace{0.2cm}  \text{or} \hspace{0.12cm} -2 \pi i T_{_C} n\\[1ex] 
\hline
\end{tabular}
\label{table:aQNFSdS}
\end{table}

\par We may compute the aQNFs of the Schwarzschild dS BH using Eq. (\ref{eq:aQNFSdS}). Within Table \ref{table:aQNFSdS} we observe that the spin-2 and spin-0 results are uniform for all $d$, as demonstrated in Refs. \cite{refNatarioSchiappa,refCardosoNatarioSchiappa}. This aQNF expression, however, cannot be solved analytically. 

\par The scalar- and vector-modes of the spin-1 aQNFs yield a common result. For $d=4,5,6$, the real part is zero and the gap is a function of the surface gravity of either the event or cosmological horizon. This is in agreement with Refs. \cite{Lopez-Ortega2006,CardosoLemosYoshida}. For example, the highly-damped QNFs of the electromagnetic scalar- and vector-modes become
\begin{equation}
\lim_{n \rightarrow \infty} \omega = 2 \pi i T_{_H} \left(n + \frac{1}{2} \right) \hspace{0.4cm}  \text{or} \hspace{0.4cm} \lim_{n \rightarrow \infty} \omega = -2 \pi i T_{_C} \left(n + \frac{1}{2} \right) 
\end{equation}
when $d=6$ and using $k_{_C} = 2 \pi T_{_C} < 0$.

\par We note with interest that the newly-computed half-integer aQNFs share a common, dimensionally-independent expression in the Schwarzschild dS BH spacetime that closely resembles that of the spin-1 case for $d=4,5,6$, $viz.$
\begin{equation}
\lim_{n \rightarrow \infty} \omega = 2 \pi i T_{_H} n \hspace{0.4cm}  \text{or} \hspace{0.4cm} \lim_{n \rightarrow \infty} \omega = -2 \pi i T_{_C} n \;.
\end{equation}
\noindent In fact, when reduced to the 4D context, the spin-1 aQNFs reflect the spin-1/2 and spin-3/2 results exactly.

\par As discussed in subsection \ref{subsubsec:STnatschiap}, the 5D case does require its own analysis and yields a slightly different expression for the aQNF. Therein, we have also demonstrated that the Schwarzschild aQNF may be extracted from Eq. (\ref{eq:aQNFSdS}) if subjected to the $\lambda \rightarrow 0^+$ limit.  

\begin{table}[t]
\centering
\caption{\textit{The aQNFs within the Schwarzschild AdS BH spacetime. While the structure of the expressions remains consistent, the argument of the $\arctan$ and the phase shift differs from dimension to dimension. Here, we provide results for $d=6$ in the rightmost column.}}
\def\arraystretch{1.22}
\begin{tabular}{|C|C|C|C|}
\hline
potential & \pm j  & \pm j^{\infty} & \displaystyle \lim_{n \rightarrow \infty} \omega x_0  \; \; (n \in \mathbb{N})\\[1ex]
\hline
V^s \; , \; V^{grav}_{_T}   & 0  & (d-1) & \multirow{2}{3.3cm}{$\frac{7\pi}{4} -\arctan \left(\frac{i}{3} \right) +n \pi$}\\
V^{grav}_V & 2 & (d-3) &\\[1ex]
\hline
V^{grav}_{_S} &  0 &   (d-5) & \frac{3\pi}{4} -\arctan \left(\frac{i}{3} \right) +n \pi \\[2.5ex]  
\hline
V^{EM}_{_S} & \displaystyle \frac{2(d-3)}{d-2} & (d-5) & \frac{\pi}{2} + \arctan \left(\frac{3+4i}{5} \right) +n \pi \\[2.5ex]
V^{EM}_{_V} & \displaystyle \frac{2}{d-2} & (d-3) & \pi + \arctan \left(\frac{3+4i}{5} \right) +n \pi \\[2.5ex] 
\hline
V^D_{\pm} &  &  1 & \frac{\pi}{2} -\arctan \left(\frac{3+4i}{5} \right) +n \pi \\[1ex] 
V^{RS_{non}}_{-} & \multirow{2}{4cm}{\hspace{1.7cm} 1}  & (d-5)& \frac{ \pi}{2} - \arctan \left(\frac{3+4i}{5} \right) +n \pi \\[1ex] 
V^{RS_{non}}_{+} \;, \; V^{RS_{TT}}_{-} &  & (d-3) & \pi - \arctan \left(\frac{3+4i}{5} \right) +n \pi \\[1ex]
V^{RS_{TT}}_{+} &  & (d-1) & \frac{3 \pi}{2} -\arctan \left(\frac{3+4i}{5} \right) +n \pi \\[1ex]
\hline
\end{tabular}
\label{table:aQNFSAdS}
\end{table}

\par For the Schwarzschild AdS BH, the aQNFs can be produced using Eq. (\ref{eq:aQNFSAdS}). There is a pronounced uniformity in the structure of these expressions: irrespective of the spin of the field, the aQNFs have a consistent form marked by an $\arctan$ and a phase shift. However, since the specific argument of the $\arctan$ and the phase shift differs from dimension to dimension, we have provided the explicit aQNF expressions for the $d=6$ case in Table \ref{table:aQNFSAdS}. 
\par The gravitational perturbations have been confirmed against the results of Refs. \cite{refNatarioSchiappa,refCardosoNatarioSchiappa,Daghigh2008SAdS}, where we note that the $i \ln(2)/2$ of Ref. \cite{Daghigh2008SAdS} is equivalent to $\arctan (i/3)$, and the sign difference is due to their use of a negative temporal dependence. As such, for the gravitational and scalar aQNFs, only the phase shift contributes to the real part of $\omega x_0 \vert_{n \rightarrow \infty}$. For fields of $s \in \{ 1/2, 1, 3/2 \}$, however, we find that the argument of this $\arctan$ incorporates real and imaginary parts, which increases the magnitude of the real part of the aQNF. Note, however, that the sign of the $\arctan$ is positive only for aQNFs of the electromagnetic field.

\par Finally, we add that there are not many analytic studies available for the aQNFs of BH spacetimes inclusive of a cosmological constant (however, see section 3.2.1 of Ref. \cite{refNatarioSchiappa} for a thorough review of the numerical results for the Schwarzschild AdS QNF), which makes these results particularly useful. However, it must be noted that the Schwarzschild AdS aQNFs are derived using the Dirichlet boundary conditions. As discussed in Ref. \cite{refnewQNMs2020}, the influence of boundary conditions in AdS BH spacetimes can have a profound effect on the computational output. Whether the application of these boundary conditions yields the most physically appropriate aQNFs requires additional consideration.

\section{\label{sec:conc} Conclusions}
\par Although highly-damped QNMs are not observable, the application of asymptotic limits allows for the isolation of real and imaginary behaviours. For our study of aQNFs, we opted to pursue the monodromy technique pioneered in Ref. \cite{refMotlNeitzke}, which can be considered as a more economical iteration of the ``phase-integral" method of Andersson and Howls \cite{refAnderssonHowls}.

\par In our explicit review of the application of the method to Schwarzschild, Reissner-Nordstr{\"o}m, and Schwarzschild (A)dS BHs, we observed that the BH family dictates the behaviour near the origin (see Table \ref{table:1}) while the cosmological constant determines the behaviour near spatial infinity (see Eq. (\ref{eq:tortads})). This is observable also from section \ref{sec:fields}, where the exploitation of the appropriate tortoise coordinate allows for the approximation of QNM potentials from the literature into the form of Eq. (\ref{eq:genpot}) dependent on a field-specific parameter $j$. As such, the computation of aQNFs via the monodromy technique remains uniform for perturbations of spin $s$: the topology of the complex plane, the contour traced, and the boundary conditions applied depend solely on the BH metric function; the contribution from the field is included once the generalised aQNF expression has been computed, via the $j$ and $j^{\infty}$ parameters.

\par With these principles in place, we computed the aQNFs of spin $s \in \{0,1/2,1,3/2,2 \}$, where all half-integer aQNFs reflect new results. As in the integer-spin cases, the $j$ and $j^{\infty}$ parameters for the half-integer perturbations were extracted through an asymptotic analysis of the QNM effective potentials. For Schwarzschild, non-extremal Reissner-Nordstr{\"o}m, and Schwarzschild dS BH spacetimes, we observed that the Dirac aQNFs emerged consistently as a purely imaginary solution proportional to the surface gravity of the horizon for all but the Schwarzschild AdS and extremal Reissner-Nordstr{\"o}m BH spacetimes. This consistency in the Dirac aQNFs is particularly interesting as it defies the general trend observed in our results {\textit{viz.}} that the final aQNF solution depends more on the BH spacetime than the spin of the perturbing field. Furthermore, we observe that the Rarita-Schwinger behaviour corresponds predominantly to that of the Dirac field (demonstrable in the cases of Schwarzschild and Schwarzschild (A)dS BHs), but matches precisely the gravitational perturbations for other BH spacetimes. Specifically, aQNFs associated with $V^{TT}_{-}$ ($\widetilde{V}^{non}_{+}$) and $V^{TT}_{+}$ ($\widetilde{V}^{non}_{-}$) corresponded to $V^{grav}_{\pm S,T}$ and $V^{grav}_{\pm V}$, respectively. The justification for these observed behaviours is not obvious and warrants further investigation.

\par Except for most Dirac aQNFs, and the spin-1 aQNFs of the Schwarzschild dS BH, we have found that the structure of the aQNF solutions is consistent for each spacetime, irrespective of the spin of the perturbing field. We observe in several cases that the real part tends to a finite value, while the imaginary part grows in fixed intervals with $n$. This can be seen for all fields of the Schwarzschild and extremal Reissner-Nordstr{\"o}m BH, and additionally for the spin-1 aQNFs of the Schwarzschild dS BH, for the spin-1/2 aQNFs in the Schwarzschild and Reissner-Nordstr{\"o}m BHs, and for the spin-3/2 aQNFs of the Schwarzschild dS BH. 

\par While a physical interpretation of our results is not immediately clear, the absence of a constant proportional to the natural logarithm of an integer in most cases weakens the already tenuous connection between aQNFs and a quantum spectrum of BHs. The BH area quantisation extracted from the real part of the aQNF is therefore not an intrinsic property of the BH \cite{refMaggiore2008}. As suggested in section V of Ref. \cite{refKonoplyaZhidenkoReview}, if a link between aQNFs and quantum gravity exists, it is likely to be more subtle than Hod's conjecture would imply. However, as argued by Maggiore \cite{refMaggiore2008}, it may be that the underlying principles of Hod's conjecture were not well applied to begin with: Bohr's correspondence principle, for example, is valid for $n \leftrightarrow n'$ transitions where $n,n'\gg 1$; Hod considered instead the excitation of a BH from the ground state. Through the use of $(\omega_0)_n$ (as defined in section \ref{sec:intro}) for the real frequency in the transition $n \rightarrow n-1$, Maggiore recovered Bekenstein's original Schwarzschild BH area irrespective of the spin of the field. As such, the consistent behaviour of $(\omega_0)_n$ within the large-$n$ limit led Maggiore to an BH area quantisation intrinsic to the Schwarzschild BH \cite{refMaggiore2008}. It would be interesting to explore whether this can be extended beyond the Schwarzschild case.
  
\par Despite the interest in aQNFs, there are relatively few examples in the literature of aQNF computations, both analytical and numerical. Consequently, there are few aQNF results available within the literature. This work then serves as a collection of known results and a point of reference against which future studies may compare. There is a pronounced scarcity in numerical checks for aQNF results: beyond Leaver's continued fraction method (which is limited in the case of extremal Reissner-Nordstr{\"o}m BHs, as discussed in section \ref{subsubsec:RNnatschiap}), few numerical attempts have been applied to aQNFs. The debate surrounding the aQNF expression for the Reissner-Nordstr{\"o}m and the unusual Dirac result obtained in this work emphasise this need for further study into numerical validation for aQNFs. In this respect, our analytical results may help guide the development of numerical techniques to approach the highly damped QNFs.  

\par Future research into the large overtone limit should seek out a justification for the observations made within this paper, particularly on the consistency of the Dirac results across various BH spacetimes. An investigation into the Rarita-Schwinger aQNFs is also necessary, to determine why the spin-3/2 aQNF behaviour resembles gravitational perturbations in certain Reissner-Nordstr{\"o}m BH contexts but Dirac perturbations in the Schwarzschild and Schwarzschild dS BH spacetimes. That this behaviour for the nonTT spin-3/2 modes holds only under the $\mu \gg \vartheta \kappa$ limit should also be assessed. The possible SUSY considerations that arise for these aQNFs of spin-3/2 and spin-2 in the extremal Reissner-Nordstr{\"o}m context also require further considerations. Here, the need for aQNF expressions corresponding to perturbing electromagnetic fields within Reissner-Nordstr{\"o}m BH spacetimes becomes especially apparent.

\par Furthermore, an extension of the aQNF investigation conducted here focused on dimensionality, in the style of Ref. \cite{Daghigh2008SAdS}, would verify explicitly the claims in the literature regarding the generalisability of the $d=6$ monodromy results reviewed in subsection \ref{subsec:aQNF}. While we are aware of the anomalous behaviour of 5D BHs, we might infer from Figs. \ref{fig:SchwarzNum}, \ref{fig:RNNum}, and \ref{fig:SAdSNum} that odd dimensions may require individual analysis. This is particularly important for BH spacetimes inclusive of the cosmological constant. Finally, whether the monodromy technique, or some variant thereof, may be applied to rotating BH spacetimes is an interesting open question.

\begin{acknowledgments}
CHC is supported in part by the Naresuan University Research Fund R2565C012. HTC is supported in part by the Ministry of Science and Technology, Taiwan, under the Grants No. MOST108-2112-M-032-002 and MOST109-2112-M-032-007. ASC is supported in part by the National Research Foundation (NRF) of South Africa. AC is supported by a Campus France Scholarship, and the NRF and Department of Science and Innovation through the SA-CERN programme. The authors would like to thank the anonymous reviewers for their thorough and insightful reports.
\end{acknowledgments}

\bibliographystyle{aipnum4-2} 
\bibliography{TheLargeNBibFile}

\begin{thebibliography}{76}%
\makeatletter
\providecommand \@ifxundefined [1]{%
 \@ifx{#1\undefined}
}%
\providecommand \@ifnum [1]{%
 \ifnum #1\expandafter \@firstoftwo
 \else \expandafter \@secondoftwo
 \fi
}%
\providecommand \@ifx [1]{%
 \ifx #1\expandafter \@firstoftwo
 \else \expandafter \@secondoftwo
 \fi
}%
\providecommand \natexlab [1]{#1}%
\providecommand \enquote  [1]{``#1''}%
\providecommand \bibnamefont  [1]{#1}%
\providecommand \bibfnamefont [1]{#1}%
\providecommand \citenamefont [1]{#1}%
\providecommand \href@noop [0]{\@secondoftwo}%
\providecommand \href [0]{\begingroup \@sanitize@url \@href}%
\providecommand \@href[1]{\@@startlink{#1}\@@href}%
\providecommand \@@href[1]{\endgroup#1\@@endlink}%
\providecommand \@sanitize@url [0]{\catcode `\\12\catcode `\$12\catcode
  `\&12\catcode `\#12\catcode `\^12\catcode `\_12\catcode `\%12\relax}%
\providecommand \@@startlink[1]{}%
\providecommand \@@endlink[0]{}%
\providecommand \url  [0]{\begingroup\@sanitize@url \@url }%
\providecommand \@url [1]{\endgroup\@href {#1}{\urlprefix }}%
\providecommand \urlprefix  [0]{URL }%
\providecommand \Eprint [0]{\href }%
\providecommand \doibase [0]{http://dx.doi.org/}%
\providecommand \selectlanguage [0]{\@gobble}%
\providecommand \bibinfo  [0]{\@secondoftwo}%
\providecommand \bibfield  [0]{\@secondoftwo}%
\providecommand \translation [1]{[#1]}%
\providecommand \BibitemOpen [0]{}%
\providecommand \bibitemStop [0]{}%
\providecommand \bibitemNoStop [0]{.\EOS\space}%
\providecommand \EOS [0]{\spacefactor3000\relax}%
\providecommand \BibitemShut  [1]{\csname bibitem#1\endcsname}%
\let\auto@bib@innerbib\@empty
\bibitem [{\citenamefont {Nollert}(1999)}]{refNollert1999}%
  \BibitemOpen
  \bibfield  {author} {\bibinfo {author} {\bibfnamefont {H.-P.}\ \bibnamefont
  {Nollert}},\ }\href {\doibase 10.1088/0264-9381/16/12/201} {\bibfield
  {journal} {\bibinfo  {journal} {Class. Quant. Grav.}\ }\textbf {\bibinfo
  {volume} {16}},\ \bibinfo {pages} {R159} (\bibinfo {year}
  {1999})}\BibitemShut {NoStop}%
\bibitem [{\citenamefont {Kokkotas}\ and\ \citenamefont
  {Schmidt}(1999)}]{refKokkotasRev}%
  \BibitemOpen
  \bibfield  {author} {\bibinfo {author} {\bibfnamefont {K.~D.}\ \bibnamefont
  {Kokkotas}}\ and\ \bibinfo {author} {\bibfnamefont {B.~G.}\ \bibnamefont
  {Schmidt}},\ }\href {\doibase 10.12942/lrr-1999-2} {\bibfield  {journal}
  {\bibinfo  {journal} {Living Rev. Rel.}\ }\textbf {\bibinfo {volume} {2}},\
  \bibinfo {pages} {2} (\bibinfo {year} {1999})},\ \Eprint
  {http://arxiv.org/abs/gr-qc/9909058} {arXiv:gr-qc/9909058} \BibitemShut
  {NoStop}%
\bibitem [{\citenamefont {Berti}, \citenamefont {Cardoso},\ and\ \citenamefont
  {Starinets}(2009)}]{refBertiCardoso}%
  \BibitemOpen
  \bibfield  {author} {\bibinfo {author} {\bibfnamefont {E.}~\bibnamefont
  {Berti}}, \bibinfo {author} {\bibfnamefont {V.}~\bibnamefont {Cardoso}}, \
  and\ \bibinfo {author} {\bibfnamefont {A.~O.}\ \bibnamefont {Starinets}},\
  }\href {\doibase 10.1088/0264-9381/26/16/163001} {\bibfield  {journal}
  {\bibinfo  {journal} {Class. Quant. Grav.}\ }\textbf {\bibinfo {volume}
  {26}},\ \bibinfo {pages} {163001} (\bibinfo {year} {2009})},\ \Eprint
  {http://arxiv.org/abs/0905.2975} {arXiv:0905.2975 [gr-qc]} \BibitemShut
  {NoStop}%
\bibitem [{\citenamefont {Konoplya}\ and\ \citenamefont
  {Zhidenko}(2011)}]{refKonoplyaZhidenkoReview}%
  \BibitemOpen
  \bibfield  {author} {\bibinfo {author} {\bibfnamefont {R.~A.}\ \bibnamefont
  {Konoplya}}\ and\ \bibinfo {author} {\bibfnamefont {A.}~\bibnamefont
  {Zhidenko}},\ }\href {\doibase 10.1103/RevModPhys.83.793} {\bibfield
  {journal} {\bibinfo  {journal} {Rev. Mod. Phys.}\ }\textbf {\bibinfo {volume}
  {83}},\ \bibinfo {pages} {793} (\bibinfo {year} {2011})},\ \Eprint
  {http://arxiv.org/abs/1102.4014} {arXiv:1102.4014 [gr-qc]} \BibitemShut
  {NoStop}%
\bibitem [{\citenamefont {Leaver}(1985)}]{refLeaver1985}%
  \BibitemOpen
  \bibfield  {author} {\bibinfo {author} {\bibfnamefont {E.~W.}\ \bibnamefont
  {Leaver}},\ }\href {\doibase 10.1098/rspa.1985.0119} {\bibfield  {journal}
  {\bibinfo  {journal} {Proc. Roy. Soc. Lond. A}\ }\textbf {\bibinfo {volume}
  {402}},\ \bibinfo {pages} {285} (\bibinfo {year} {1985})}\BibitemShut
  {NoStop}%
\bibitem [{\citenamefont {Leaver}(1986{\natexlab{a}})}]{refLeaver1986a}%
  \BibitemOpen
  \bibfield  {author} {\bibinfo {author} {\bibfnamefont {E.~W.}\ \bibnamefont
  {Leaver}},\ }\href {\doibase 10.1063/1.527130} {\bibfield  {journal}
  {\bibinfo  {journal} {J. Math. Phys}\ }\textbf {\bibinfo {volume} {27}},\
  \bibinfo {pages} {1238} (\bibinfo {year} {1986}{\natexlab{a}})}\BibitemShut
  {NoStop}%
\bibitem [{\citenamefont {Leaver}(1986{\natexlab{b}})}]{refLeaver1986b}%
  \BibitemOpen
  \bibfield  {author} {\bibinfo {author} {\bibfnamefont {E.~W.}\ \bibnamefont
  {Leaver}},\ }\href {\doibase 10.1103/PhysRevD.34.384} {\bibfield  {journal}
  {\bibinfo  {journal} {Phys. Rev. D}\ }\textbf {\bibinfo {volume} {34}},\
  \bibinfo {pages} {384} (\bibinfo {year} {1986}{\natexlab{b}})}\BibitemShut
  {NoStop}%
\bibitem [{\citenamefont {Cardoso}\ and\ \citenamefont
  {Lemos}(2001)}]{refCardosoLemos2001}%
  \BibitemOpen
  \bibfield  {author} {\bibinfo {author} {\bibfnamefont {V.}~\bibnamefont
  {Cardoso}}\ and\ \bibinfo {author} {\bibfnamefont {J.~P.~S.}\ \bibnamefont
  {Lemos}},\ }\href {\doibase 10.1103/PhysRevD.64.084017} {\bibfield  {journal}
  {\bibinfo  {journal} {Phys. Rev. D}\ }\textbf {\bibinfo {volume} {64}},\
  \bibinfo {pages} {084017} (\bibinfo {year} {2001})}\BibitemShut {NoStop}%
\bibitem [{\citenamefont {Cardoso}, \citenamefont {Lemos},\ and\ \citenamefont
  {Yoshida}(2004)}]{CardosoLemosYoshida}%
  \BibitemOpen
  \bibfield  {author} {\bibinfo {author} {\bibfnamefont {V.}~\bibnamefont
  {Cardoso}}, \bibinfo {author} {\bibfnamefont {J.~P.~S.}\ \bibnamefont
  {Lemos}}, \ and\ \bibinfo {author} {\bibfnamefont {S.}~\bibnamefont
  {Yoshida}},\ }\href {\doibase 10.1103/PhysRevD.69.044004} {\bibfield
  {journal} {\bibinfo  {journal} {Phys. Rev. D}\ }\textbf {\bibinfo {volume}
  {69}},\ \bibinfo {pages} {044004} (\bibinfo {year} {2004})},\ \Eprint
  {http://arxiv.org/abs/gr-qc/0309112} {arXiv:gr-qc/0309112} \BibitemShut
  {NoStop}%
\bibitem [{\citenamefont {Berti}\ and\ \citenamefont
  {Kokkotas}(2003{\natexlab{a}})}]{Berti2003}%
  \BibitemOpen
  \bibfield  {author} {\bibinfo {author} {\bibfnamefont {E.}~\bibnamefont
  {Berti}}\ and\ \bibinfo {author} {\bibfnamefont {K.~D.}\ \bibnamefont
  {Kokkotas}},\ }\href {\doibase 10.1103/PhysRevD.68.044027} {\bibfield
  {journal} {\bibinfo  {journal} {Phys. Rev. D}\ }\textbf {\bibinfo {volume}
  {68}},\ \bibinfo {pages} {044027} (\bibinfo {year} {2003}{\natexlab{a}})},\
  \Eprint {http://arxiv.org/abs/hep-th/0303029} {arXiv:hep-th/0303029}
  \BibitemShut {NoStop}%
\bibitem [{\citenamefont {Andersson}\ and\ \citenamefont
  {Linn\ae{}us}(1992)}]{refAnderssonLinnaeus}%
  \BibitemOpen
  \bibfield  {author} {\bibinfo {author} {\bibfnamefont {N.}~\bibnamefont
  {Andersson}}\ and\ \bibinfo {author} {\bibfnamefont {S.}~\bibnamefont
  {Linn\ae{}us}},\ }\href {\doibase 10.1103/PhysRevD.46.4179} {\bibfield
  {journal} {\bibinfo  {journal} {Phys. Rev. D}\ }\textbf {\bibinfo {volume}
  {46}},\ \bibinfo {pages} {4179} (\bibinfo {year} {1992})}\BibitemShut
  {NoStop}%
\bibitem [{\citenamefont {Andersson}(1993)}]{refAndersson1993}%
  \BibitemOpen
  \bibfield  {author} {\bibinfo {author} {\bibfnamefont {N.}~\bibnamefont
  {Andersson}},\ }\href {\doibase https://doi.org/10.1088/0264-9381/10/6/001}
  {\bibfield  {journal} {\bibinfo  {journal} {Class. Quant. Gravity}\ }\textbf
  {\bibinfo {volume} {10}},\ \bibinfo {pages} {L61} (\bibinfo {year}
  {1993})}\BibitemShut {NoStop}%
\bibitem [{\citenamefont {Andersson}\ and\ \citenamefont
  {Howls}(2004)}]{refAnderssonHowls}%
  \BibitemOpen
  \bibfield  {author} {\bibinfo {author} {\bibfnamefont {N.}~\bibnamefont
  {Andersson}}\ and\ \bibinfo {author} {\bibfnamefont {C.~J.}\ \bibnamefont
  {Howls}},\ }\href {\doibase 10.1088/0264-9381/21/6/021} {\bibfield  {journal}
  {\bibinfo  {journal} {Class. Quant. Grav.}\ }\textbf {\bibinfo {volume}
  {21}},\ \bibinfo {pages} {1623} (\bibinfo {year} {2004})},\ \Eprint
  {http://arxiv.org/abs/gr-qc/0307020} {arXiv:gr-qc/0307020} \BibitemShut
  {NoStop}%
\bibitem [{\citenamefont {Motl}(2003)}]{refMotl}%
  \BibitemOpen
  \bibfield  {author} {\bibinfo {author} {\bibfnamefont {L.}~\bibnamefont
  {Motl}},\ }\href {\doibase 10.4310/ATMP.2002.v6.n6.a3} {\bibfield  {journal}
  {\bibinfo  {journal} {Adv. Theor. Math. Phys.}\ }\textbf {\bibinfo {volume}
  {6}},\ \bibinfo {pages} {1135} (\bibinfo {year} {2003})},\ \Eprint
  {http://arxiv.org/abs/gr-qc/0212096} {arXiv:gr-qc/0212096} \BibitemShut
  {NoStop}%
\bibitem [{\citenamefont {Motl}\ and\ \citenamefont
  {Neitzke}(2003)}]{refMotlNeitzke}%
  \BibitemOpen
  \bibfield  {author} {\bibinfo {author} {\bibfnamefont {L.}~\bibnamefont
  {Motl}}\ and\ \bibinfo {author} {\bibfnamefont {A.}~\bibnamefont {Neitzke}},\
  }\href {\doibase 10.4310/ATMP.2003.v7.n2.a4} {\bibfield  {journal} {\bibinfo
  {journal} {Adv. Theor. Math. Phys.}\ }\textbf {\bibinfo {volume} {7}},\
  \bibinfo {pages} {307} (\bibinfo {year} {2003})},\ \Eprint
  {http://arxiv.org/abs/hep-th/0301173} {arXiv:hep-th/0301173} \BibitemShut
  {NoStop}%
\bibitem [{\citenamefont {Nat{\'a}rio}\ and\ \citenamefont
  {Schiappa}(2004)}]{refNatarioSchiappa}%
  \BibitemOpen
  \bibfield  {author} {\bibinfo {author} {\bibfnamefont {J.}~\bibnamefont
  {Nat{\'a}rio}}\ and\ \bibinfo {author} {\bibfnamefont {R.}~\bibnamefont
  {Schiappa}},\ }\href {\doibase 10.4310/ATMP.2004.v8.n6.a4} {\bibfield
  {journal} {\bibinfo  {journal} {Adv. Theor. Math. Phys.}\ }\textbf {\bibinfo
  {volume} {8}},\ \bibinfo {pages} {1001} (\bibinfo {year} {2004})},\ \Eprint
  {http://arxiv.org/abs/hep-th/0411267} {arXiv:hep-th/0411267} \BibitemShut
  {NoStop}%
\bibitem [{\citenamefont {Das}\ and\ \citenamefont
  {Shankaranarayanan}(2005)}]{Das2005}%
  \BibitemOpen
  \bibfield  {author} {\bibinfo {author} {\bibfnamefont {S.}~\bibnamefont
  {Das}}\ and\ \bibinfo {author} {\bibfnamefont {S.}~\bibnamefont
  {Shankaranarayanan}},\ }\href {\doibase 10.1088/0264-9381/22/3/L01}
  {\bibfield  {journal} {\bibinfo  {journal} {Class. Quant. Grav.}\ }\textbf
  {\bibinfo {volume} {22}},\ \bibinfo {pages} {L7} (\bibinfo {year} {2005})},\
  \Eprint {http://arxiv.org/abs/hep-th/0410209} {arXiv:hep-th/0410209}
  \BibitemShut {NoStop}%
\bibitem [{\citenamefont {Ghosh}, \citenamefont {Shankaranarayanan},\ and\
  \citenamefont {Das}(2006)}]{Ghosh2006}%
  \BibitemOpen
  \bibfield  {author} {\bibinfo {author} {\bibfnamefont {A.}~\bibnamefont
  {Ghosh}}, \bibinfo {author} {\bibfnamefont {S.}~\bibnamefont
  {Shankaranarayanan}}, \ and\ \bibinfo {author} {\bibfnamefont
  {S.}~\bibnamefont {Das}},\ }\href {\doibase 10.1088/0264-9381/23/6/003}
  {\bibfield  {journal} {\bibinfo  {journal} {Class. Quant. Grav.}\ }\textbf
  {\bibinfo {volume} {23}},\ \bibinfo {pages} {1851} (\bibinfo {year}
  {2006})},\ \Eprint {http://arxiv.org/abs/hep-th/0510186}
  {arXiv:hep-th/0510186} \BibitemShut {NoStop}%
\bibitem [{\citenamefont {Birmingham}(2003)}]{Birmingham2003}%
  \BibitemOpen
  \bibfield  {author} {\bibinfo {author} {\bibfnamefont {D.}~\bibnamefont
  {Birmingham}},\ }\href {\doibase 10.1016/j.physletb.2003.07.041} {\bibfield
  {journal} {\bibinfo  {journal} {Phys. Lett. B}\ }\textbf {\bibinfo {volume}
  {569}},\ \bibinfo {pages} {199} (\bibinfo {year} {2003})}\BibitemShut
  {NoStop}%
\bibitem [{\citenamefont {Daghigh}\ and\ \citenamefont
  {Kunstatter}(2005)}]{Daghigh2005generic}%
  \BibitemOpen
  \bibfield  {author} {\bibinfo {author} {\bibfnamefont {R.~G.}\ \bibnamefont
  {Daghigh}}\ and\ \bibinfo {author} {\bibfnamefont {G.}~\bibnamefont
  {Kunstatter}},\ }\href {\doibase 10.1088/0264-9381/22/19/020} {\bibfield
  {journal} {\bibinfo  {journal} {Class. Quant. Grav.}\ }\textbf {\bibinfo
  {volume} {22}},\ \bibinfo {pages} {4113} (\bibinfo {year} {2005})},\ \Eprint
  {http://arxiv.org/abs/gr-qc/0505044} {arXiv:gr-qc/0505044} \BibitemShut
  {NoStop}%
\bibitem [{\citenamefont {Daghigh}\ and\ \citenamefont
  {Kunstatter}(2006)}]{Daghigh2006universal}%
  \BibitemOpen
  \bibfield  {author} {\bibinfo {author} {\bibfnamefont {R.~G.}\ \bibnamefont
  {Daghigh}}\ and\ \bibinfo {author} {\bibfnamefont {G.}~\bibnamefont
  {Kunstatter}},\ }\href {\doibase 10.1139/p06-030} {\bibfield  {journal}
  {\bibinfo  {journal} {Can. J. Phys.}\ }\textbf {\bibinfo {volume} {84}},\
  \bibinfo {pages} {473} (\bibinfo {year} {2006})},\ \Eprint
  {http://arxiv.org/abs/gr-qc/0507019} {arXiv:gr-qc/0507019} \BibitemShut
  {NoStop}%
\bibitem [{\citenamefont {Daghigh}\ \emph {et~al.}(2006)\citenamefont
  {Daghigh}, \citenamefont {Kunstatter}, \citenamefont {Ostapchuk},\ and\
  \citenamefont {Bagnulo}}]{Daghigh2006RNsmallQ}%
  \BibitemOpen
  \bibfield  {author} {\bibinfo {author} {\bibfnamefont {R.~G.}\ \bibnamefont
  {Daghigh}}, \bibinfo {author} {\bibfnamefont {G.}~\bibnamefont {Kunstatter}},
  \bibinfo {author} {\bibfnamefont {D.}~\bibnamefont {Ostapchuk}}, \ and\
  \bibinfo {author} {\bibfnamefont {V.}~\bibnamefont {Bagnulo}},\ }\href
  {\doibase 10.1088/0264-9381/23/17/002} {\bibfield  {journal} {\bibinfo
  {journal} {Class. Quant. Grav.}\ }\textbf {\bibinfo {volume} {23}},\ \bibinfo
  {pages} {5101} (\bibinfo {year} {2006})},\ \Eprint
  {http://arxiv.org/abs/gr-qc/0604073} {arXiv:gr-qc/0604073} \BibitemShut
  {NoStop}%
\bibitem [{\citenamefont {Daghigh}\ and\ \citenamefont
  {Green}(2008)}]{Daghigh2007eRN}%
  \BibitemOpen
  \bibfield  {author} {\bibinfo {author} {\bibfnamefont {R.~G.}\ \bibnamefont
  {Daghigh}}\ and\ \bibinfo {author} {\bibfnamefont {M.~D.}\ \bibnamefont
  {Green}},\ }\href {\doibase 10.1088/0264-9381/25/5/055001} {\bibfield
  {journal} {\bibinfo  {journal} {Class. Quant. Grav.}\ }\textbf {\bibinfo
  {volume} {25}},\ \bibinfo {pages} {055001} (\bibinfo {year} {2008})},\
  \Eprint {http://arxiv.org/abs/0708.1333} {arXiv:0708.1333 [gr-qc]}
  \BibitemShut {NoStop}%
\bibitem [{\citenamefont {Babb}, \citenamefont {Daghigh},\ and\ \citenamefont
  {Kunstatter}(2011)}]{Daghigh2011QMBH}%
  \BibitemOpen
  \bibfield  {author} {\bibinfo {author} {\bibfnamefont {J.}~\bibnamefont
  {Babb}}, \bibinfo {author} {\bibfnamefont {R.}~\bibnamefont {Daghigh}}, \
  and\ \bibinfo {author} {\bibfnamefont {G.}~\bibnamefont {Kunstatter}},\
  }\href {\doibase 10.1103/PhysRevD.84.084031} {\bibfield  {journal} {\bibinfo
  {journal} {Phys. Rev. D}\ }\textbf {\bibinfo {volume} {84}},\ \bibinfo
  {pages} {084031} (\bibinfo {year} {2011})},\ \Eprint
  {http://arxiv.org/abs/1106.4357} {arXiv:1106.4357 [gr-qc]} \BibitemShut
  {NoStop}%
\bibitem [{\citenamefont {Cho}(2006)}]{Cho2006}%
  \BibitemOpen
  \bibfield  {author} {\bibinfo {author} {\bibfnamefont {H.~T.}\ \bibnamefont
  {Cho}},\ }\href {\doibase 10.1103/PhysRevD.73.024019} {\bibfield  {journal}
  {\bibinfo  {journal} {Phys. Rev. D}\ }\textbf {\bibinfo {volume} {73}},\
  \bibinfo {pages} {024019} (\bibinfo {year} {2006})},\ \Eprint
  {http://arxiv.org/abs/gr-qc/0512052} {arXiv:gr-qc/0512052} \BibitemShut
  {NoStop}%
\bibitem [{\citenamefont {L{\'{o}}pez-Ortega}(2006)}]{Lopez-Ortega2006}%
  \BibitemOpen
  \bibfield  {author} {\bibinfo {author} {\bibfnamefont {A.}~\bibnamefont
  {L{\'{o}}pez-Ortega}},\ }\href {\doibase 10.1007/s10714-006-0358-2}
  {\bibfield  {journal} {\bibinfo  {journal} {Gen. Rel. Grav.}\ }\textbf
  {\bibinfo {volume} {38}},\ \bibinfo {pages} {1747} (\bibinfo {year}
  {2006})},\ \Eprint {http://arxiv.org/abs/gr-qc/0605034} {arXiv:gr-qc/0605034}
  \BibitemShut {NoStop}%
\bibitem [{\citenamefont {Casals}\ and\ \citenamefont
  {Ottewill}(2018)}]{Casals2018}%
  \BibitemOpen
  \bibfield  {author} {\bibinfo {author} {\bibfnamefont {M.}~\bibnamefont
  {Casals}}\ and\ \bibinfo {author} {\bibfnamefont {A.~C.}\ \bibnamefont
  {Ottewill}},\ }\href {\doibase 10.1103/PhysRevD.97.024048} {\bibfield
  {journal} {\bibinfo  {journal} {Phys. Rev. D}\ }\textbf {\bibinfo {volume}
  {97}},\ \bibinfo {pages} {024048} (\bibinfo {year} {2018})},\ \Eprint
  {http://arxiv.org/abs/1606.03423} {arXiv:1606.03423 [gr-qc]} \BibitemShut
  {NoStop}%
\bibitem [{\citenamefont {Dreyer}(2003)}]{Dreyer2003}%
  \BibitemOpen
  \bibfield  {author} {\bibinfo {author} {\bibfnamefont {O.}~\bibnamefont
  {Dreyer}},\ }\href {\doibase 10.1103/PhysRevLett.90.081301} {\bibfield
  {journal} {\bibinfo  {journal} {Phys. Rev. Lett.}\ }\textbf {\bibinfo
  {volume} {90}},\ \bibinfo {pages} {081301} (\bibinfo {year} {2003})},\
  \Eprint {http://arxiv.org/abs/gr-qc/0211076} {arXiv:gr-qc/0211076}
  \BibitemShut {NoStop}%
\bibitem [{\citenamefont {Maggiore}(2008)}]{refMaggiore2008}%
  \BibitemOpen
  \bibfield  {author} {\bibinfo {author} {\bibfnamefont {M.}~\bibnamefont
  {Maggiore}},\ }\href {\doibase 10.1103/PhysRevLett.100.141301} {\bibfield
  {journal} {\bibinfo  {journal} {Phys. Rev. Lett.}\ }\textbf {\bibinfo
  {volume} {100}},\ \bibinfo {pages} {141301} (\bibinfo {year} {2008})},\
  \Eprint {http://arxiv.org/abs/0711.3145} {arXiv:0711.3145 [gr-qc]}
  \BibitemShut {NoStop}%
\bibitem [{\citenamefont {Skakala}\ and\ \citenamefont
  {Visser}(2010)}]{Skakala2010}%
  \BibitemOpen
  \bibfield  {author} {\bibinfo {author} {\bibfnamefont {J.}~\bibnamefont
  {Skakala}}\ and\ \bibinfo {author} {\bibfnamefont {M.}~\bibnamefont
  {Visser}},\ }\href {\doibase 10.1103/PhysRevD.81.125023} {\bibfield
  {journal} {\bibinfo  {journal} {Phys. Rev. D}\ }\textbf {\bibinfo {volume}
  {81}},\ \bibinfo {pages} {125023} (\bibinfo {year} {2010})},\ \Eprint
  {http://arxiv.org/abs/1007.4039} {arXiv:1007.4039 [gr-qc]} \BibitemShut
  {NoStop}%
\bibitem [{\citenamefont {Hod}(1998)}]{Hod1998}%
  \BibitemOpen
  \bibfield  {author} {\bibinfo {author} {\bibfnamefont {S.}~\bibnamefont
  {Hod}},\ }\href {\doibase 10.1103/PhysRevLett.81.4293} {\bibfield  {journal}
  {\bibinfo  {journal} {Phys. Rev. Lett.}\ }\textbf {\bibinfo {volume} {81}},\
  \bibinfo {pages} {4293} (\bibinfo {year} {1998})},\ \Eprint
  {http://arxiv.org/abs/gr-qc/9812002} {arXiv:gr-qc/9812002} \BibitemShut
  {NoStop}%
\bibitem [{\citenamefont {Bekenstein}(1972)}]{refBekenstein1972}%
  \BibitemOpen
  \bibfield  {author} {\bibinfo {author} {\bibfnamefont {J.~D.}\ \bibnamefont
  {Bekenstein}},\ }\href {\doibase 10.1007/BF02757029} {\bibfield  {journal}
  {\bibinfo  {journal} {Lett. Nuovo Cim.}\ }\textbf {\bibinfo {volume} {4}},\
  \bibinfo {pages} {737} (\bibinfo {year} {1972})}\BibitemShut {NoStop}%
\bibitem [{\citenamefont {Mukhanov}(1986)}]{refMukhanov1986}%
  \BibitemOpen
  \bibfield  {author} {\bibinfo {author} {\bibfnamefont {V.~F.}\ \bibnamefont
  {Mukhanov}},\ }\href@noop {} {\bibfield  {journal} {\bibinfo  {journal} {JETP
  Lett.}\ }\textbf {\bibinfo {volume} {44}},\ \bibinfo {pages} {63} (\bibinfo
  {year} {1986})}\BibitemShut {NoStop}%
\bibitem [{\citenamefont {Bekenstein}\ and\ \citenamefont
  {Mukhanov}(1995)}]{refBekenstein1995}%
  \BibitemOpen
  \bibfield  {author} {\bibinfo {author} {\bibfnamefont {J.~D.}\ \bibnamefont
  {Bekenstein}}\ and\ \bibinfo {author} {\bibfnamefont {V.~F.}\ \bibnamefont
  {Mukhanov}},\ }\href {\doibase 10.1016/0370-2693(95)01148-J} {\bibfield
  {journal} {\bibinfo  {journal} {Phys. Lett. B}\ }\textbf {\bibinfo {volume}
  {360}},\ \bibinfo {pages} {7} (\bibinfo {year} {1995})},\ \Eprint
  {http://arxiv.org/abs/gr-qc/9505012} {arXiv:gr-qc/9505012} \BibitemShut
  {NoStop}%
\bibitem [{\citenamefont {Nollert}(1993)}]{refNollert1993}%
  \BibitemOpen
  \bibfield  {author} {\bibinfo {author} {\bibfnamefont {H.-P.}\ \bibnamefont
  {Nollert}},\ }\href {\doibase 10.1103/PhysRevD.47.5253} {\bibfield  {journal}
  {\bibinfo  {journal} {Phys. Rev. D}\ }\textbf {\bibinfo {volume} {47}},\
  \bibinfo {pages} {5253} (\bibinfo {year} {1993})}\BibitemShut {NoStop}%
\bibitem [{\citenamefont {Cardoso}\ and\ \citenamefont
  {Lemos}(2003)}]{refCardosoLemos2003}%
  \BibitemOpen
  \bibfield  {author} {\bibinfo {author} {\bibfnamefont {V.}~\bibnamefont
  {Cardoso}}\ and\ \bibinfo {author} {\bibfnamefont {J.~P.~S.}\ \bibnamefont
  {Lemos}},\ }\href {\doibase 10.1103/PhysRevD.67.084020} {\bibfield  {journal}
  {\bibinfo  {journal} {Phys. Rev. D}\ }\textbf {\bibinfo {volume} {67}},\
  \bibinfo {pages} {084020} (\bibinfo {year} {2003})}\BibitemShut {NoStop}%
\bibitem [{\citenamefont {Cardoso}, \citenamefont {Natario},\ and\
  \citenamefont {Schiappa}(2004)}]{refCardosoNatarioSchiappa}%
  \BibitemOpen
  \bibfield  {author} {\bibinfo {author} {\bibfnamefont {V.}~\bibnamefont
  {Cardoso}}, \bibinfo {author} {\bibfnamefont {J.}~\bibnamefont {Natario}}, \
  and\ \bibinfo {author} {\bibfnamefont {R.}~\bibnamefont {Schiappa}},\ }\href
  {\doibase 10.1063/1.1812828} {\bibfield  {journal} {\bibinfo  {journal} {J.
  Math. Phys.}\ }\textbf {\bibinfo {volume} {45}},\ \bibinfo {pages} {4698}
  (\bibinfo {year} {2004})},\ \Eprint {http://arxiv.org/abs/hep-th/0403132}
  {arXiv:hep-th/0403132} \BibitemShut {NoStop}%
\bibitem [{\citenamefont {Daghigh}\ and\ \citenamefont
  {Green}(2009)}]{Daghigh2008SAdS}%
  \BibitemOpen
  \bibfield  {author} {\bibinfo {author} {\bibfnamefont {R.~G.}\ \bibnamefont
  {Daghigh}}\ and\ \bibinfo {author} {\bibfnamefont {M.~D.}\ \bibnamefont
  {Green}},\ }\href {\doibase 10.1088/0264-9381/26/12/125017} {\bibfield
  {journal} {\bibinfo  {journal} {Class. Quant. Grav.}\ }\textbf {\bibinfo
  {volume} {26}},\ \bibinfo {pages} {125017} (\bibinfo {year} {2009})},\
  \Eprint {http://arxiv.org/abs/0808.1596} {arXiv:0808.1596 [gr-qc]}
  \BibitemShut {NoStop}%
\bibitem [{\citenamefont {Arnold}\ and\ \citenamefont
  {Szepietowski}(2013)}]{Arnold2013}%
  \BibitemOpen
  \bibfield  {author} {\bibinfo {author} {\bibfnamefont {P.}~\bibnamefont
  {Arnold}}\ and\ \bibinfo {author} {\bibfnamefont {P.}~\bibnamefont
  {Szepietowski}},\ }\href {\doibase 10.1103/PhysRevD.88.086002} {\bibfield
  {journal} {\bibinfo  {journal} {Phys. Rev. D}\ }\textbf {\bibinfo {volume}
  {88}},\ \bibinfo {pages} {086002} (\bibinfo {year} {2013})},\ \Eprint
  {http://arxiv.org/abs/1308.0341} {arXiv:1308.0341 [hep-th]} \BibitemShut
  {NoStop}%
\bibitem [{\citenamefont {Arnold}, \citenamefont {Szepietowski},\ and\
  \citenamefont {Vaman}(2014)}]{Arnold2014}%
  \BibitemOpen
  \bibfield  {author} {\bibinfo {author} {\bibfnamefont {P.}~\bibnamefont
  {Arnold}}, \bibinfo {author} {\bibfnamefont {P.}~\bibnamefont
  {Szepietowski}}, \ and\ \bibinfo {author} {\bibfnamefont {D.}~\bibnamefont
  {Vaman}},\ }\href {\doibase 10.1103/PhysRevD.89.046001} {\bibfield  {journal}
  {\bibinfo  {journal} {Phys. Rev. D}\ }\textbf {\bibinfo {volume} {89}},\
  \bibinfo {pages} {046001} (\bibinfo {year} {2014})},\ \Eprint
  {http://arxiv.org/abs/1311.6409} {arXiv:1311.6409 [hep-th]} \BibitemShut
  {NoStop}%
\bibitem [{\citenamefont {Medved}(2008)}]{Medved2008}%
  \BibitemOpen
  \bibfield  {author} {\bibinfo {author} {\bibfnamefont {A.~J.~M.}\
  \bibnamefont {Medved}},\ }\href {\doibase 10.1088/0264-9381/25/20/205014}
  {\bibfield  {journal} {\bibinfo  {journal} {Class. Quant. Grav.}\ }\textbf
  {\bibinfo {volume} {25}},\ \bibinfo {pages} {205014} (\bibinfo {year}
  {2008})},\ \Eprint {http://arxiv.org/abs/0804.4346} {arXiv:0804.4346 [gr-qc]}
  \BibitemShut {NoStop}%
\bibitem [{\citenamefont {Daghigh}\ \emph {et~al.}(2020)\citenamefont
  {Daghigh}, \citenamefont {Green}, \citenamefont {Morey},\ and\ \citenamefont
  {Kunstatter}}]{Daghigh2020_ScalarPertRegBH}%
  \BibitemOpen
  \bibfield  {author} {\bibinfo {author} {\bibfnamefont {R.~G.}\ \bibnamefont
  {Daghigh}}, \bibinfo {author} {\bibfnamefont {M.~D.}\ \bibnamefont {Green}},
  \bibinfo {author} {\bibfnamefont {J.~C.}\ \bibnamefont {Morey}}, \ and\
  \bibinfo {author} {\bibfnamefont {G.}~\bibnamefont {Kunstatter}},\ }\href
  {\doibase 10.1103/PhysRevD.102.104040} {\bibfield  {journal} {\bibinfo
  {journal} {Phys. Rev. D}\ }\textbf {\bibinfo {volume} {102}},\ \bibinfo
  {pages} {104040} (\bibinfo {year} {2020})},\ \Eprint
  {http://arxiv.org/abs/2009.02367} {arXiv:2009.02367 [gr-qc]} \BibitemShut
  {NoStop}%
\bibitem [{\citenamefont {Kodama}\ and\ \citenamefont
  {Ishibashi}(2003)}]{refIKschwarz1}%
  \BibitemOpen
  \bibfield  {author} {\bibinfo {author} {\bibfnamefont {H.}~\bibnamefont
  {Kodama}}\ and\ \bibinfo {author} {\bibfnamefont {A.}~\bibnamefont
  {Ishibashi}},\ }\href {\doibase 10.1143/PTP.110.701} {\bibfield  {journal}
  {\bibinfo  {journal} {Prog. Theor. Phys.}\ }\textbf {\bibinfo {volume}
  {110}},\ \bibinfo {pages} {701} (\bibinfo {year} {2003})},\ \Eprint
  {http://arxiv.org/abs/hep-th/0305147} {arXiv:hep-th/0305147} \BibitemShut
  {NoStop}%
\bibitem [{\citenamefont {Ishibashi}\ and\ \citenamefont
  {Kodama}(2003)}]{refIKschwarz2}%
  \BibitemOpen
  \bibfield  {author} {\bibinfo {author} {\bibfnamefont {A.}~\bibnamefont
  {Ishibashi}}\ and\ \bibinfo {author} {\bibfnamefont {H.}~\bibnamefont
  {Kodama}},\ }\href {\doibase 10.1143/PTP.110.901} {\bibfield  {journal}
  {\bibinfo  {journal} {Prog. Theor. Phys.}\ }\textbf {\bibinfo {volume}
  {110}},\ \bibinfo {pages} {901} (\bibinfo {year} {2003})},\ \Eprint
  {http://arxiv.org/abs/hep-th/0305185} {arXiv:hep-th/0305185} \BibitemShut
  {NoStop}%
\bibitem [{\citenamefont {Kodama}\ and\ \citenamefont
  {Ishibashi}(2004)}]{refIKrn}%
  \BibitemOpen
  \bibfield  {author} {\bibinfo {author} {\bibfnamefont {H.}~\bibnamefont
  {Kodama}}\ and\ \bibinfo {author} {\bibfnamefont {A.}~\bibnamefont
  {Ishibashi}},\ }\href {\doibase 10.1143/PTP.111.29} {\bibfield  {journal}
  {\bibinfo  {journal} {Prog. Theor. Phys.}\ }\textbf {\bibinfo {volume}
  {111}},\ \bibinfo {pages} {29} (\bibinfo {year} {2004})},\ \Eprint
  {http://arxiv.org/abs/hep-th/0308128} {arXiv:hep-th/0308128} \BibitemShut
  {NoStop}%
\bibitem [{\citenamefont {Ishibashi}\ and\ \citenamefont
  {Kodama}(2011)}]{refIKchap6}%
  \BibitemOpen
  \bibfield  {author} {\bibinfo {author} {\bibfnamefont {A.}~\bibnamefont
  {Ishibashi}}\ and\ \bibinfo {author} {\bibfnamefont {H.}~\bibnamefont
  {Kodama}},\ }\href {\doibase 10.1143/PTPS.189.165} {\bibfield  {journal}
  {\bibinfo  {journal} {Prog. Theor. Phys. Suppl.}\ }\textbf {\bibinfo {volume}
  {189}},\ \bibinfo {pages} {165} (\bibinfo {year} {2011})},\ \Eprint
  {http://arxiv.org/abs/1103.6148} {arXiv:1103.6148 [hep-th]} \BibitemShut
  {NoStop}%
\bibitem [{\citenamefont {Regge}\ and\ \citenamefont {Wheeler}(1957)}]{refRW}%
  \BibitemOpen
  \bibfield  {author} {\bibinfo {author} {\bibfnamefont {T.}~\bibnamefont
  {Regge}}\ and\ \bibinfo {author} {\bibfnamefont {J.~A.}\ \bibnamefont
  {Wheeler}},\ }\href {\doibase 10.1103/PhysRev.108.1063} {\bibfield  {journal}
  {\bibinfo  {journal} {Phys. Rev.}\ }\textbf {\bibinfo {volume} {108}},\
  \bibinfo {pages} {1063} (\bibinfo {year} {1957})}\BibitemShut {NoStop}%
\bibitem [{\citenamefont {Zerilli}(1974)}]{ZerMon1}%
  \BibitemOpen
  \bibfield  {author} {\bibinfo {author} {\bibfnamefont {F.~J.}\ \bibnamefont
  {Zerilli}},\ }\href {\doibase 10.1103/PhysRevD.9.860} {\bibfield  {journal}
  {\bibinfo  {journal} {Phys. Rev. D}\ }\textbf {\bibinfo {volume} {9}},\
  \bibinfo {pages} {860} (\bibinfo {year} {1974})}\BibitemShut {NoStop}%
\bibitem [{\citenamefont {Moncrief}(1974{\natexlab{a}})}]{ZerMon2}%
  \BibitemOpen
  \bibfield  {author} {\bibinfo {author} {\bibfnamefont {V.}~\bibnamefont
  {Moncrief}},\ }\href {\doibase 10.1103/PhysRevD.9.2707} {\bibfield  {journal}
  {\bibinfo  {journal} {Phys. Rev. D}\ }\textbf {\bibinfo {volume} {9}},\
  \bibinfo {pages} {2707} (\bibinfo {year} {1974}{\natexlab{a}})}\BibitemShut
  {NoStop}%
\bibitem [{\citenamefont {Moncrief}(1974{\natexlab{b}})}]{ZerMon3}%
  \BibitemOpen
  \bibfield  {author} {\bibinfo {author} {\bibfnamefont {V.}~\bibnamefont
  {Moncrief}},\ }\href {\doibase 10.1103/PhysRevD.10.1057} {\bibfield
  {journal} {\bibinfo  {journal} {Phys. Rev. D}\ }\textbf {\bibinfo {volume}
  {10}},\ \bibinfo {pages} {1057} (\bibinfo {year}
  {1974}{\natexlab{b}})}\BibitemShut {NoStop}%
\bibitem [{\citenamefont {Moncrief}(1975)}]{ZerMon4}%
  \BibitemOpen
  \bibfield  {author} {\bibinfo {author} {\bibfnamefont {V.}~\bibnamefont
  {Moncrief}},\ }\href {\doibase 10.1103/PhysRevD.12.1526} {\bibfield
  {journal} {\bibinfo  {journal} {Phys. Rev. D}\ }\textbf {\bibinfo {volume}
  {12}},\ \bibinfo {pages} {1526} (\bibinfo {year} {1975})}\BibitemShut
  {NoStop}%
\bibitem [{\citenamefont {Decanini}, \citenamefont {Folacci},\ and\
  \citenamefont {Raffaelli}(2011)}]{refDecanini2011}%
  \BibitemOpen
  \bibfield  {author} {\bibinfo {author} {\bibfnamefont {Y.}~\bibnamefont
  {Decanini}}, \bibinfo {author} {\bibfnamefont {A.}~\bibnamefont {Folacci}}, \
  and\ \bibinfo {author} {\bibfnamefont {B.}~\bibnamefont {Raffaelli}},\ }\href
  {\doibase 10.1103/PhysRevD.84.084035} {\bibfield  {journal} {\bibinfo
  {journal} {Phys. Rev. D}\ }\textbf {\bibinfo {volume} {84}},\ \bibinfo
  {pages} {084035} (\bibinfo {year} {2011})},\ \Eprint
  {http://arxiv.org/abs/1108.5076} {arXiv:1108.5076 [gr-qc]} \BibitemShut
  {NoStop}%
\bibitem [{\citenamefont {Chen}, \citenamefont {Cho},\ and\ \citenamefont
  {Cornell}(2020)}]{refnewQNMs2020}%
  \BibitemOpen
  \bibfield  {author} {\bibinfo {author} {\bibfnamefont {C.-H.}\ \bibnamefont
  {Chen}}, \bibinfo {author} {\bibfnamefont {H.-T.}\ \bibnamefont {Cho}}, \
  and\ \bibinfo {author} {\bibfnamefont {A.~S.}\ \bibnamefont {Cornell}},\
  }\href {\doibase 10.1016/j.cjph.2020.08.015} {\bibfield  {journal} {\bibinfo
  {journal} {Chin. J. Phys.}\ }\textbf {\bibinfo {volume} {67}},\ \bibinfo
  {pages} {646} (\bibinfo {year} {2020})},\ \Eprint
  {http://arxiv.org/abs/2004.05806} {arXiv:2004.05806 [gr-qc]} \BibitemShut
  {NoStop}%
\bibitem [{\citenamefont {Chen}\ \emph {et~al.}(2021)\citenamefont {Chen},
  \citenamefont {Cho}, \citenamefont {Chrysostomou},\ and\ \citenamefont
  {Cornell}}]{refOurLargeL}%
  \BibitemOpen
  \bibfield  {author} {\bibinfo {author} {\bibfnamefont {C.-H.}\ \bibnamefont
  {Chen}}, \bibinfo {author} {\bibfnamefont {H.-T.}\ \bibnamefont {Cho}},
  \bibinfo {author} {\bibfnamefont {A.}~\bibnamefont {Chrysostomou}}, \ and\
  \bibinfo {author} {\bibfnamefont {A.~S.}\ \bibnamefont {Cornell}},\
  }\href@noop {} {\  (\bibinfo {year} {2021})},\ \Eprint
  {http://arxiv.org/abs/2103.07777} {arXiv:2103.07777 [gr-qc]} \BibitemShut
  {NoStop}%
\bibitem [{\citenamefont {Daghigh}\ and\ \citenamefont
  {Green}(2012)}]{Daghigh2012WKBvalid}%
  \BibitemOpen
  \bibfield  {author} {\bibinfo {author} {\bibfnamefont {R.~G.}\ \bibnamefont
  {Daghigh}}\ and\ \bibinfo {author} {\bibfnamefont {M.~D.}\ \bibnamefont
  {Green}},\ }\href {\doibase 10.1103/PhysRevD.85.127501} {\bibfield  {journal}
  {\bibinfo  {journal} {Phys. Rev. D}\ }\textbf {\bibinfo {volume} {85}},\
  \bibinfo {pages} {127501} (\bibinfo {year} {2012})},\ \Eprint
  {http://arxiv.org/abs/1112.5397} {arXiv:1112.5397 [gr-qc]} \BibitemShut
  {NoStop}%
\bibitem [{\citenamefont {Wick}(1954)}]{Wick1954}%
  \BibitemOpen
  \bibfield  {author} {\bibinfo {author} {\bibfnamefont {G.~C.}\ \bibnamefont
  {Wick}},\ }\href {\doibase 10.1103/PhysRev.96.1124} {\bibfield  {journal}
  {\bibinfo  {journal} {Phys. Rev.}\ }\textbf {\bibinfo {volume} {96}},\
  \bibinfo {pages} {1124} (\bibinfo {year} {1954})}\BibitemShut {NoStop}%
\bibitem [{\citenamefont {Abramowitz}\ and\ \citenamefont
  {Stegun}(1964)}]{refmathbible}%
  \BibitemOpen
  \bibfield  {author} {\bibinfo {author} {\bibfnamefont {M.}~\bibnamefont
  {Abramowitz}}\ and\ \bibinfo {author} {\bibfnamefont {I.~A.}\ \bibnamefont
  {Stegun}},\ }\href@noop {} {\emph {\bibinfo {title} {Handbook of Mathematical
  Functions}}},\ \bibinfo {edition} {6th}\ ed.\ (\bibinfo  {publisher} {U.S.
  Government Printing Office},\ \bibinfo {address} {Washington D.C.},\ \bibinfo
  {year} {1964})\BibitemShut {NoStop}%
\bibitem [{\citenamefont {Berti}\ and\ \citenamefont
  {Kokkotas}(2003{\natexlab{b}})}]{refBertiKokkotas}%
  \BibitemOpen
  \bibfield  {author} {\bibinfo {author} {\bibfnamefont {E.}~\bibnamefont
  {Berti}}\ and\ \bibinfo {author} {\bibfnamefont {K.~D.}\ \bibnamefont
  {Kokkotas}},\ }\href {\doibase 10.1103/PhysRevD.67.064020} {\bibfield
  {journal} {\bibinfo  {journal} {Phys. Rev. D}\ }\textbf {\bibinfo {volume}
  {67}},\ \bibinfo {pages} {064020} (\bibinfo {year} {2003}{\natexlab{b}})},\
  \Eprint {http://arxiv.org/abs/gr-qc/0301052} {arXiv:gr-qc/0301052}
  \BibitemShut {NoStop}%
\bibitem [{\citenamefont {Berti}(2004)}]{Berti2004}%
  \BibitemOpen
  \bibfield  {author} {\bibinfo {author} {\bibfnamefont {E.}~\bibnamefont
  {Berti}},\ }\href@noop {} {\bibfield  {journal} {\bibinfo  {journal} {Conf.
  Proc. C}\ }\textbf {\bibinfo {volume} {0405132}},\ \bibinfo {pages} {145}
  (\bibinfo {year} {2004})},\ \Eprint {http://arxiv.org/abs/gr-qc/0411025}
  {arXiv:gr-qc/0411025} \BibitemShut {NoStop}%
\bibitem [{\citenamefont {Onozawa}\ \emph {et~al.}(1996)\citenamefont
  {Onozawa}, \citenamefont {Mishima}, \citenamefont {Okamura},\ and\
  \citenamefont {Ishihara}}]{refOnozawa1996}%
  \BibitemOpen
  \bibfield  {author} {\bibinfo {author} {\bibfnamefont {H.}~\bibnamefont
  {Onozawa}}, \bibinfo {author} {\bibfnamefont {T.}~\bibnamefont {Mishima}},
  \bibinfo {author} {\bibfnamefont {T.}~\bibnamefont {Okamura}}, \ and\
  \bibinfo {author} {\bibfnamefont {H.}~\bibnamefont {Ishihara}},\ }\href
  {\doibase 10.1103/PhysRevD.53.7033} {\bibfield  {journal} {\bibinfo
  {journal} {Phys. Rev. D}\ }\textbf {\bibinfo {volume} {53}},\ \bibinfo
  {pages} {7033} (\bibinfo {year} {1996})},\ \Eprint
  {http://arxiv.org/abs/gr-qc/9603021} {arXiv:gr-qc/9603021} \BibitemShut
  {NoStop}%
\bibitem [{\citenamefont {Bekenstein}(1974)}]{refBekenstein1974}%
  \BibitemOpen
  \bibfield  {author} {\bibinfo {author} {\bibfnamefont {J.~D.}\ \bibnamefont
  {Bekenstein}},\ }\href {\doibase 10.1007/BF02762768} {\bibfield  {journal}
  {\bibinfo  {journal} {Lett. Nuovo Cim.}\ }\textbf {\bibinfo {volume} {11}},\
  \bibinfo {pages} {467} (\bibinfo {year} {1974})}\BibitemShut {NoStop}%
\bibitem [{\citenamefont {Onozawa}\ \emph {et~al.}(1997)\citenamefont
  {Onozawa}, \citenamefont {Okamura}, \citenamefont {Mishima},\ and\
  \citenamefont {Ishihara}}]{Onozawa1996}%
  \BibitemOpen
  \bibfield  {author} {\bibinfo {author} {\bibfnamefont {H.}~\bibnamefont
  {Onozawa}}, \bibinfo {author} {\bibfnamefont {T.}~\bibnamefont {Okamura}},
  \bibinfo {author} {\bibfnamefont {T.}~\bibnamefont {Mishima}}, \ and\
  \bibinfo {author} {\bibfnamefont {H.}~\bibnamefont {Ishihara}},\ }\href
  {\doibase 10.1103/PhysRevD.55.R4529} {\bibfield  {journal} {\bibinfo
  {journal} {Phys. Rev. D}\ }\textbf {\bibinfo {volume} {55}},\ \bibinfo
  {pages} {1} (\bibinfo {year} {1997})}\BibitemShut {NoStop}%
\bibitem [{\citenamefont {Kallosh}, \citenamefont {Rahmfeld},\ and\
  \citenamefont {Wong}(1998)}]{Kallosh1997}%
  \BibitemOpen
  \bibfield  {author} {\bibinfo {author} {\bibfnamefont {R.}~\bibnamefont
  {Kallosh}}, \bibinfo {author} {\bibfnamefont {J.}~\bibnamefont {Rahmfeld}}, \
  and\ \bibinfo {author} {\bibfnamefont {W.~K.}\ \bibnamefont {Wong}},\ }\href
  {\doibase 10.1103/PhysRevD.57.1063} {\bibfield  {journal} {\bibinfo
  {journal} {Phys. Rev. D}\ }\textbf {\bibinfo {volume} {57}},\ \bibinfo
  {pages} {1063} (\bibinfo {year} {1998})},\ \Eprint
  {http://arxiv.org/abs/hep-th/9706048} {arXiv:hep-th/9706048} \BibitemShut
  {NoStop}%
\bibitem [{\citenamefont {Zerilli}(1970)}]{refZerilli}%
  \BibitemOpen
  \bibfield  {author} {\bibinfo {author} {\bibfnamefont {F.~J.}\ \bibnamefont
  {Zerilli}},\ }\href {\doibase 10.1103/PhysRevD.2.2141} {\bibfield  {journal}
  {\bibinfo  {journal} {Phys. Rev. D}\ }\textbf {\bibinfo {volume} {2}},\
  \bibinfo {pages} {2141} (\bibinfo {year} {1970})}\BibitemShut {NoStop}%
\bibitem [{\citenamefont {Crispino}, \citenamefont {Higuchi},\ and\
  \citenamefont {Matsas}(2001)}]{refCrispinoHiguchiMatsas}%
  \BibitemOpen
  \bibfield  {author} {\bibinfo {author} {\bibfnamefont {L.~C.~B.}\
  \bibnamefont {Crispino}}, \bibinfo {author} {\bibfnamefont {A.}~\bibnamefont
  {Higuchi}}, \ and\ \bibinfo {author} {\bibfnamefont {G.~E.~A.}\ \bibnamefont
  {Matsas}},\ }\href {\doibase 10.1103/PhysRevD.80.029906} {\bibfield
  {journal} {\bibinfo  {journal} {Phys. Rev. D}\ }\textbf {\bibinfo {volume}
  {63}},\ \bibinfo {pages} {124008} (\bibinfo {year} {2001})},\ \bibinfo {note}
  {[Erratum: Phys.Rev.D 80, 029906 (2009)]}\BibitemShut {NoStop}%
\bibitem [{\citenamefont {Cooper}, \citenamefont {Khare},\ and\ \citenamefont
  {Sukhatme}(1995)}]{refSUSYpot}%
  \BibitemOpen
  \bibfield  {author} {\bibinfo {author} {\bibfnamefont {F.}~\bibnamefont
  {Cooper}}, \bibinfo {author} {\bibfnamefont {A.}~\bibnamefont {Khare}}, \
  and\ \bibinfo {author} {\bibfnamefont {U.}~\bibnamefont {Sukhatme}},\ }\href
  {\doibase https://doi.org/10.1016/0370-1573(94)00080-M} {\bibfield  {journal}
  {\bibinfo  {journal} {Physics Reports}\ }\textbf {\bibinfo {volume} {251}},\
  \bibinfo {pages} {267} (\bibinfo {year} {1995})}\BibitemShut {NoStop}%
\bibitem [{\citenamefont {Chen}\ \emph {et~al.}(2015)\citenamefont {Chen},
  \citenamefont {Cho}, \citenamefont {Cornell}, \citenamefont {Harmsen},\ and\
  \citenamefont {Naylor}}]{refRS15}%
  \BibitemOpen
  \bibfield  {author} {\bibinfo {author} {\bibfnamefont {C.~H.}\ \bibnamefont
  {Chen}}, \bibinfo {author} {\bibfnamefont {H.~T.}\ \bibnamefont {Cho}},
  \bibinfo {author} {\bibfnamefont {A.~S.}\ \bibnamefont {Cornell}}, \bibinfo
  {author} {\bibfnamefont {G.}~\bibnamefont {Harmsen}}, \ and\ \bibinfo
  {author} {\bibfnamefont {W.}~\bibnamefont {Naylor}},\ }\href {\doibase
  10.6122/CJP.20150511} {\bibfield  {journal} {\bibinfo  {journal} {Chin. J.
  Phys.}\ }\textbf {\bibinfo {volume} {53}},\ \bibinfo {pages} {110101}
  (\bibinfo {year} {2015})},\ \Eprint {http://arxiv.org/abs/1504.02579}
  {arXiv:1504.02579 [gr-qc]} \BibitemShut {NoStop}%
\bibitem [{\citenamefont {Chen}\ \emph {et~al.}(2016)\citenamefont {Chen},
  \citenamefont {Cho}, \citenamefont {Cornell},\ and\ \citenamefont
  {Harmsen}}]{refRSSchwarz16}%
  \BibitemOpen
  \bibfield  {author} {\bibinfo {author} {\bibfnamefont {C.~H.}\ \bibnamefont
  {Chen}}, \bibinfo {author} {\bibfnamefont {H.~T.}\ \bibnamefont {Cho}},
  \bibinfo {author} {\bibfnamefont {A.~S.}\ \bibnamefont {Cornell}}, \ and\
  \bibinfo {author} {\bibfnamefont {G.}~\bibnamefont {Harmsen}},\ }\href
  {\doibase 10.1103/PhysRevD.94.044052} {\bibfield  {journal} {\bibinfo
  {journal} {Phys. Rev. D}\ }\textbf {\bibinfo {volume} {94}},\ \bibinfo
  {pages} {044052} (\bibinfo {year} {2016})},\ \Eprint
  {http://arxiv.org/abs/1605.05263} {arXiv:1605.05263 [gr-qc]} \BibitemShut
  {NoStop}%
\bibitem [{\citenamefont {Chen}\ \emph {et~al.}(2018)\citenamefont {Chen},
  \citenamefont {Cho}, \citenamefont {Cornell}, \citenamefont {Harmsen},\ and\
  \citenamefont {Ngcobo}}]{refRSRN18}%
  \BibitemOpen
  \bibfield  {author} {\bibinfo {author} {\bibfnamefont {C.~H.}\ \bibnamefont
  {Chen}}, \bibinfo {author} {\bibfnamefont {H.~T.}\ \bibnamefont {Cho}},
  \bibinfo {author} {\bibfnamefont {A.~S.}\ \bibnamefont {Cornell}}, \bibinfo
  {author} {\bibfnamefont {G.}~\bibnamefont {Harmsen}}, \ and\ \bibinfo
  {author} {\bibfnamefont {X.}~\bibnamefont {Ngcobo}},\ }\href {\doibase
  10.1103/PhysRevD.97.024038} {\bibfield  {journal} {\bibinfo  {journal} {Phys.
  Rev. D}\ }\textbf {\bibinfo {volume} {97}},\ \bibinfo {pages} {024038}
  (\bibinfo {year} {2018})},\ \Eprint {http://arxiv.org/abs/1710.08024}
  {arXiv:1710.08024 [gr-qc]} \BibitemShut {NoStop}%
\bibitem [{\citenamefont {Chen}\ \emph {et~al.}(2019)\citenamefont {Chen},
  \citenamefont {Cho}, \citenamefont {Cornell},\ and\ \citenamefont
  {Harmsen}}]{refRSSAdS19}%
  \BibitemOpen
  \bibfield  {author} {\bibinfo {author} {\bibfnamefont {C.-H.}\ \bibnamefont
  {Chen}}, \bibinfo {author} {\bibfnamefont {H.-T.}\ \bibnamefont {Cho}},
  \bibinfo {author} {\bibfnamefont {A.~S.}\ \bibnamefont {Cornell}}, \ and\
  \bibinfo {author} {\bibfnamefont {G.~E.}\ \bibnamefont {Harmsen}},\ }\href
  {\doibase 10.1103/PhysRevD.100.104018} {\bibfield  {journal} {\bibinfo
  {journal} {Phys. Rev. D}\ }\textbf {\bibinfo {volume} {100}},\ \bibinfo
  {pages} {104018} (\bibinfo {year} {2019})},\ \Eprint
  {http://arxiv.org/abs/1907.11856} {arXiv:1907.11856 [gr-qc]} \BibitemShut
  {NoStop}%
\bibitem [{\citenamefont {Cho}\ \emph {et~al.}(2007)\citenamefont {Cho},
  \citenamefont {Cornell}, \citenamefont {Doukas},\ and\ \citenamefont
  {Naylor}}]{refDirac07}%
  \BibitemOpen
  \bibfield  {author} {\bibinfo {author} {\bibfnamefont {H.~T.}\ \bibnamefont
  {Cho}}, \bibinfo {author} {\bibfnamefont {A.~S.}\ \bibnamefont {Cornell}},
  \bibinfo {author} {\bibfnamefont {J.}~\bibnamefont {Doukas}}, \ and\ \bibinfo
  {author} {\bibfnamefont {W.}~\bibnamefont {Naylor}},\ }\href {\doibase
  10.1103/PhysRevD.75.104005} {\bibfield  {journal} {\bibinfo  {journal} {Phys.
  Rev. D}\ }\textbf {\bibinfo {volume} {75}},\ \bibinfo {pages} {104005}
  (\bibinfo {year} {2007})},\ \Eprint {http://arxiv.org/abs/hep-th/0701193}
  {arXiv:hep-th/0701193} \BibitemShut {NoStop}%
\bibitem [{\citenamefont {Chakrabarti}(2009)}]{refChakrabarti}%
  \BibitemOpen
  \bibfield  {author} {\bibinfo {author} {\bibfnamefont {S.~K.}\ \bibnamefont
  {Chakrabarti}},\ }\href {\doibase 10.1140/epjc/s10052-009-1026-y} {\bibfield
  {journal} {\bibinfo  {journal} {Eur. Phys. J. C}\ }\textbf {\bibinfo {volume}
  {61}},\ \bibinfo {pages} {477} (\bibinfo {year} {2009})},\ \Eprint
  {http://arxiv.org/abs/0809.1004} {arXiv:0809.1004 [gr-qc]} \BibitemShut
  {NoStop}%
\bibitem [{\citenamefont {Kallosh}\ \emph {et~al.}(1992)\citenamefont
  {Kallosh}, \citenamefont {Linde}, \citenamefont {Ortin}, \citenamefont
  {Peet},\ and\ \citenamefont {Van~Proeyen}}]{Kallosh1992SUSYcosmic}%
  \BibitemOpen
  \bibfield  {author} {\bibinfo {author} {\bibfnamefont {R.}~\bibnamefont
  {Kallosh}}, \bibinfo {author} {\bibfnamefont {A.~D.}\ \bibnamefont {Linde}},
  \bibinfo {author} {\bibfnamefont {T.}~\bibnamefont {Ortin}}, \bibinfo
  {author} {\bibfnamefont {A.~W.}\ \bibnamefont {Peet}}, \ and\ \bibinfo
  {author} {\bibfnamefont {A.}~\bibnamefont {Van~Proeyen}},\ }\href {\doibase
  10.1103/PhysRevD.46.5278} {\bibfield  {journal} {\bibinfo  {journal} {Phys.
  Rev. D}\ }\textbf {\bibinfo {volume} {46}},\ \bibinfo {pages} {5278}
  (\bibinfo {year} {1992})},\ \Eprint {http://arxiv.org/abs/hep-th/9205027}
  {arXiv:hep-th/9205027} \BibitemShut {NoStop}%
\bibitem [{\citenamefont {Kallosh}(1992)}]{Kallosh1992SUSYBH}%
  \BibitemOpen
  \bibfield  {author} {\bibinfo {author} {\bibfnamefont {R.}~\bibnamefont
  {Kallosh}},\ }\href {\doibase 10.1016/0370-2693(92)90482-J} {\bibfield
  {journal} {\bibinfo  {journal} {Phys. Lett. B}\ }\textbf {\bibinfo {volume}
  {282}},\ \bibinfo {pages} {80} (\bibinfo {year} {1992})},\ \Eprint
  {http://arxiv.org/abs/hep-th/9201029} {arXiv:hep-th/9201029} \BibitemShut
  {NoStop}%
\bibitem [{\citenamefont {Gibbons}\ and\ \citenamefont
  {Hull}(1982)}]{GibbonsextRN}%
  \BibitemOpen
  \bibfield  {author} {\bibinfo {author} {\bibfnamefont {G.~W.}\ \bibnamefont
  {Gibbons}}\ and\ \bibinfo {author} {\bibfnamefont {C.~M.}\ \bibnamefont
  {Hull}},\ }\href {\doibase 10.1016/0370-2693(82)90751-1} {\bibfield
  {journal} {\bibinfo  {journal} {Phys. Lett. B}\ }\textbf {\bibinfo {volume}
  {109}},\ \bibinfo {pages} {190} (\bibinfo {year} {1982})}\BibitemShut
  {NoStop}%
\bibitem [{\citenamefont {Hartle}\ and\ \citenamefont
  {Hawking}(1972)}]{HartleHawkingextRN}%
  \BibitemOpen
  \bibfield  {author} {\bibinfo {author} {\bibfnamefont {J.~B.}\ \bibnamefont
  {Hartle}}\ and\ \bibinfo {author} {\bibfnamefont {S.~W.}\ \bibnamefont
  {Hawking}},\ }\href {\doibase 10.1007/BF01645696} {\bibfield  {journal}
  {\bibinfo  {journal} {Commun. Math. Phys.}\ }\textbf {\bibinfo {volume}
  {26}},\ \bibinfo {pages} {87} (\bibinfo {year} {1972})}\BibitemShut {NoStop}%
\end{thebibliography}%

\end{document}